\documentclass[useAMS,usenatbib]{mn2e}

\input{psfig.sty}
\usepackage{graphicx}
\usepackage{epsfig}
\usepackage{amssymb}
\usepackage{amsmath}
\usepackage{amsfonts}
\usepackage{txfonts}
\usepackage{multirow}
\usepackage{afterpage}
\usepackage{multirow}
\usepackage{color}

\definecolor{mypink1}{rgb}{0.858, 0.188, 0.478}
\definecolor{mygreen1}{rgb}{0.258, 0.788, 0.878}
\definecolor{myorange1}{rgb}{0.5, 0.2, 0.2}

\title[Combined-Probe Simulations]{Cosmological Simulations for Combined-Probe Analyses:  Covariance and Neighbour-Exclusion Bias.}
\author[Joachim Harnois-D\'{e}raps et al.]{J. Harnois-D\'{e}raps$^{1,2}$\thanks{email: jharno@roe.ac.uk},  A. Amon$^1$, A. Choi$^3$, V. Demchenko$^1$, C. Heymans$^1$, 
\newauthor A. Kannawadi$^4$, R. Nakajima$^5$, E.  Sirks$^1$, L.  van Waerbeke$^{2}$, Yan-Chuan Cai$^1$, B. Giblin$^1$, 
\newauthor  H. Hildebrandt$^5$, H. Hoekstra$^4$, L. Miller$^6$ \&  T. Tr\"{o}ster$^{1,2}$.
\\
$^{1}$Scottish Universities Physics Alliance, Institute for Astronomy, University of Edinburgh, Blackford Hill, Scotland, UK\\
$^{2}$Department of Physics and Astronomy, University of British Columbia, V6T 1Z1, B.C., Canada\\
$^{3}$Center for Cosmology and AstroParticle Physics, The Ohio State University, 191 West Woodruff Avenue, Columbus, OH 43210, USA\\
$^{4}$Leiden Observatory, Leiden University, P.O. Box 9513, 2300RA Leiden, The Netherlands\\
$^{5}$Argelander-Institut f\"ur Astronomie, Auf dem H\"ugel 71, 53121 Bonn, Germany\\
$^{6}$Dept of Physics, University of Oxford, Denys Wilkinson Building, Keble Road, Oxford, OX1 3RH, U.K.
}

\begin{document}

\date{\today}

\pagerange{\pageref{firstpage}--\pageref{lastpage}} \pubyear{2013}

\maketitle

\label{firstpage}

\begin{abstract}
We present a public suite of weak lensing mock data, extending the Scinet Light Cone Simulations (SLICS) to simulate cross-correlation analyses with different cosmological probes.  These mocks include KiDS-450- and LSST-like lensing data, cosmic microwave background lensing maps and simulated spectroscopic surveys that emulate the GAMA, BOSS and 2dFLenS galaxy surveys. With 844 independent realisations, our mocks are optimised for combined-probe covariance estimation, which we illustrate for the case of a joint measurement involving cosmic shear, galaxy-galaxy lensing  and galaxy clustering from KiDS-450 and BOSS data.  With their high spatial resolution, the SLICS are also optimal for predicting the signal for novel lensing estimators, for the validation of analysis pipelines, and for testing a range of systematic effects such as the impact of neighbour-exclusion bias on the measured tomographic cosmic shear signal.  For surveys like KiDS and DES, where the rejection of neighbouring galaxies occurs within $\sim 2$ arcseconds, we show that the measured cosmic shear signal will be biased low, but by less than a percent on the angular scales that are typically used in cosmic shear analyses.  
The amplitude of the neighbour-exclusion bias doubles in deeper, LSST-like data.
The simulation products described in this paper are made available at http://slics.roe.ac.uk/.




\end{abstract} 



\section{Introduction}

The standard model of cosmology has been highly successful in describing a number of observations,
including fluctuations in the cosmic microwave background \citep[e.g.][]{2014JCAP...04..014D, 2016A&A...594A..13P},  and baryonic acoustic oscillations in galaxy surveys \citep[e.g.][]{2011MNRAS.415.2892B, 2012MNRAS.427.2132P, 2017MNRAS.470.2617A}. The technique of weak gravitational lensing has recently seen 
rapid progress, resulting in the early results from the Kilo Degree Survey (KiDS) and  the Dark Energy Survey (DES) presented in \citet[][H17 hereafter]{KiDS450} and \textcolor{black}{\citet{DES1_Troxel}}, respectively.
Based on the measurement of correlations between the shapes of distant galaxies that are produced by a foreground matter distribution,
the weak lensing signal is a key probe of dark matter and structure formation \citep[see][for a review]{2001PhR...340..291B}.

To reach its full potential, \textcolor{black}{this} technique must address a number of systematic effects  \citep{2013MNRAS.429..661M, 2017arXiv171003235M}, 
many of which are associated with the fact that the scales probed by the signal reside in the non-linear regime of gravitational collapse. 
Complications therefore arise due to non-linear dynamics, which generate important deviations from linear predictions and 
produce non-Gaussian features in the matter distribution that can affect likelihood analyses. Many of these challenges can be overcome with 
numerical $N$-body simulations, which accurately capture the gravitational physics over the scales relevant to the weak lensing measurement. 
These calculations are expensive to carry out and require vast resources on supercomputers, but their scientific outcome is rich
and their applications numerous and central to many aspects of weak lensing analysis:

\begin{table*}
   \centering
   \caption{ Properties of some of the weak lensing simulations that are publicly available:   $L_{\rm box}$ is the \textcolor{black}{comoving} side length of the simulation box, 
   $m_{\rm p}$ is the particle mass, $N_{\rm cosmo}$ is the number of cosmologies available, $N_{\rm sim}$ is the number of realisations, and    $A_{\rm tot}$ is the total area,
   combining all cosmologies and realisations.
   The second cosmology covered by the Millennium Simulation is obtained by post-processing 
  the  first one \citep{2010MNRAS.405..143A} hence is not independent. 
  The particle mass varied with cosmology in the DH10 simulations, while both $m_{\rm p}$ and $L_{\rm box}$ varied with redshift in the HSC simulations, in 11 steps between $z=0$ and $z=3$. 
  The 108 realisations of the HSC mocks are not fully independent: 18 light-cones are produced from each of the 6 truly independent volumes. \textcolor{black}{$N_{\rm sim}=932$ for the SLICS comic-shear and CMB lensing data, and 844 for the full set of probes.}}
   \tabcolsep=0.11cm
      \begin{tabular}{@{} lccccccc @{}} 
      \hline
     & SLICS & HSC & DH10 & CLONE & MICE-GC & Millennium \\ 
     \hline
      \multirow{2}{*}{$L_{\rm box}$ ($h^{-1}$Mpc)}&  \multirow{2}{*}{505}&450 ($z\sim0$)&  \multirow{2}{*}{140}& 231.1 ($z>2$)&\multirow{2}{*}{3072} &\multirow{2}{*}{500} \\ 
                                                   &&4950 ($z\sim3$) && 147.0 ($z<2$)& & & \\ 
                                                         \hline
   \multirow{2}{*}{$m_{\rm p}$ ($h^{-1}M_{\odot}$)}& \multirow{2}{*}{2.88$\times10^{9}$}&  8.2$\times10^8$ $(z\sim0)$& $6.51\times10^9 (\Omega_{\rm m} = 0.07)$& 8.94$\times 10^8$ ($z>2$)&  \multirow{2}{*}{2.93$\times10^{10}$} & \multirow{2}{*}{8.6$\times10^8$}\\
    & & 1.1$\times10^{12}$ $(z\sim3)$ &  $5.74 \times10^{10}(\Omega_{\rm m} = 0.62)$ & 2.30$\times10^{8}$  ($z<2$)& && \\
        \hline
     $N_{\rm cosmo}$ &3 &  1& 158 &1& 1 &2 \\
     $N_{\rm sim}$ & \textcolor{black}{844}& 108 &  192 &185& 1& 1\\
\hline
     $A_{\rm tot}$ (deg$^2$)& \textcolor{black}{8.44$\times 10^4$}& 4.45$\times 10^6$ &  6912 &2.37$\times 10^3$& 1.03$\times 10^4$&  1024 \\
            \hline
   \end{tabular}
    \label{table:Nbody}
    \end{table*}

{\it \textcolor{black}{1-}Modelling --} 
Numerical cosmological simulations are required in the modelling of weak lensing signals for which theoretical predictions are either not available or not accurate enough.
Modern prediction tools such as {\small HALOFIT} \citep{Smith03, Takahashi2012}, the Cosmic Emulator \citep{Heitmann2013}, \textcolor{black}{HMCode \citep{MeadFit}} or the Mira-Titan project \citep{MiraTitan} are all based on large  suites of $N$-body simulations in which the input cosmology parameters were varied. The science objectives were, in these cases, to construct high-precision predictions of the two-point correlation functions (for the collisionless dark matter)
that extend deep into the non-linear regime of large-scale-structure formation.
In the context of weak lensing, these serve to model the cosmic shear signal.
On top of this, there are complementary lensing measurements that are particularly sensitive to non-linear structures 
and which contain additional information, such as the lensing peak statistics \citep[\textcolor{black}{e.g.}][]{2015PhRvD..91f3507L, 2015MNRAS.450.2888L, 2016MNRAS.463.3653K, 2017arXiv170907678M},
the lensing of under- and over-dense regions \citep{2017arXiv171005162F, 2017arXiv171005045G, 2018arXiv180500562B}, or clipped lensing \citep{2016MNRAS.456..278S, Giblin18}, which in some cases completely rely on simulations to estimate the expected signal. It is worth mentioning here  that an important part of the modelling comes from the presence (or the absence) of massive neutrinos, modified gravity and baryon feedback. However, these effects are outside the scope of this paper, 
and can be dealt with separately with analytic halo models \citep[e.g.][]{2016MNRAS.459.1468M} or hydrodynamical simulations \citep[e.g.][]{Semboloni11, 2017arXiv171202411M, 2018arXiv180108559C}. 

{\it \textcolor{black}{2-}Validation of estimators --} 
In addition to their key role in the modelling of weak lensing observables, simulations can be post-processed into mock catalogues that are constructed  to match a number of  properties of the input data. In that form, the mocks serve to test, calibrate and optimise different estimation techniques, 
and can tell us how these respond to different observational effects that can be added by hand (e.g. survey masking, photometric error, PSF residuals).
Another usage is to test the sensitivity of different measurement techniques to known systematic contamination.
\textcolor{black}{This is of particular importance when developing new weak lensing estimators, e.g. clipped lensing,}
\textcolor{black}{and study their response to secondary physical effects such as source clustering.}

{\it \textcolor{black}{3-}Covariance matrix estimation --}
Weak lensing analyses are carried out on correlated data points, which means that an accurate assessment of the uncertainty on the measurements requires a full covariance matrix. 
The ideal way to measure this relies on a large ensemble 
of independent $N$-body simulations at each of the cosmologies that are being sampled along the Monte Carlo Markov Chain parameter sampler (MCMC). 
Given the requirement that the number of simulations per ensemble must significantly exceed the dimension of the data vector, 
this scenario requires computing  resources far exceeding those currently available.
Alternative techniques have been used instead to estimate the covariance matrices of weak lensing observables, 
including `internal'  estimates such as jack-knife or bootstrap resampling of the data, 
analytic calculations  \citep[see][for example]{Takada2009a, cosmolike}, lognormal realisations \citep{2011A&A...536A..85H}, or approximate gravity solvers such as ICE-COLA \citep{ICE-COLA}. 
Another approach is to run an ensemble of full $N$-body simulations at a single cosmology and ignore the variation of the covariance with cosmology. 
Hybrid techniques are also possible, where for example one can use fast  Gaussian approximations to promote a matrix with some cosmological dependence 
\citep[see][for example]{Kilbinger2013}.
Each of these techniques have pros and cons, and the best choice for a given measurement will strike a compromise that minimises the impact on the final parameter inference. 
Generally, internal estimates become inaccurate at large scales, lognormal and approximate methods do not reproduce exactly the non-linear structures,  
analytical calculations need to be validated against ensembles of $N$-body simulations to begin with, and simulation-based estimates are themselves subjected
to the missing ``Super Sample Covariance'' (SSC) term \citep{SSC}. Undoubtedly, even at a fixed cosmology, the ensemble approach offers a valuable tool to estimate covariance matrices, which is the central focus of this  paper.


{\it \textcolor{black}{4-}Likelihood modelling -- } 
Weak lensing analyses are mostly carried out under the assumption
that the underlying data are distributed according to a multivariate Gaussian function. The likelihood that describes such idealised data
can be expressed analytically, however little is known about the accuracy of this assumption. In fact, this is expected to break
in the highly non-linear regime, and there are even hints that this could already be a source of systematic error in the interpretation
of the current weak lensing data  \citep{2016MNRAS.456L.132S}. There is a need to study extensions to the current method, and numerical simulations can serve to test non-Gaussian likelihood models  \citep{2017arXiv171204923S, 2018arXiv180306348H}. {These assumptions and their numerical implementation can be tested in a full mock analysis,
where it can be verified whether the likelihood analysis can recover the input cosmology \citep{2018arXiv180309795M}.}

There is a range of  public mock data sets designed to serve the weak lensing community, each having their strengths and limitations\footnote{This is not an \textcolor{black}{exhaustive} list of all public mock weak lensing data, but instead a subset that shows the diversity of the available tools. }. 
We present a few of them here, and summarise in Table \ref{table:Nbody} some of the key properties that affect their performance 
at estimating weak lensing covariance matrices.
Ray-tracing through the  Millennium Simulation \citep{Millennium,Hilbert2009a}, for instance, has yielded a rich science outcome, however there is only one realisation (and two cosmologies). 
The MICE-GC  simulation \citep{Fosalba2013}  is particularly useful for the volume  it covers, but again there is a single realisation available.
Complementary to these are the \citet[][DH10 hereafter]{DH10} simulations, which probe 158 different cosmologies in the [$\sigma_8 - \Omega_{\rm m}$] plane, 
and additionally contain an ensemble of 37 realisations for the main cosmology. Compared to the other simulations, this large suite was constructed with smaller volumes and at a lower mass resolution. 

The CLONE catalogue \citep{Harnois-Deraps2012a} was specifically tailored for data quality assessment and covariance estimation in weak lensing data analyses of the Canada-France-Hawaii Telescope Lensing Survey \citep{Heymans2012}.
With 185 realisations, the CLONE probes very small scales, but also suffers from small volumes (the box sizes are 231 and 147 $h^{-1}$Mpc on the side, depending on the redshift) at a level that \textcolor{black}{is now inadequate for the current generation of lensing surveys}. 

An ensemble of 108  full-sky weak lensing mock data has also been produced by \citet{HSCmocks}
and made publicly available, combined with a release of dark matter halo catalogues and CMB lensing maps. 
These simulation products are designed for the Hyper Suprime Camera (HSC) weak lensing survey, 
but can serve broader science cases. 
Being full sky, these `HSC' mocks are well suited to test estimators acting on 
spherical coordinates, such as curved-sky map reconstruction algorithms. 
While there are 108 realisations in the release, 
these mocks are not statistically independent, having `recycled' a smaller number of truly independent $N$-body realisations. 
It has been shown that such recycling has little impact on the cosmic shear covariance matrix \citep{Petri16},
however its effect on higher order statistics and likelihood modelling is still unknown. 
 The finite mass resolution of these simulations can be limiting for some applications, since the  minimal halo mass that they form gradually varies from  $1\times10^{12}~h^{-1}M_{\odot}$ at $z\sim0.3$ to  $5\times10^{13}~h^{-1}M_{\odot}$ haloes at $z\sim3$ (see their figure 3). 
This is insufficient to describe many galaxy populations that reside in less massive systems, but can serve to model low-redshift luminous red galaxies (LRG), which are hosted in $1\times10^{12}~h^{-1}M_{\odot}$ haloes (see Section \ref{subsubsec:CMASS} and Fig. \ref{fig:MassFunction}). 
According to these limitations, a  $z\sim0.7$ LRG sample based on these HSC mocks would be missing its least massive members.
 However, their large volumes make these HSC mocks particularly suitable for the evaluation of the SSC term.

The SLICS  \citep[Scinet LIght Cone Simulations, described in][HvW hereafter]{2015MNRAS.450.2857H}  were designed as a massive upgrade of the CLONE.
With a volume of $L_{\rm box}$=505 $h^{-1}$Mpc on the side, they significantly reduce the limitations caused by the finite box size, thereby allowing data analyses that include larger angular scales (\textcolor{black}{the cosmic shear signal is valid out to 2 degrees}, as opposed to about half a degree in the CLONE). 
They resolve structure deep within the non-linear regime, and the large  ensemble size supports longer data vectors without introducing high levels of noise in the covariance matrix. 
The SLICS were first tailored for the RCS Lensing Survey \citep{RCSLenS}, and later reprocessed for the  cosmic shear analysis presented in H17, which is based on the first 450 square degrees of the KiDS data. This flexibility is one of the highlights of numerical simulations: once the lensing data have been 
computed and stored on disk, it is relatively inexpensive to reproduce the properties of many different surveys.


This paper presents a significant expansion of the SLICS  suite from its original version,  
%
with a focus on cross-correlation science.
On top of the weak lensing mass and shear planes introduced in HvW15, we present here the KiDS-450- and the LSST-like `source' catalogues, which emulate the two photometric surveys they are named after.  We also describe the backbone dark matter halo catalogues as well   as three  mock  `lens' galaxy catalogues that reproduce properties of the CMASS and LOWZ LRG samples \citep{2016MNRAS.455.1553R} that are part of the Baryon Oscillation Spectroscopic Survey (BOSS), and 
 the denser galaxy sample from the Galaxy And Mass Assembly  spectroscopic survey \citep[][GAMA hereafter]{GAMA}. We construct an additional set of galaxy catalogues (KiDS-HOD and LSST-like HOD) specially designed to study systematic and selection effects related to source-lens coupling \citep{2011A&A...528A..51H, 2015ApJ...803...46Y}, 
and finally supplement the light cones  with simulated lensing maps of the Cosmic Microwave Background (CMB).
As a direct application, we construct a combined-probe data vector that incorporates cosmic shear, 
galaxy-galaxy lensing and galaxy clustering and \textcolor{black}{present} the full covariance matrix.  

Many of these simulation products already served in cosmological analyses: the cross-correlation of weak lensing  with {\it Planck} lensing \citep{2016MNRAS.460..434H, 2017MNRAS.471.1619H},  cosmic shear (H17),  
peak statistics  \citep{2017arXiv170907678M},  combined-probe analyses with redshift-space distortions \citep{2017arXiv170706627J, Amon2017}
and galaxy clustering \citep{2017arXiv170605004V}, clipped lensing \citep{Giblin18} and  density-split statistics  \citep{2018arXiv180500562B}. 
The first part of this paper therefore serves as a reference for those interested in the different SLICS products, 
where we detail their design, performance and limitations.


%
%
%

In the second part of this paper, we revisit the {\it neighbour-exclusion bias},  a subtle selection effect first reported in \citet{2011A&A...528A..51H}
and revisited by \citet{DES-SV_MacCrann}, sourced from the fact that {objects with close neighbours are more common in regions with foreground clusters than with foreground voids.
Positions and shapes are more difficult to extract for these objects, hence they are typically rejected or down-weighted in weak lensing analyses.}
This selection therefore preferentially down-samples regions with the highest density of  foreground galaxies, which also correspond to regions that yield the  highest lensing signal. This is a form  of source-lens coupling unrelated to the photometric uncertainty or contamination by cluster members, 
and which affects the cosmic shear signal over a wide range of scales. 
We first investigate this neighbour-exclusion bias in the context of a weak lensing survey at  KiDS depth,
including tomographic decomposition, different levels of {close-pairs exclusion}, and two different strategies to deal with them, 
then extend this measurement to LSST depth.

This paper is structured as follow. We review   the configuration of the N-body runs, our strategy to extract lensing maps and dark matter halos in Section \ref{sec:darkmatter}. We then describe our different galaxy catalogues in  Section \ref{sec:galaxies}, 
we list the caveats and limits that are known to affect the numerical products, 
and conclude the first part of this paper by presenting the combined-probe covariance matrix in  Section \ref{sec:combined_probe}.
We next investigate the neighbour-exclusion bias in Section \ref{subsec:close_pairs}, and conclude in Section \ref{sec:conclusion}.
 We finally present complementary information about some of the mock products in the Appendices.

\section{Dark Matter Light Cones}
\label{sec:darkmatter}

 \begin{table*}
   \centering
   \caption{Lens and source redshift planes used to construct our past light cones. These are obtained by stacking half boxes, each $252.5 ~h^{-1}\mbox{Mpc}$ thick, 
   from the observer out to $z_{\rm max} \sim 3.0$. The lens planes lie at the centre of the projected volumes, and the `natural' source planes correspond to  the back of each half box.}
   \tabcolsep=0.11cm
      \begin{tabular}{@{} lllllllllllllllllll @{}} 
      \hline
    $z_{\rm l}$ & 0.042 &  0.130 &   0.221    &0.317 &  0.418 &   0.525    &0.640 &  0.764    &0.897  &  1.041  &  1.199    &1.373  &  1.562  &  1.772  &   2.007   & 2.269 &   2.565    &2.899\\
  \hline
  $z_{\rm s}$& 0.086  &  0.175  & 0.268 &   0.366   & 0.471    &0.582  &  0.701   & 0.829  &  0.968    &1.118   & 1.283  &  1.464   & 1.664 &   1.886   &  2.134 &   2.412   & 2.727 &   3.084\\
 \hline
   \end{tabular}
    \label{table:redshifts}
    \end{table*}

\subsection{The $N$-body calculations}

The SLICS are based on a series of  $1025$ $N$-body simulations produced by the high performance gravity solver  {\small CUBEP$^3$M} \citep{2013MNRAS.436..540H}. They were first presented in HvW15, and we report here some of the key properties.
The fiducial cosmology adopts the best fit  WMAP9 + BAO + SN parameters \citep{2013ApJS..208...19H}, namely:
$\Omega_{\rm m} = 0.2905$, $\Omega_\Lambda = 0.7095$, $\Omega_{\rm b} = 0.0473$, $h = 0.6898$, $\sigma_8 = 0.826$ and $n_{\rm s} = 0.969$.
\textcolor{black}{This choice lies close to the mid-point between the cosmic shear and the {\it Planck} best-fit values in the $[\sigma_8 - \Omega_{\rm m}]$ plane.}
Each run follows $1536^3$ particles inside a grid cube of \textcolor{black}{comoving side length}  $L_{\rm box} = 505 ~h^{-1}\mbox{Mpc}$ and nc = 3072 grid cells on the side, starting from a set of initial conditions at $z_i = 120$ obtained via the Zel'dovich approximation. The $N$-body code computes  the non-linear evolution of these collisionless particles down to $z =0$ and generates on-the-fly  the halo catalogues and mass sheets required for a full light cone construction 
(see Sections \ref{subsec:planes} and \ref{subsec:haloes}). By construction, this setup  makes no distinction between baryons and dark matter, and ignores the impact of  massive neutrinos. 

The complete SLICS series consists of a core `Large Ensemble' (the SLICS-LE suite) of 932 fully independent realisations, 
augmented with 5 runs in  which the gravitational force is resolved to smaller scales \citep[with the extended particle-particle mode described in][]{2013MNRAS.436..540H}.
These extra runs make up the SLICS-HR suite, which served for convergence tests of the SLICS-LE.
We also produced an additional 73 runs at $\sigma_8=0.861$, 
and 15 with $\sigma_8 =0.817$ and $n_{\rm s} = 0.960$.
Although restricted in their sampling of the parameter space, these runs enable some sensitivity tests to differences in  cosmology.
This paper solely focusses on the development of simulation products performed in the large ensemble, which we hereafter refer to as the `SLICS simulations'. 

Each of the SLICS realisations required 64 {\small MPI} processes, each running either 8  or 16 {\small CPU}s in a {\small OPENMP} parallelisation mode, for a total of 512-1024 cores depending on the machines. The real runtime to reach $z=0$ on the Compute Canada SciNet-GPC and Westgrid-Orcinus clusters 
 (intel x86 processors) was about 30 hours per simulation, depending on the architecture, on the network usage,
and on the level of non-linear structures formed inside the cosmological volume. {\small CUBEP$^3$M} does not explicitly enforce load balance across the compute nodes, hence a super-structure forming inside one node will require more time to resolve, effectively slowing down all nodes. With  six phase-space elements per particle at 4 bytes each, a single particle dump takes up 87Gb of disk space. Given our need for multiple redshift checkpoints for over 1000 realisations, storing the particle data was not an option. Once halo catalogues and mass sheets were generated, the particles were deleted (with the exception of the SLICS-HR suite,
for which the particle data will be made available upon request).

The particle mass is set to $2.88\times10^{9} ~h^{-1}M_{\odot}$, thereby resolving dark matter haloes below $10^{11} ~h^{-1}M_{\odot}$
and structure formation deep in the non-linear regime. 
The three-dimensional dark matter power spectrum, $P(k)$,  agrees within 2\% with the  SLICS-HR 
as well as with the predictions from the Extended Cosmic Emulator \citep{Heitmann2013} for Fourier modes  $k<2.0 ~h{\rm Mpc}^{-1}$ (figure 6 of HvW15).
 Higher $k$-modes (corresponding to smaller scales) are affected by finite force/mass resolution,
such that at $k=5.0$ $(10.0) ~h{\rm Mpc}^{-1}$, the simulated $P(k)$ from the SLICS is 15\%  (50\%) lower than the emulator, \textcolor{black}{which achieves 5\% precision up to $k=10 ~h$Mpc$^{-1}$}.
This resolution limit inevitably propagates into the light cone, which then also impacts the projected measurements such as the  
shear two-point correlation function or the convergence power spectrum (see figures 1 and 7 in HvW15). 
\textcolor{black}{As always, mass resolution needs to be considered when deciding on the scales at which the cosmic shear results from SLICS are reliable;
this is further discussed in HvW15 and in Sec. \ref{subsec:V0}.} 





\subsection{Gravitational lenses}
\label{subsec:planes}

We construct flat sky weak lensing maps with the multiple-plane tiling technique \citep[in many aspects similar to][]{Vale2003a},
in which convergence and shear maps are extracted from a series of \textcolor{black}{18} mass sheets under the Born approximation. 
When the simulation reaches pre-selected \textcolor{black}{lens redshifts, $z_{\rm l}$}, the particles from half the cosmological volume are projected along the shorter dimension on two-dimensional grids of $12,288^2$ pixels  following a `cloud in cell' (CIC) interpolation scheme \citep{1981csup.book.....H}. This process is repeated for the three Cartesian axes, 
however we keep on  disk only one of these mass planes per redshift following a regular sequence (e.g. $xy$, $xz$, $yz$, $xy$...). 
The redshifts of these planes, reported as $z_{\rm l}$ in Table \ref{table:redshifts},  are chosen such that the half volumes continuously fill the space from $z=0$ to $z=3$. This requires  18 planes in the adopted cosmology.
Starting from the observer at $z=0$, the first mass plane corresponds to the projection of the comoving volume in the range [$0$ -- $252.5\ h^{-1}\mbox{Mpc}$],
which we assign to its centre (at $126.25  ~h^{-1}\mbox{Mpc}$, or $z_{\rm l} = 0.042$); the second plane projects the volume  [$252.5$ -- $505\ h^{-1}\mbox{Mpc}$],
 also assigned to its centre (at $378.75\ h^{-1}\mbox{Mpc}$, or $z_{\rm l} = 0.130$), and so on for all 18 planes. We turn these density maps into over-density maps
 by subtracting off the mean.

We carve out our light cones\footnote{Note that the setup described here has changed since HvW15,
in which the light cones had an opening angle  of $60$ deg$^2$ with $6000^2$ pixels. } by shooting rays on a regular grid of $7745^2$ pixels with an opening angle  of $100$ deg$^2$, 
which corresponds to the angular extension of the simulation box at redshift $z = 1.36$.
We extend the light cones up to $z=3$ by using periodic boundary conditions to fill in regions of the mass sheets that fall outside the volume. 
The light cone over-density mass maps, which we label $\delta_{\rm 2D}({\boldsymbol \theta}, z_{\rm l})$,  are obtained from a linear interpolation of the mass over-density sheets 
onto the mock pixels ${\boldsymbol \theta}$ after randomly shifting the origins. This translation, together with the sequential change of the projection axis mentioned above,
are designed to minimize the repetition of structure across redshift when constructing a light cone from a single $N$-body run.

\begin{figure*}
\begin{center}
\includegraphics[width=7.0in]{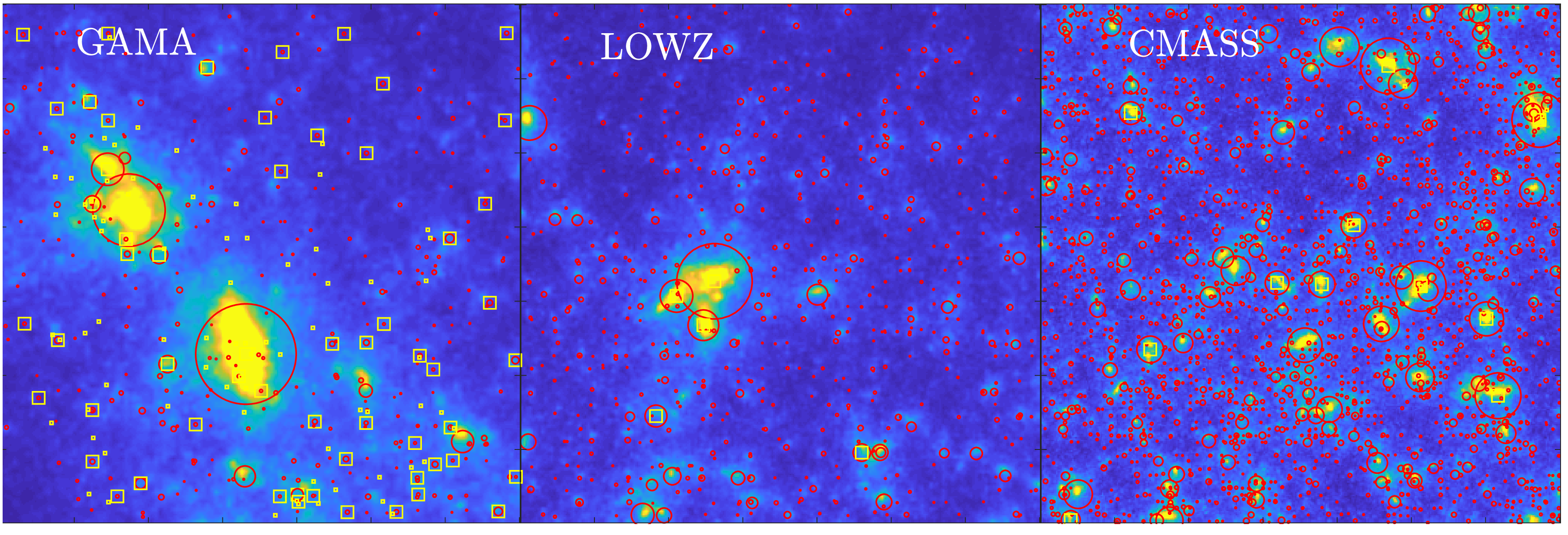}
\caption{Sample of the different simulation products presented in this paper. The background color maps represent 256 $h^{-1}$Mpc of projected dark matter,
the red circles show the dark matter haloes with sizes scaling with  their mass, the large and small yellow squares show the central and satellite galaxies, respectively. 
The left panels shows the GAMA galaxies
centred at redshift $z=0.221$, the central panel shows the LOWZ galaxies centred at $z=0.317$, while the right panel shows the CMASS galaxies, centred at $z = 0.640$. \textcolor{black}{These three mock galaxy samples are described in Secs. \ref{subsubsec:CMASS} -  \ref{subsubsec:GAMA}}.  
The side length of the three panels each subtend half a degree. }
\label{fig:delta_map}
\end{center}
\end{figure*}

Samples of these mass over-density maps are presented in Fig. \ref{fig:delta_map}.
One direct consequence of this procedure is that correlations in the matter field are explicitly broken between boxes.
This is important to note when measuring three-dimensional quantities within the SLICS light cones.
 
Given a discrete set of thin lenses at comoving distance $\chi_{\rm l}$ and a discrete source distribution $n(z)$ given in bins of width $\Delta \chi_{\rm s}$, 
we construct convergence maps $\kappa({\boldsymbol \theta})$ from a weighted sum over the mass planes (equation 6 in HvW15):
\begin{eqnarray}
 \kappa( {\boldsymbol \theta} ) = \frac{3 H_{0}^{2} \Omega_{\rm m}}{2 c^2} \sum_{\chi_{\rm l}=0}^{\chi_{\rm H}} \delta_{\rm 2D}({\boldsymbol \theta},\chi_{\rm l}) (1 + z_{\rm l})  \chi_{\rm l}
         \bigg[\sum_{\chi_{\rm s} = \chi_{\rm l}}^{\chi_{\rm H}} n(\chi_{\rm s})\frac{\chi_{\rm s} - \chi_{\rm l}}{\chi_{\rm s}} {\Delta}\chi_{\rm s}  \bigg] \Delta \chi_{\rm l},          \label{eq:kappa_disc}
\end{eqnarray}
where $\chi_{\rm H}$ is the comoving distance to the horizon, $H_0$ is the value of the Hubble parameter \textcolor{black}{today}, $c$ is the speed of light,
$n(\chi) = n(z){\rm d}\chi / {\rm d}z$ and $\Delta \chi_{\rm l} =  L_{\rm box}/{\rm nc} $.
Each of the lens redshifts is associated with a `natural' source redshift $z_{\rm s}$ that corresponds to an infinitely thin plane located just behind the half box, also listed in Table \ref{table:redshifts}.
We take advantage of the fact that  these require no interpolation along the redshift direction
and construct 18 convergence maps per light cone, assuming $n(z) = \delta(z - z_{\rm s})$.
For each of these natural source redshift planes, we also compute shear maps $\gamma_{1,2}(\boldsymbol \theta)$ with fast Fourier transforms \citep[see][for details on our numerical implementation]{Harnois-Deraps2012a}.
These lensing maps are described in HvW15,
where one can find a comparison between different prediction models for the matter power spectrum (figure 6 therein) and shear two-point correlation functions (figure 1); we refer the reader to this paper for more details about such comparisons. It is also shown therein that the variance of lensing observables converges with the Gaussian predictions at large angular scales, which reinforces our confidence that residual correlations between different mass sheets from the same light-cone can be safely ignored.

\subsubsection{CMB lensing maps}

For each of the light cones, we also produced convergence maps that extend to $z_{\rm s} = 1100$, which were described and used in 
\citet{2016MNRAS.460..434H} for the validation of combined-probe measurement techniques involving cosmic microwave background lensing data. These $\kappa_{\rm CMB}$ maps were constructed in a hybrid scheme: a single set of 10 mass planes were generated from linear theory to fill the volume between $3.0 < z < 1100$. They were first smoothed to reduce shot noise, then placed at the back end of each of  the 
main SLICS light cones, enabling ray-tracing up to the CMB for all lines-of-sight.

The fact that the same back-end volume is used for each of the $\kappa_{\rm CMB}$ maps  effectively couples the maps across different lines-of-sight, 
which means that the covariance matrix of the auto-spectrum (or auto correlation function) of these $\kappa_{\rm CMB}$ maps will be wrong.
However, these maps are primarily constructed for the study of combined probes, hence any cross-correlation measurement with $z<3.0$  mock data
will only see the main SLICS light cone hence the covariance will not be affected by this. 

We additionally produced a series of $\kappa_{\rm CMB}$ maps that reproduce the {\it Planck} lensing measurements, which we obtained by
adding noise maps with the noise spectrum given by in the data release\footnote{{\it Planck} lensing package: pla.esac.esa.int/pla/\#cosmology},
followed by a Fourier filtering procedure that removes the $\ell > 2048$ modes, as in the data \citep{PlanckCollaboration2016}.
These maps are constructed with the same foreground matter fields hence can serve for estimator validation and covariance estimation in cross-correlation
analyses involving the {\it Planck} lensing  data. 


\subsubsection{Data products: lensing maps}
\label{subsubsec:data:maps}

For all \textcolor{black}{932} light cones, we provide the following lensing maps:

\begin{itemize}
\item $\delta_{\rm 2D}(\chi_{\rm l},{\boldsymbol \theta})$ for the 18 lens planes ($z_{\rm l}$) listed in  Table \ref{table:redshifts}
\item $\gamma_{1,2}(\boldsymbol \theta)$ for the 18 source planes ($z_{\rm s}$) listed in  Table \ref{table:redshifts}
\item Noise-free $\kappa_{\rm CMB}(\boldsymbol \theta)$ convergence maps 
\item {\it Planck}-like $\kappa_{\rm CMB}(\boldsymbol \theta)$ convergence maps 
\end{itemize}
These are all flat-sky, 100 deg$^2$ maps with $7745^2$ pixels, stored in FITS format. 
The mass maps can be used to re-create convergence and shear maps with any redshift distribution if needed, 
while the shear maps can be populated with a galaxy catalogue of arbitrary  $n(z)$ in the range [0.0, 3.0] and used to assign shear to each object.


\subsection{Dark matter halo catalogues}
\label{subsec:haloes}

\begin{figure}
\begin{center}
\includegraphics[width=3.2in]{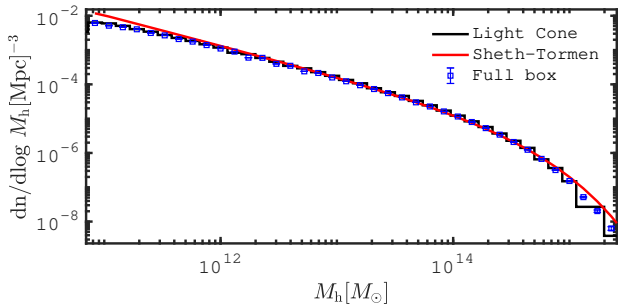}
\caption{\textcolor{black}{The halo mass function at $z=0.22$ in the full simulation box and in the light cone, compared to predictions from \citet{sheth2001a}. 
Error bars show the error on the mean, obtained from 100 lines-of-sight. The agreement is similar at other redshifts.}}
\label{fig:dndm}
\end{center}
\end{figure}

\begin{figure}
\begin{center}
\includegraphics[width=3.2in]{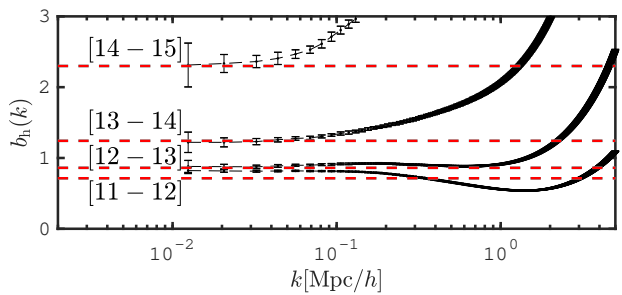}
\caption{Halo bias  in the mocks for redshift  $z=0.042$ in four wide mass bins,
labelled in the figure in units of $M_{\odot}$. 
Poisson shot noise is not subtracted, and the error is on the mean, estimated from 100 realizations. 
 Shown with the red dashed lines are the linear bias predictions from \citet{Tinker2010a}. 
}
\label{fig:bias}
\end{center}
\end{figure}

\begin{figure}
\begin{center}
\includegraphics[width=3.2in]{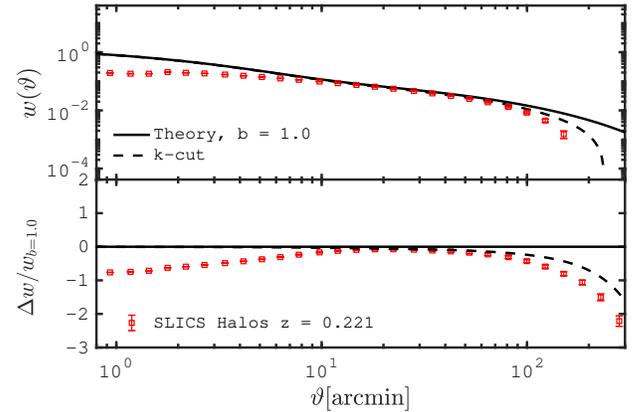}
\caption{({\it upper:}) Angular correlation function measured from all haloes combined in the range $z \in [0.175  - 0.268]$, compared with non-linear predictions with $b_{\rm h} = 1.0$. 
The dashed curve includes a cut in $k$-modes larger than the simulation box from the SLICS.
The errors bars show the error on the mean, obtained here from 100 realisations.
({\it lower:}) Fractional error with respect to the predictions without the cut in $k$-modes.}
\label{fig:w_theta_halo}
\end{center}
\end{figure}

Dark matter haloes serve as the skeleton for the galaxy population algorithms used in this paper (Sections \ref{subsec:V2}-\ref{subsubsec:KiDS}), 
hence we document their key properties in this section. We identify haloes using a spherical over-density algorithm  \citep[detailed in][]{2013MNRAS.436..540H},
which first assigns particles onto the fine simulation grid, then looks for maxima and ranks them in descending order according to their peak height.
The halo finder then grows a series of spherical shells over each maximum until the total over-density (with respect to the cosmological background) falls under the threshold of 178.0,
in accordance with the top-hat spherical collapse model. Particles within the collapse radius are then re-examined in order to extract a number of halo properties, 
including the halo mass, the position of its centre-of-mass and of its peak, the velocity dispersion for all three dimensions, 
its angular momentum and inertia matrix. 
We reject haloes with  less than 20 particles, which introduces a low-mass cut-off in the reconstructed halo catalogue at 
$M_{\rm h,min} = 5.76\times10^{10} ~h^{-1}M_{\odot}$. In this process, particles cannot contribute to more than one halo.

The mass function of these haloes reproduces the results expected from predictions by \citet{sheth2001a}, as shown in Fig. \ref{fig:dndm}. 
We also show in Fig. \ref{fig:bias} the  halo bias $b_{\rm h}$ at $z=0.042$ for 4 mass bins. 
This quantity was extracted from the simulation by computing the power spectrum of the halo catalogues, 
$P_{\rm halo}(M_{\rm h},k,z) = \langle |\delta_{{\rm halo}, M_{\rm h}}(k,z)|^2\rangle$, and that of the particle data,  
$P(k,z) = \langle |\delta(k,z)|^2\rangle$. The halo density $\delta_{{\rm halo}, M_{\rm h}}(x,z)$ is constructed by placing haloes in mass bin $M_{\rm h}$ and redshift $z$ on a $3072^3$ grid, which is Fourier transformed, squared, and angle-averaged to obtain the halo power spectrum. We repeat this procedure with the full particle data to obtain $\delta(k,z)$ and $P(k,z)$, and extract 
the bias via the relation $b^2_{\rm h}(M_{\rm h},z,k) = P_{\rm halo}(k,z,M_{\rm h})/P(k,z)$. 
Note that this numerical computation provides only the two-halo term contribution to the power spectrum, which is enough to estimate the linear bias.
The one-halo term would require sub-halo catalogues, which we have not constructed.
In this calculation, the particle and halo \textcolor{black}{mass assignment scheme was corrected for by dividing the power spectra by the window function \citep{1981csup.book.....H}, 
but the shot noise was not subtracted. }

Looking at the linear regime ($k<0.05 ~h$Mpc$^{-1}$), we clearly see that the most massive haloes are the highest-biased tracers of the underlying dark matter field, 
and that  haloes in the mass range $[10^{11} - 10^{13}] ~h^{-1}M_{\odot}$ have a bias lower than 1.0. 
Our measurements are in excellent agreement  with the predictions from the spherical collapse model of \citet{Tinker2010a}  
for  the largest three mass bins plotted in Fig. \ref{fig:bias},
however the  $[10^{11} - 10^{12}] ~M_{\odot}$ haloes exhibit a bias that is 14\% higher than the predictions, 
($b_{\rm h} =0.82$ in the mocks, compared to the predicted value of 0.72). 
The size of these deviation is similar to the differences between linear bias models \citep[e.g.][]{MoWhite1996, sheth2001a, ShethTormen2001} 
which means that our  halo clustering agrees well with the models within the theoretical accuracy.
The linear bias approximation holds well at large scales ($k<0.1 ~h{\rm Mpc}^{-1}$ for haloes with $M_{\rm h}<10^{14} M_{\odot}$,
 smaller $k$-modes for heavier haloes).
The bias $b_{\rm h}(k)$ in all mass bins deviates from the horizontal at $k>0.2 ~h{\rm Mpc}^{-1}$, in part because of the  shot noise, 
in part because of the non-linear bias (which we do not attempt to model in this paper).
We note, however, that the shape of the non-linear bias heavily depends on the halo mass: 
whereas the bias of haloes with $M_{\rm h} > 10^{12} ~h^{-1}M_{\odot}$ is flat at large scales then exhibits a sharp increase at high $k$-modes,
the bias of lighter haloes first drops  between $k= 0.2$ and $2.0 ~h{\rm Mpc}^{-1}$, then follows a steep ascent at higher $k$.
Similar shapes and mass dependences of the non-linear bias were recently reported in  \citet{2017arXiv171102677S}.

The requirement we have for producing a large ensemble of simulations comes at a cost, 
such that some key ingredients often found in other recent halo catalogues are omitted here.
For instance, and as mentioned previously, there is no sub-halo information  available, and since the particle data are not stored, these catalogues cannot be further improved 
with a more sophisticated halo finder.
In addition, merger trees were not generated,  which limits the use of semi-analytic algorithms to populate these haloes with galaxies.
Finally, there is no phase-space cleaning included in the halo-finding routine, which reduces the accuracy of the inertia matrix and angular momentum
measured from these haloes. These limitations have a negligible impact on cosmic shear measurements based on 
these mocks, but may affect some analyses that rely on these properties, for example implementing 
intrinsic galaxy alignments or studying environmental dependences. 

 We show in Fig. \ref{fig:w_theta_halo}  the angular correlation function  of the light cone haloes in the redshift range 
$0.175 < z < 0.268$, 
measured with the   \citet{1993ApJ...412...64L} estimator: 
\begin{eqnarray}
w(\vartheta) = \frac{\rm DD - 2DR + RR}{\rm RR},
\label{eq:w_theta}
\end{eqnarray}
where DD, RR and DR refer to the pair counts of the data-data, random-random and data-random, respectively, as a function of separation angle $\vartheta$. These quantities are measured with {\small TREECORR} \citep{TreeCorr} and split in 50 logarithmically-spaced bins spanning $0.01 < \vartheta < 300$ arcmin.
Shown in red is the clustering measurement obtained from 100 lines-of-sight, compared with theoretical predictions obtained from {\small CosmoSIS}\footnote{CosmoSIS: https://bitbucket.org/joezuntz/cosmosis/wiki/Home} \citep{CosmoSIS} 
with a bias of  $b_{\rm h}=1.0$ and the SLICS input cosmology. 
Throughout this paper, all clustering measurements are extracted from the same number of independent realisations ($N_{\rm sim}=100$).
This number was chosen because it is large enough to provide accurate estimates of the signals in the full sample,  
while the error bars in the figures remain visible and useful.
We show the errors on the mean (i.e. the $1\sigma$ scatter between the measurements, divided by $\sqrt{100}$)
in order to highlight the small residual discrepancies with the predictions.

As seen in Fig. \ref{fig:w_theta_halo}, the linear bias for this sample of haloes is on average close to 1.0 for $\vartheta > 10$ arcmin, 
but the measured amplitude undershoots this constant bias model at smaller separations.
This drop is caused by the fact that a large fraction of this sample consists of haloes with mass $M_{\rm h} < 10^{12} ~h^{-1}M_{\odot}$, as seen from the mass function in Fig. \ref{fig:dndm}, and the non-linear bias of this same sample decreases towards small scales (or towards high-$k$, see Fig. \ref{fig:bias}).
The sharp increases seen in the halo bias at very high $k$-modes is not seen in $w(\vartheta)$ since it mostly consists of shot noise.
A full mass-dependent, redshift-dependent, non-linear bias model would be required to improve the match between theory and measurements
in Fig.  \ref{fig:w_theta_halo},
which is beyond the scope of this paper.
The dashed black curve  shows the theoretical prediction for $w(\vartheta)$ after the theory matter power spectrum has been set to zero for $k$-modes probing scales larger than 
the simulation box. This resembles the finite-box effect observed in $w(\vartheta)$ beyond 100 arcmin,
although the match is not perfect. 
For this measurement to be accurate, it is critical to construct random catalogues that properly capture the properties 
of the survey in absence of clustering, mainly its depth and mask. We discuss this further in the context of our light cone geometry in Section \ref{subsubsec:randoms}.

Note that these halo catalogues serve as the input in the construction of galaxy catalogues based on Halo Occupation Distributions (HOD), which we describe in Section \ref{subsec:V2}. 

\subsubsection{Data products: halo catalogues}
\label{subsubsec:data:halos}

For each dark matter halo, we store, in FITS format:
the position of the halo, the pixel it corresponds to in the lensing maps, 
the mass, the centre-of-mass velocity, the  velocity dispersion, the angular momentum, the inertia matrix, and the rank\footnote{This halo property refers to its rank in a mass-ordered halo catalogue, where the lowest rank corresponds to the most massive halo.}
within the full volume simulation  (i.e. before extracting the light cone).
The catalogues of haloes that populate each of the light cones will be made available upon request.

We note here that the haloes are not available for all simulations, notably due to an unfortunate disk failure that caused a loss of many catalogues.
For this reasons, the haloes and HOD galaxies are available for 844 lines -of-sight out of the 932 for which we have mass and shear planes. 


\section{Mock Galaxy catalogues}
\label{sec:galaxies}

The mock data described in this paper have already found a number of applications in the analysis of large-scale structure and/or weak lensing data,
which required fine preparation of the simulation products. 
To achieve this, we use different techniques to add galaxies  in the light cones, tailored to different science targets. 
In particular we:
\begin{enumerate}
\item{enforce a redshift distribution of source galaxies $n(z)$ and a number density $n_{\rm gal}$ that matches the KiDS-450 data, with galaxies put at random positions in the light cone.
This represents our baseline mock `source' galaxy sample in this paper, as it is designed to estimate covariance matrices for cosmic shear analyses with KiDS-450 data. 
We also produce a second version with a higher galaxy density, 
and a third version, this time with LSST-like densities and $n(z)$.
Details are provided in Section \ref{subsec:V0} and Appendix \ref{subsubsec:LSST} respectively;}
\item{generate galaxy positions, $n(z)$ and $n_{\rm gal}$ from  HOD prescriptions. 
This is our main strategy to generate mock galaxies matching different spectroscopic surveys (i.e. CMASS, LOWZ,  GAMA),
used as `lens' targets in combined-probe measurements.
We also generate two additional HOD-based mock surveys, at KiDS and LSST depth, including lensing and photometric information. 
These are described in Sections \ref{subsec:V2} - \ref{subsubsec:LSST-HOD};}
\item{generate another lensing source galaxy catalogue based on  (i) but placing galaxies at positions chosen such as to produce a galaxy density field with a known bias,
which is theoretically simpler to model than the HOD catalogues from (ii).
This can be particularly useful when one needs to include simple  source clustering, or test linear bias models as in \citet{2017arXiv170605004V}.
In particular, it requires a sampling of the mass sheets $\delta_{\rm 2D}(\chi_{\rm l},{\boldsymbol \theta})$, as detailed in Appendix \ref{subsec:V04}. 
These mocks are not a part of the release, but we provide the code to reproduce these catalogues from the shear and mass maps; }
\item{place mock galaxies at the positions of observed galaxies in the KiDS-450 survey. 
This naturally enforces the $n(z)$ and spatially varying $n_{\rm gal}$ of the data,
 which are required  for analyses that are sensitive to these properties, including 
the peak statistics analysis of \citet{2017arXiv170907678M}. See Appendix \ref{subsec:V1} for more details.}
\end{enumerate}
This is not an exhaustive list of all possibilities, but covers many of the commonly used galaxy inpainting techniques.
The following sections describe the main strategies -- (i) and (ii) from the list above -- by which source and lens galaxies 
are assigned to our  simulations. 

\subsection{Mock KiDS-450 source galaxies}
\label{subsec:V0}

In this method, galaxies are placed at random angular coordinates on the 100 deg$^2$ light cone, with number density and redshift distribution matching a pre-specified $n_{\rm gal}$ and $n(z)$. This method is general and can be used to emulate any weak lensing survey.
We show here an application of this technique to the KiDS-450 data described in H17, and present in  Appendix \ref{subsubsec:LSST} a similar emulation for a LSST-like lensing survey that follows the specifications listed in \citet{LSSTgal}.
 
\begin{figure}
\begin{center}
\hspace{-5mm}
\includegraphics[width=2.85in]{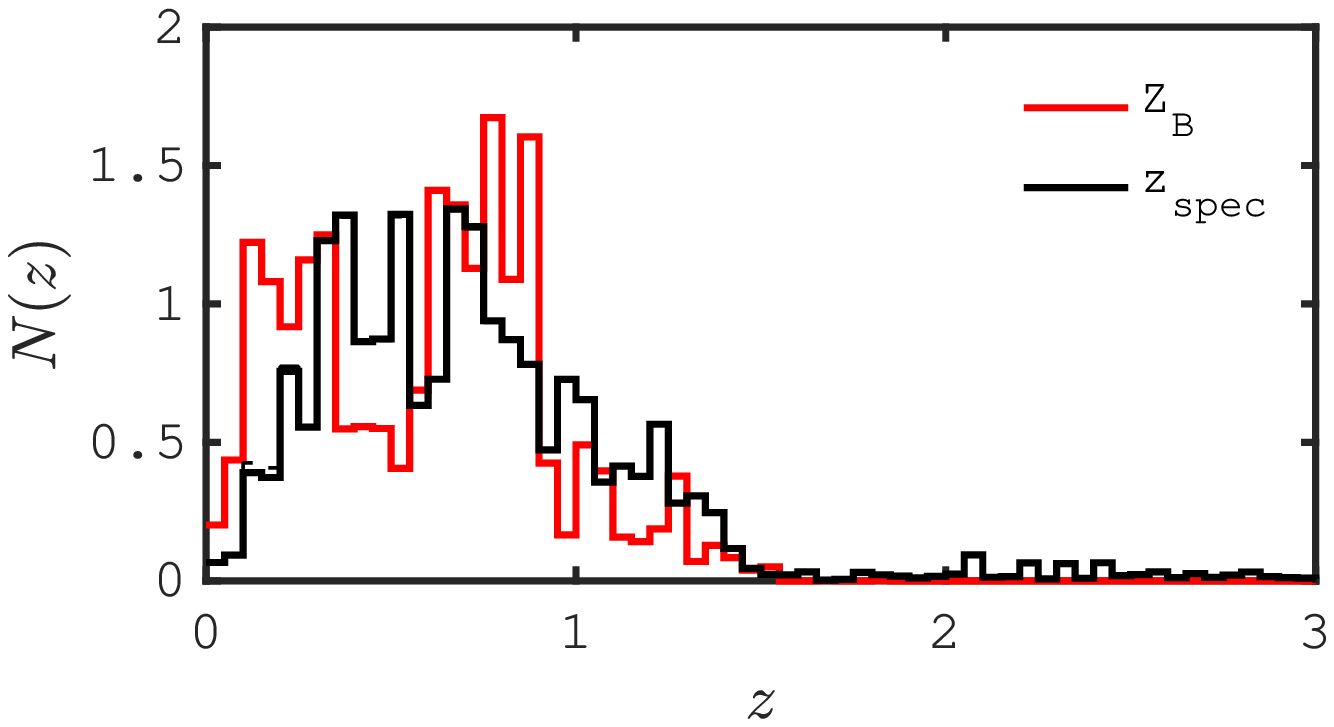}
\includegraphics[width=2.8in]{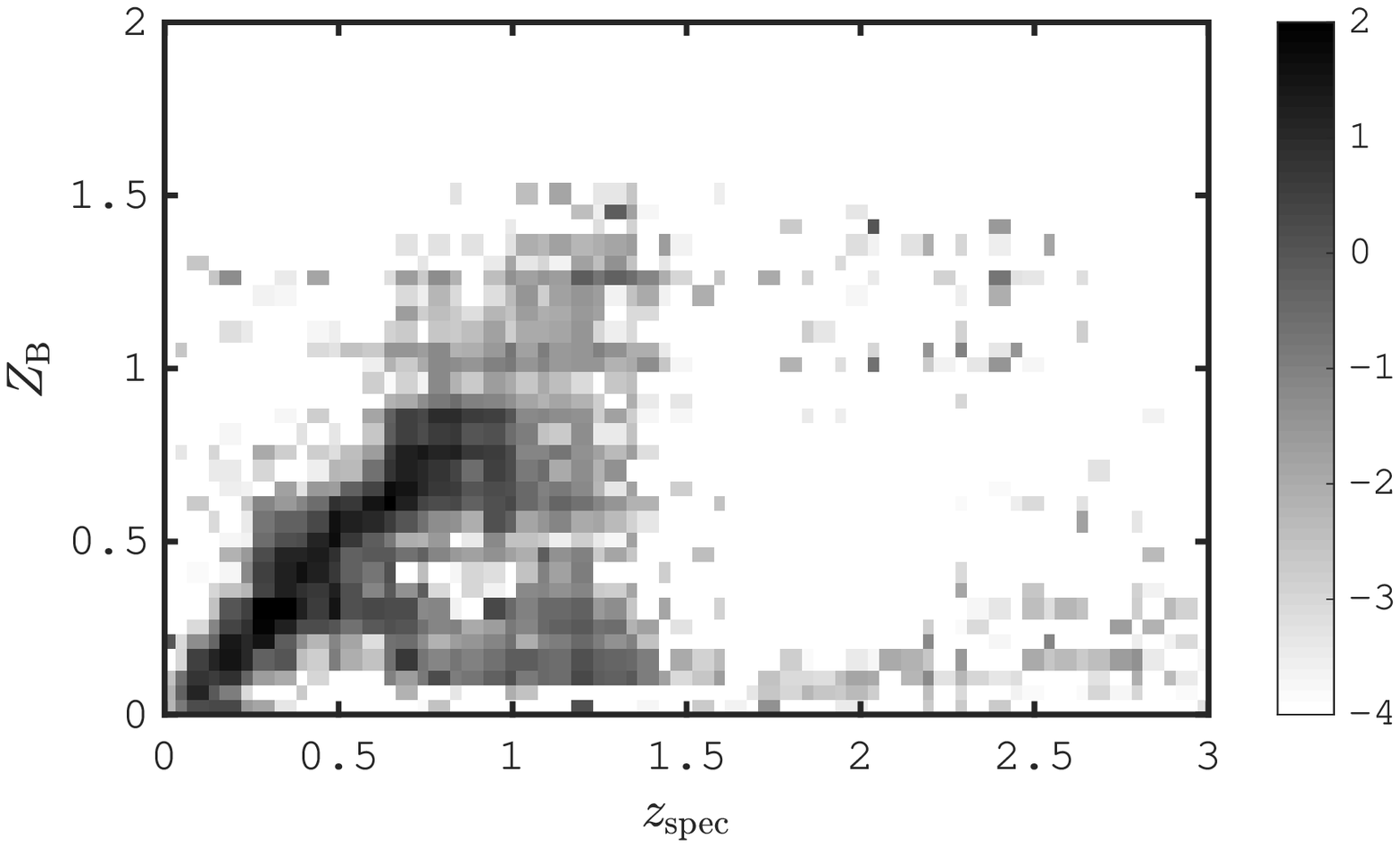}
\caption{({\it upper:})  Estimate of the source redshift distribution  in the KiDS-450 mocks, described in Section \ref{subsec:V0} and
shown with the black line. This reproduces the `DIR' \textcolor{black}{$n(z)$} in \citet{KiDS450} and is included in the mocks as the $z_{\rm spec}$ column. The red line shows the $Z_{\rm B}$ distribution in the mocks, which is used to split the samples into tomographic bins.
({\it lower:}) Joint PDF between $Z_{\rm B}$ and $z_{\rm spec}$ constructed from the matched sample.
The grey scale shows the number of objects per matrix element in log scale. }
\label{fig:nz_w}
\end{center}
\end{figure}

The mock creation starts with the choice of a redshift distribution and galaxy density.
\textcolor{black}{We populated the mocks with $n_{\rm gal} = 8.53$ gal/arcmin$^2$, matching the effective galaxy density of KiDS.  The raw galaxy number density is almost double this value but the galaxies are then weighted in any subsequent analysis.  The effective galaxy number density is the equivalent number density of galaxies with unit weight that have the same noise properties as the weighted analysis \cite[see section 3.5 of][for further discussion]{2015MNRAS.454.3500K}.  We use the $n(z)$ calibrated using the `DIR' method of H17,
 identified as the most accurate of the four different methods applied on the KiDS-450 data.}
 It is based on a reweighted spectroscopically-matched sub-sample of the KiDS-450 data that covers 2 deg$^2$, for which we can measure both the photometric and spectroscopic redshifts. 
Photometric redshifts in KiDS are estimated from the maximum of the probability distribution obtained from the photo-$z$ code {\small BPZ} \citep{BPZ},
referred to as $Z_{\rm B}$. In data and mock analyses, this quantity is used to define tomographic bins, but does not enter in the estimation of the $n(z)$. 
We show in the upper panel of Fig. \ref{fig:nz_w} a comparison between the DIR $n(z)$ and  the  $Z_{\rm B}$ distributions measured from these KiDS-450 mocks. 
Given a $z_{\rm spec}$, a photometric redshift is  assigned to each mock \textcolor{black}{galaxy} by drawing $Z_{\rm B}$ from a joint PDF,  $P(Z_{\rm B} | z_{\rm spec})$,  constructed from the \textcolor{black}{reweighted} matched sample (see the \textcolor{black}{lower} panel of Fig. \ref{fig:nz_w}).

Although  $n(z)$, $n_{\rm gal}$ and $P(Z_{\rm B} | z_{\rm spec})$ are the same in the mock as in the data, subtle effects
inherent to the DIR method \textcolor{black}{cause the level of  agreement to reduce after selections in $Z_{\rm B}$ are made}. 
Indeed, Table \ref{table:n_eff_comp} shows that some of the tomographic 
bins in the KiDS-450 data have more galaxies than in the mocks, and some less.
This is caused by sampling variance that affects the DIR method, covering only a small area that might not be fully representative of  the full data set.
The residual difference with full data set propagates into the mocks and causes this mismatch in galaxy density.
One way around this is  to construct mocks with higher densities and to down-sample them to match exactly the $n_{\rm gal}$ from the data. For this reason, we produced a second set of mocks, the KiDS-450-dense,  in which the number density was increased to 13.0 gal/arcmin$^2$.
After tomographic decompositions, there are more galaxies in the mocks than in the data in all bins; 
one can then down-sample the mocks to match exactly the $n_{\rm gal}$ per tomographic bin.
Another strategy is to produce mock catalogues for each tomographic bin, matching the $n(z)$ and $n_{\rm gal}$ therein.
This is the approach we used for the LSST-like mocks, which are described in Appendix \ref{subsubsec:LSST},
but in this case the choice of tomographic decomposition can no longer be changed.

Once  galaxies are assigned their coordinates and spectroscopic redshifts, we next compute the lensing information.
The weak lensing shear components $\gamma_{1,2}$ are linearly interpolated at the galaxy coordinates and redshift from the shear planes described in Section \ref{subsec:planes}. 
Note that the interpolation is only done along the redshift direction, not in the pixel direction.
In other words, galaxies at the same redshift falling within the same pixel are assigned the same shear. 
This could easily be modified, but introduces a calculation overhead and only affects the weak lensing measurements at scales below 0.2 arcmin,
where limitations in the mass resolution dominate the systematic effects in the mocks.

In addition to the cosmological shear, the observed ellipticity is included in the catalogue and is computed from:
\textcolor{black}{
\begin{eqnarray}
\epsilon^{\rm obs} = \frac{ \epsilon^{\rm int} + \gamma}{1 + \epsilon^{\rm int}\gamma^*} + \eta \approx \frac{ \epsilon^{\rm n} + \gamma}{1 + \epsilon^{\rm n}\gamma^*}
\label{eq:eps_obs}
\end{eqnarray}
where $\epsilon$, $\eta$ and $\gamma$ are complex numbers (i.e. $\gamma = \gamma_1 + i\gamma_2$).  $\epsilon^{\rm int}$ is the intrinsic ellipticity of the galaxy which is sheared by $\gamma$.   The observed ellipticity $\epsilon^{\rm obs}$ is also subject to measurement noise $\eta$.  For this mock we choose to not distinguish between intrinsic and measurement shape noise, and make an approximation by including both the intrinsic and measurement shape noise into one pre-sheared noisy ellipticity $\epsilon^{\rm n}$ which is assigned by drawing random numbers from a Gaussian distribution with width $\sigma = 0.29$ per component, consistent with the weighted observed ellipticity distribution of the KiDS data. }
%
%
The Gaussian is truncated such that $\left(\epsilon^{\rm int}_{1}\right)^2 + \left(\epsilon^{\rm int}_{2}\right)^2 \le 1$.
The resulting \textcolor{black}{noisy} shape distribution is uncorrelated with the properties of galaxies such as colour, measured shape weights, galaxy type, size or brightness.
This is of course a simplification of the reality, but it is not believed to be important for the primary goal of these simulations,
plus it can easily be modified if needed in the future.
Table \ref{table:cat_columns} summarises the catalogue content for these KiDS-450 source mocks.



%
%

\begin{figure*}
\begin{center}
\includegraphics[width=3.1in]{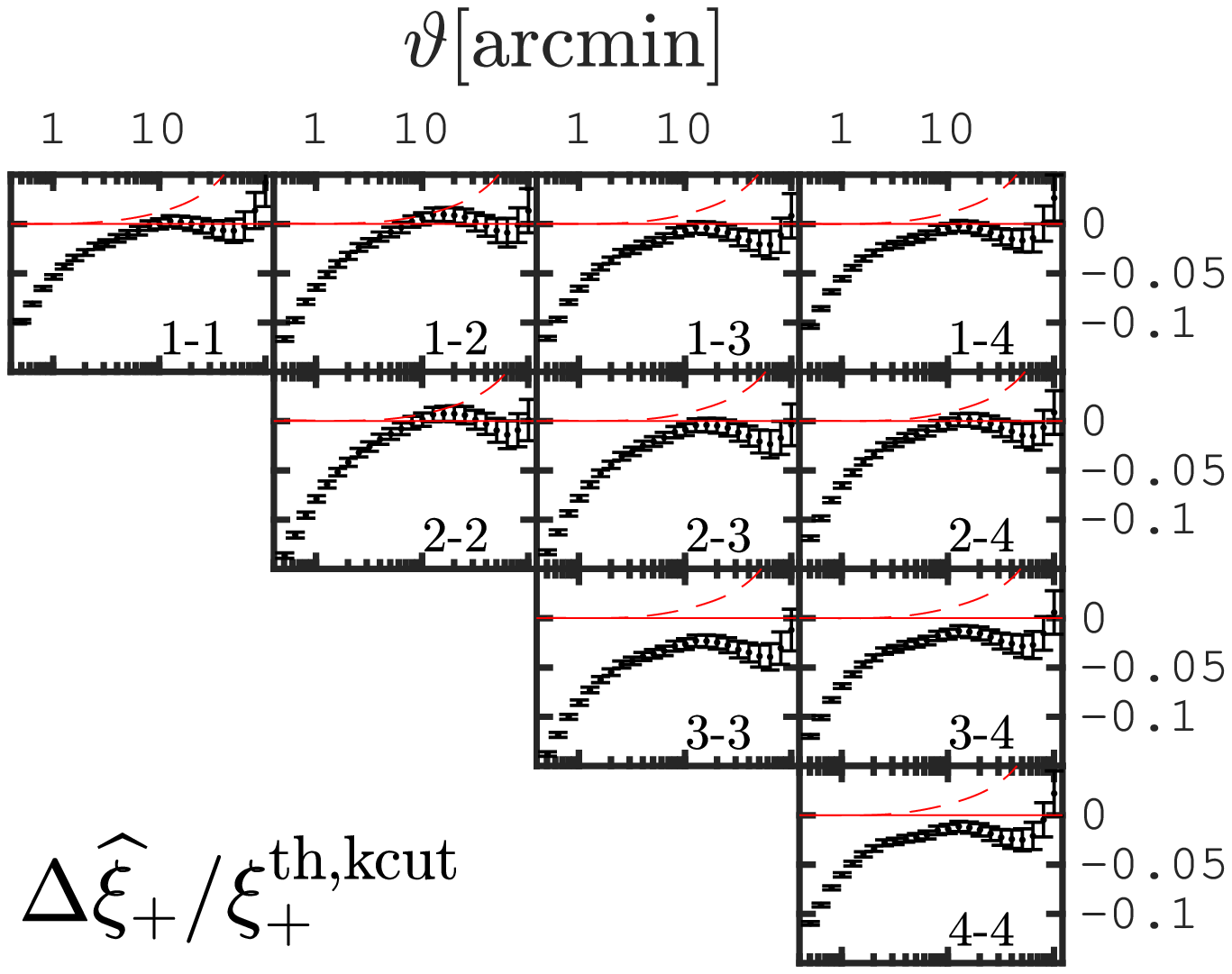}
\hspace{10mm}
\includegraphics[width=3.0in]{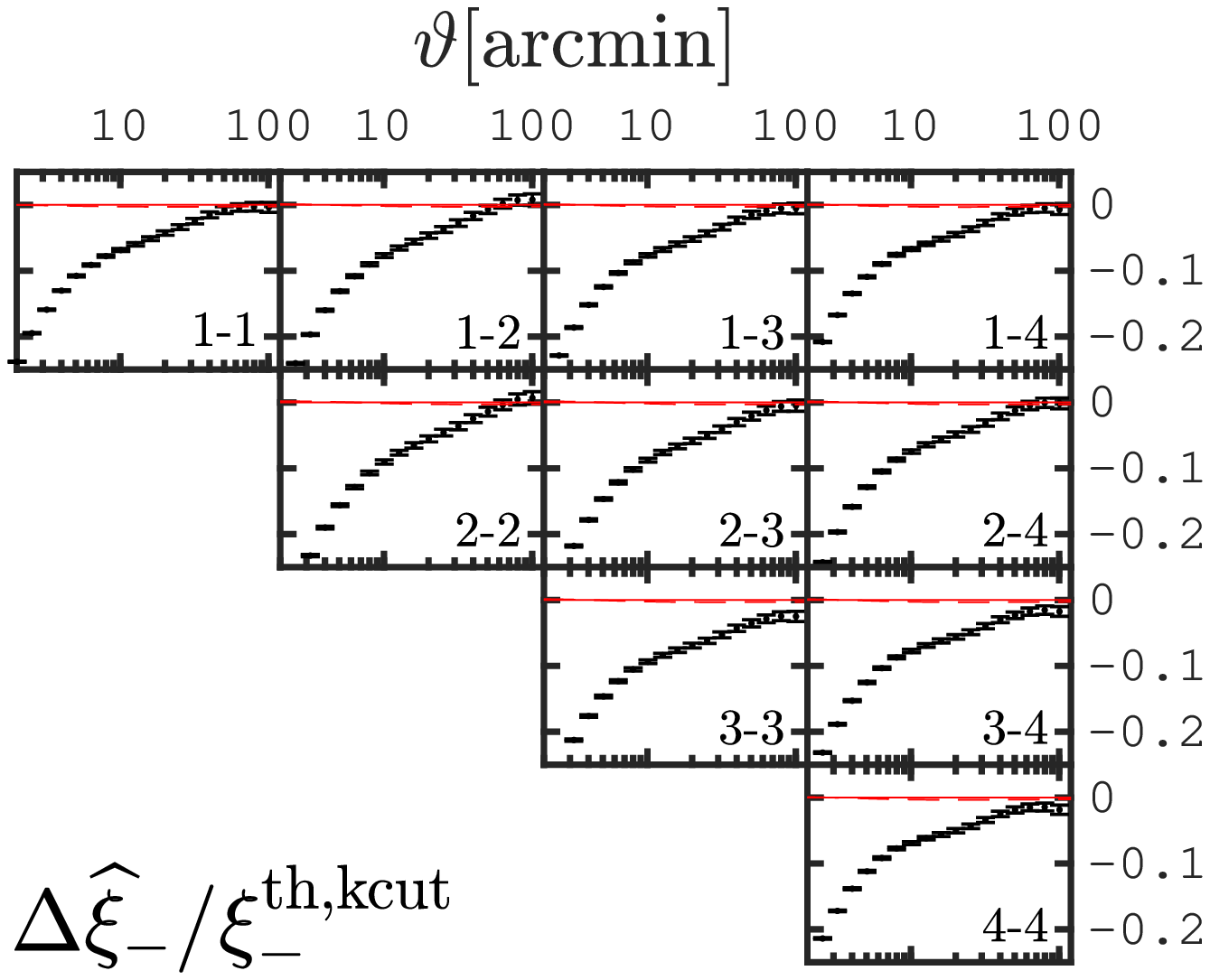}
\caption{Cosmic shear measured from all combinations of the four tomographic bins from the KiDS-450 mocks, ignoring shape noise.
The $y$-axis shows $\widehat{\xi_{\pm}}/\xi_{\pm}-1$, the fractional difference between the measurements $\widehat{\xi}_+$ (left) and $\widehat{\xi}_-$ (right) from the mocks and the predictions $\xi_{\pm}$.  The finite box effect (solid red) is present in the mocks and modelled in these predictions: we set the theoretical matter power spectrum to zero for $k$-modes corresponding to scales larger than the simulation box. Removing this effect results in the red dashed lines.
The $x$-axis shows the opening angle $\vartheta$ in arcmin.
Error bars show the error on the mean, here computed from 932 lines-of-sight to highlight the accuracy of the lensing signal extracted from these mocks. 
The tomographic bins are labeled on the sub-panels, where for example the notation 1-2 refers to the cosmic shear signal measured between bins selected with $Z_{\rm B} \in [0.1-0.3]$ and $Z_{\rm B} \in [0.3-0.5]$.}
\label{fig:KiDS_xip}
\end{center}
\end{figure*}

The shear two-point correlation functions $\xi_{\pm}$ of the SLICS were presented in HvW15 for the case where all galaxies
are placed at a single source redshift. We show here the measurement from the KiDS-450 mocks, which have instead a broad
redshift distribution, and have been split into the same tomographic bins as in the KiDS-450 cosmic shear analysis. 
We applied cuts on $Z_{\rm B}$  to create four bins, with $Z_{\rm B} \in $ [0.1--0.3], [0.3--0.5], [0.5--0.7] and [0.7--0.9], 
each of which by construction has a redshift distribution that matches the corresponding DIR-estimated $n(z)$.  

We compute the two-point correlation function between tomographic bins $\alpha$ and $\beta$ with {\small ATHENA} \citep{Schneider2002a}, estimated from\footnote{ATHENA: www.cosmostat.org/software/athena/}: 
\begin{eqnarray}
\xi_{\pm}^{\alpha \beta}(\vartheta)={\sum_{i,j} w_i w_j \left[e_{\rm t}^i e_{\rm t}^j \pm e_{\times}^i e_{\times}^j \right]\over \sum_{i,j} w_i w_j},
\label{eq:xi_estimator}
\end{eqnarray}
where the sum extends over all galaxy pairs `$(i,j)$' separated by a position angle in the range [$\vartheta \pm \Delta \vartheta/2]$ on the simulated sky.
The bin width has uniform logarithmic intervals, with  log$_{10}\Delta \vartheta = 0.1$.
The quantities $e_{{\rm t},\times}$ are the tangential and cross components of the ellipticities, while the weights $w_i$ capture the quality 
of the shape measurement of the object $i$. For the rest of the paper, these weights are all set to unity;
however, it is possible to assign different values based on other galaxy properties. 
Galaxies $i$ and $j$ are drawn from redshift bins $\alpha$ and $\beta$, respectively.

The results are shown in Fig. \ref{fig:KiDS_xip}
for all tomographic combinations, and ignoring shape noise (i.e. $\epsilon^{\rm n}$ is set to 0 in Eq. \ref{eq:eps_obs}).
These measurements are compared to theoretical predictions obtained from {\small NICAEA} \citep{Nicaea}, 
a public numerical package that rapidly  computes  accurate cosmological statistics\footnote{NICAEA: www.cosmostat.org/software/nicaea/}. The input predictions for the matter power spectrum 
are computed from the revised {\small HALOFIT} code \citep{Takahashi2012}.
We recover the results presented in HvW15, namely that  \textcolor{black}{ the angular scales larger than 1 arcmin
in $\xi_+$ are generally accurate to better than 5\% when forward-modelling the finite box effects; smaller scales suffer from limits in particle mass resolution}. 

The covariance matrix of $\xi_{\pm}(\vartheta)$ extracted from the SLICS was also presented in HvW15 and in H17, 
and we refer the reader to these two papers for more details.
In short, the covariance matrix was shown to reconnect with the Gaussian predictions
at large angular scales that are mostly sensitive to the linear regime of structure formation,
while significant non-Gaussian features are present at smaller scales. 
The full covariance is in general agreement with halo-model-based predictions.

\begin{table}
   \centering
       \caption{KiDS-450 source mocks: Comparison between $n_{\rm gal}$ in the main mocks, the dense mocks and the data,
       after splitting the catalogues in the four tomographic bins with $Z_{\rm B}$ \citep[see][]{KiDS450}.
       Numbers are in units of gal/arcmin$^2$. Although there is some discrepancy in the number density, these mocks exactly reproduce the DIR $n(z)$ in each bin, 
       and their  shape noise has been set to $\sigma = 0.29$ per component.} 
   \begin{tabular}{ccccc} 
  \hline
  \multirow{2}{*}{$Z_{\rm B}$ cut}                             &   Data      &     \multicolumn{2}{c}{Mocks} \\
   & KiDS-450 &   KiDS-450  & KiDS-450-dense \\ 
\hline
0.1-0.3&          2.354   &   2.098&  3.197   \\
0.3-0.5&          1.856   &   2.062&  3.144   \\
0.5-0.7&          1.830   &   1.968 &  2.995     \\
0.7-0.9&          1.493   &   1.419 &  2.169     \\
0.9-10 &          0.813   &   0.690  &  1.050     \\
\hline
no cut  &            8.53   &    8.53  & 13.0 \\
  \hline
    \end{tabular}
   \label{table:n_eff_comp}
\end{table}


\subsection{Halo occupation distribution}
\label{subsec:V2}


As demonstrated by recent analyses from KiDS and DES, constraints on cosmological parameters are further improved when cosmic shear measurements
are supplemented with galaxy-galaxy lensing measurements and clustering measurements extracted from overlapping surveys
\citep{2017arXiv170605004V, 2017arXiv170706627J, 2017arXiv170801530D}. These measurements, often referred to as $3\times2$-point combined-probes, 
have a higher constraining power, provided that one can accurately estimate the covariance matrix of the full data vectors, including the cross-terms (see Section \ref{sec:combined_probe}).

In this section, we describe the construction of simulation products that are designed to estimate such matrices, tailored for combined-probe measurements
based on the CMASS (see Section \ref{subsubsec:CMASS}), LOWZ (Section \ref{subsubsec:LOWZ}) and  GAMA (Section \ref{subsubsec:GAMA}) spectroscopic surveys. 
We  aim to match observations of the foreground lens clustering and of the galaxy-galaxy lensing signals involving these three samples, 
and we achieve this by first producing mock lens catalogues of similar redshift distributions,  galaxy densities and galaxy biases. 

We produce mock galaxy catalogues from HOD models, which are statistical descriptions of the data  that assign a galaxy population to host dark matter haloes
solely based on their mass. Every HOD model is  calibrated  to reproduce key properties of the survey it attempts to recreate.
 For the LOWZ and CMASS mock lenses, 
we use the prescription of \citet{2017MNRAS.465.4853A}, with minor modifications to the best fit parameters. 
The GAMA  mocks are based on a hybrid technique that mixes the prescriptions of \citet{2013MNRAS.430..767C} and of \citet{2017arXiv170106581S}.
For the  KiDS-HOD mock (Section \ref{subsubsec:KiDS}, distinct from the KiDS-450 mocks described in Section \ref{subsec:V0}) 
and the LSST-like HOD mock (Section \ref{subsubsec:LSST-HOD}, distinct from the LSST-like source mocks described in Appendix \ref{subsubsec:LSST})
we extend the GAMA HOD to $z=1.5$ and $3.0$, respectively. 
All these different HOD prescriptions share some common ingredients and methods, which we describe here.

%

Based on its mass, each halo is assigned a mean number of  central galaxies,  $\langle N_{\rm cen} \rangle$, which varies from zero to one,
and a mean satellite number $\langle N_{\rm sat} \rangle$. 
The sum of these two quantities gives the mean number of galaxies per halo, and we ensure that haloes with no centrals have no satellites.
Central galaxies  are pasted  at the  location of the halo peak, while satellites are distributed following 
a spherically symmetric NFW profile \citep{1997ApJ...490..493N}. This is not the most sophisticated  method to populate satellites, as we ignore possible relations between their positions and the anisotropic shape of the dark matter halo, the merging history, etc. Note also that we have not included any scatter in the $c(M)$ relation into our mocks.
This is fine since our purposes here are to validate estimators, to evaluate covariance matrices and to create a relatively realistic environment that is well controlled on which to test analysis pipelines. We therefore argue that our choice of satellite assignment scheme does not  introduce significant additional bias
for the science cases of interest. Even more, if we used a different profile, we would then run into an inconsistency problem because  the HOD models were  calibrated on data assuming NFW profiles. We therefore leave investigations of this type for future work.

 A key ingredient that enters the profile is the concentration parameter $c$, 
which strongly correlates with the halo mass. Many models exist for this $c(M)$ relation,
and we use the models that were used in the original HOD prescriptions that  we are reproducing.
Specifically, we use the  \citet{Bullock2001} relation for the CMASS and LOWZ HOD \citep[as in][]{2017MNRAS.465.4853A}, 
and the \citet{2008MNRAS.391.1940M} relation for the GAMA HOD \citep[as in][]{2013MNRAS.430..767C}. 
We further scale these relations by a free multiplicative factor to improve the match of the clustering measurements with the data. 
Note that it is challenging to construct an HOD model where this match is achieved at all scales, while preserving the redshift distribution and the galaxy density.
Our final choice of parameters reach a compromise between all these quantities.

Of interest for combined-probe programmes is the fact that these foreground lens samples emulate spectroscopic data, for which we can measure redshift-space distortions (RSD).
The RSD are based on the measurement of the Doppler shift caused by the peculiar velocities of the galaxies, which induces anisotropies in the observed large-scale structures in a manner that can be related to the underlying cosmological parameters \citep[see][for a review]{RSD_review}.
This phenomenon therefore contains additional cosmological information that nicely complements cosmic shear measurements, as recently seen in \citet{2017arXiv170706627J}.  
We implement the effect of RSD in our mock data by assigning a peculiar velocity (along the line-of-sight) to every galaxy.
The radial position in redshift space is therefore given by a distortion term $\Delta \chi$ acting on the line-of-sight coordinate:
\begin{eqnarray}
\chi_{\rm RSD} = \chi + \Delta \chi =  \chi + \frac{v_{\rm pec}}{a(z)H(z)},
\label{eq:RSD}
\end{eqnarray}
where $v_{\rm pec}$  is the peculiar velocity of the galaxy, and $H(z)$ is the redshift-dependent expansion parameter. 
For central galaxies, $v_{\rm pec}$ is obtained directly from the centre of mass velocity of the host halo (projected on the line-of-sight),
while for the satellites it is drawn from a Gaussian distribution with  mean set to the centre of mass halo velocity, and with variance given by the line-of-sight component of the velocity dispersion,
provided by our halo-finder.
The redshift-space position $\chi_{\rm RSD}$  is finally converted to redshift assuming our fiducial cosmology, and written in the catalogue
as  $z_{\rm spec}^{\rm s}$.  We do not use this quantity elsewhere in this paper, but make it available in the catalogues for applications based on RSD. 

The following sections (Section \ref{subsubsec:CMASS} -- Section \ref{subsubsec:LSST-HOD}) contain the description of the HOD models tailored for the different mock spectroscopic surveys.


\subsection{Mock CMASS lens galaxies}
\label{subsubsec:CMASS}

\begin{table*}
   \centering
       \caption{Organisation of the different mock source catalogues (KiDS-450 and LSST-like), lens catalogues (CMASS, LOWZ and GAMA) and hybrid catalogues (KiDS-HOD and LSST-like HOD) described and used in this paper.
       The difference between `ray-tracing' and `clustering' coordinates is explained in Appendix \ref{subsec:coordinates}. 
       Note that the order of the entries in this table and in the mocks may differ. Also,  for each light cone, the $(x,y)_{\rm ray-tracing}$ positions cover 10$\times$10  deg$^2$ in flat sky coordinates, 
       hence are best described by a square patch placed at the equator (dec = 0) where the difference with the curved sky 
       coordinates is minimal. }

   \begin{tabular}{rcccccc} 
  \hline
  \multirow{2}{*}{content} &  \multirow{2}{*}{units} & KiDS-450 &CMASS &  \multirow{2}{*}{GAMA} & KiDS-HOD &  \multirow{2}{*}{description}\\
                                       &                                     & +LSST-like sources& + LOWZ &             & + LSST-like HOD & \\
\hline
 $M_{\rm h}$  & $~h^{-1}M_{\odot}$ & no&  yes  & yes & yes & halo mass\\
   Halo ID &  & no& yes  & yes & yes & ID of the host dark matter halo\\
     $N_{\rm sat}$&    & no& yes  & yes & yes & number of satellites (central only)\\
  d$x_{\rm sat}$ &  \multirow{3}{*}{\Bigg\} $h^{-1}$kpc}  & no& yes  & yes & yes& \multirow{3}{*}{\Bigg\} distances to the central galaxy (satellites only)} \\
  d$y_{\rm sat}$ & & no& yes  & yes & yes  & \\
  d$z_{\rm sat}$ & & no& yes  & yes & yes &  \\
\hline
  $x_{\rm ray-tracing}$ &   \multirow{4}{*}{\Bigg\} arcmin} & yes& yes  & yes & yes &   \multirow{2}{*}{\Bigg\} coordinates for lensing}  \\
  $y_{\rm ray-tracing}$ &   & yes& yes  & yes & yes &  \\
  $x_{\rm clustering}$ &   & no& yes  & yes & yes &  \multirow{2}{*}{\Bigg\} coordinates for clustering} \\
  $y_{\rm clustering}$ &     & no& yes  & yes & yes & \\
  \hline
  $z_{\rm spec}$ &   & yes& yes  & yes & yes & cosmological redshift \\
  $z_{\rm spec}^{\rm s}$ &   & no& yes  & yes & yes & observed spectroscopic redshift\\
     $Z_{\rm B}$ &  & yes& no  & no & yes & photometric redshift \\
    \hline
  $M_r$ &  & no& no  & yes & yes & absolute $r$-band magnitude\\
  $m_r$ &  & no& no  & yes & yes & apparent $r$-band magnitude\\
    $M_{\star}$ & $~h^{-2}M_{\odot}$ & no& no  & yes & no & stellar mass\\
    \hline
   $\gamma_1$ &  & yes& no  & no & yes &  \multirow{2}{*}{\Bigg\} cosmic shear}\\
    $\gamma_2$ &   & yes& no  & no & yes & \\
    $\epsilon_{1}^{\rm obs}$& & yes& no  & no & yes &  \multirow{2}{*}{\Bigg\} observed ellipticity}\\
   $\epsilon_{2}^{\rm obs}$ &  & yes& no  & no & yes &  \\
\hline
$N_{\rm sim}$ &  & 932 & 844 & 844 & 120 & number of independent realisations \\ 
\hline
    \end{tabular}
   \label{table:cat_columns}
\end{table*}

\begin{figure}
\begin{center}
\includegraphics[width=3.2in]{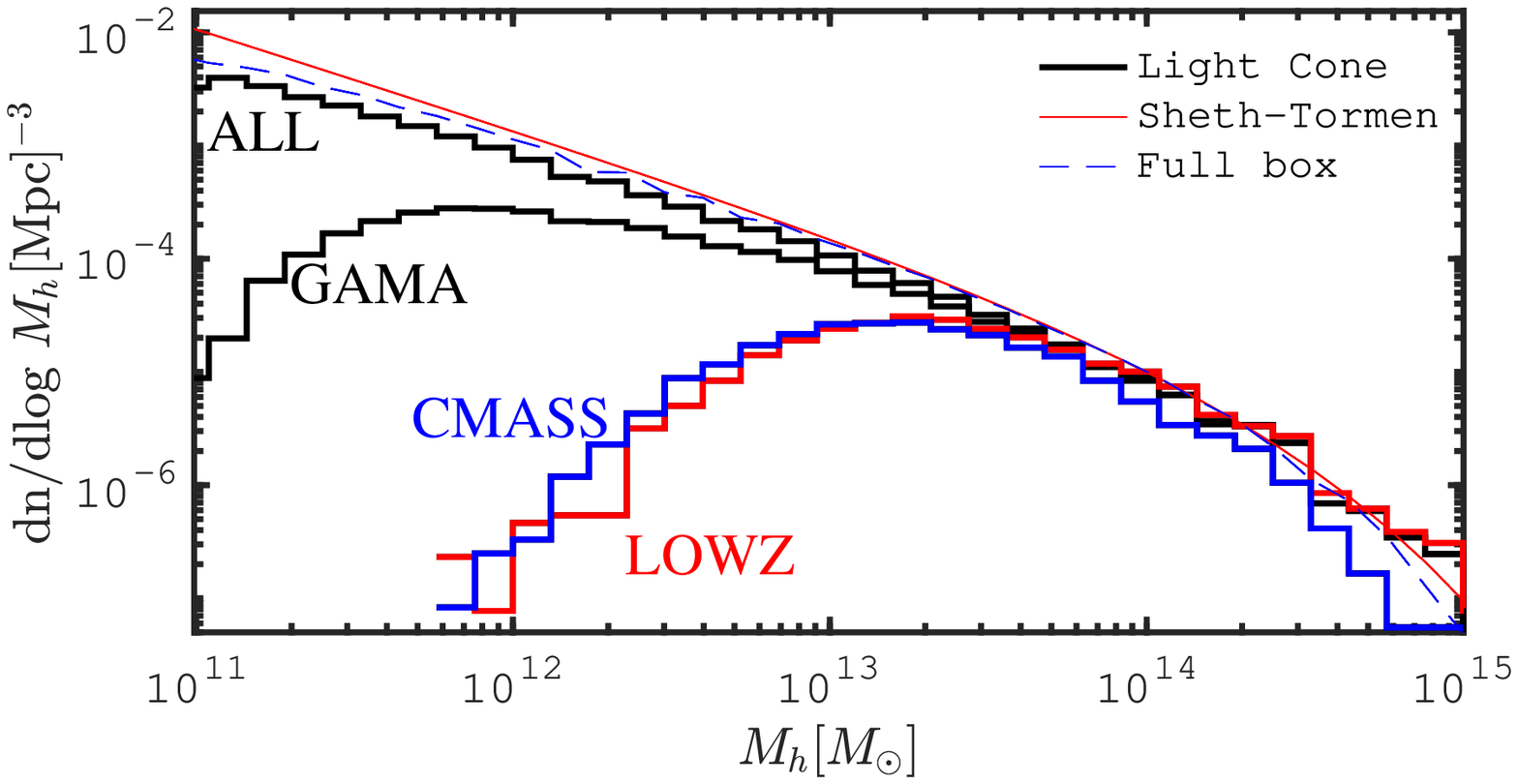}
\vspace{10mm}
\includegraphics[width=3.15in]{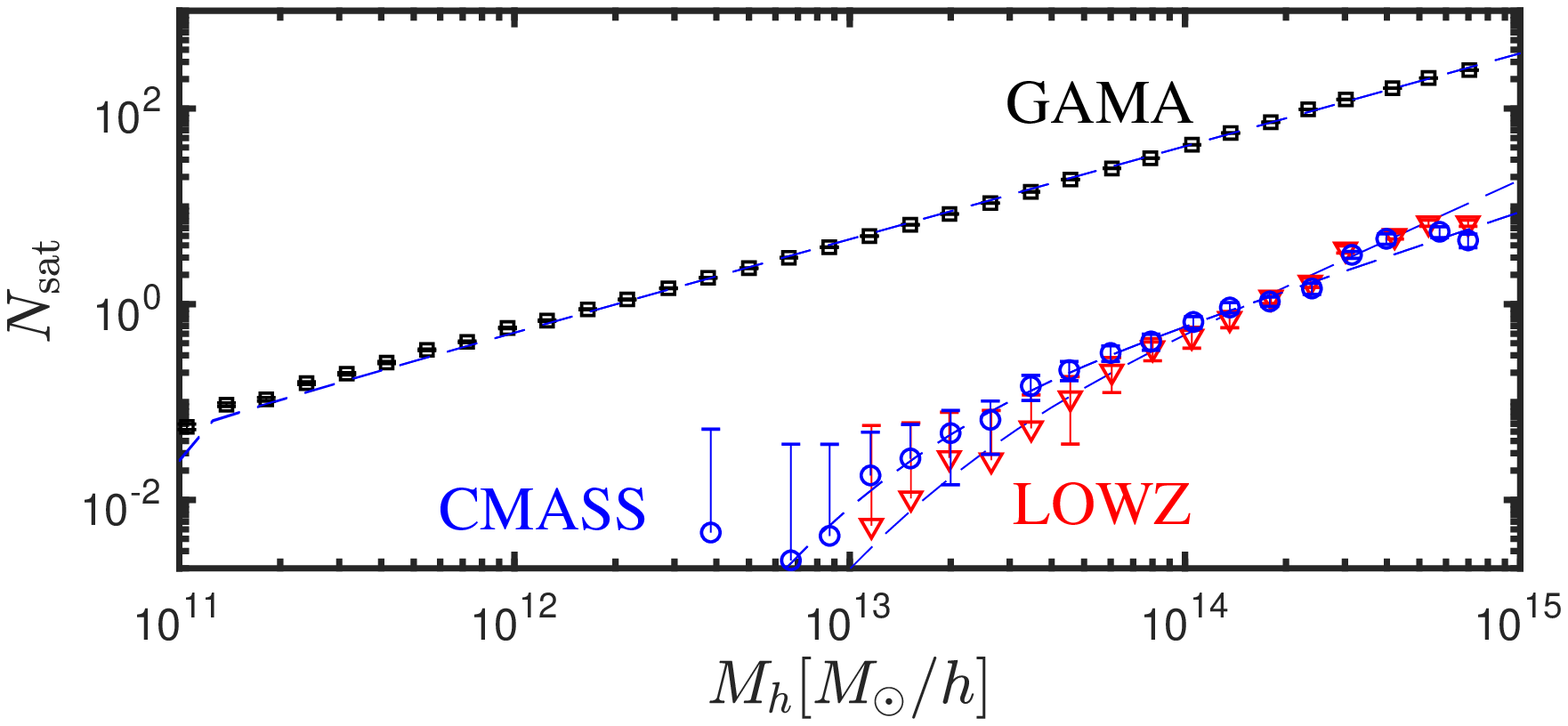}
\caption{({\it upper:}) Galaxy mass function in the GAMA, CMASS and LOWZ mocks, compared to the halo mass function of the mocks at $z=0.4$ (dashed blue).
Also shown is the GAMA mass function before the $m_r < 19.8$ magnitude selection cut, labelled `ALL' here, as it closely traces the underlying halo mass function. 
({\it lower:}) Number of satellites per haloes in the GAMA (black squares), CMASS (blue circles) and LOWZ (red triangles) mocks, compared 
to their input HODs. }
\label{fig:MassFunction}
\end{center}
\end{figure}


\begin{figure*}
\begin{center}
\hspace{-8mm}
\includegraphics[width=2.5in]{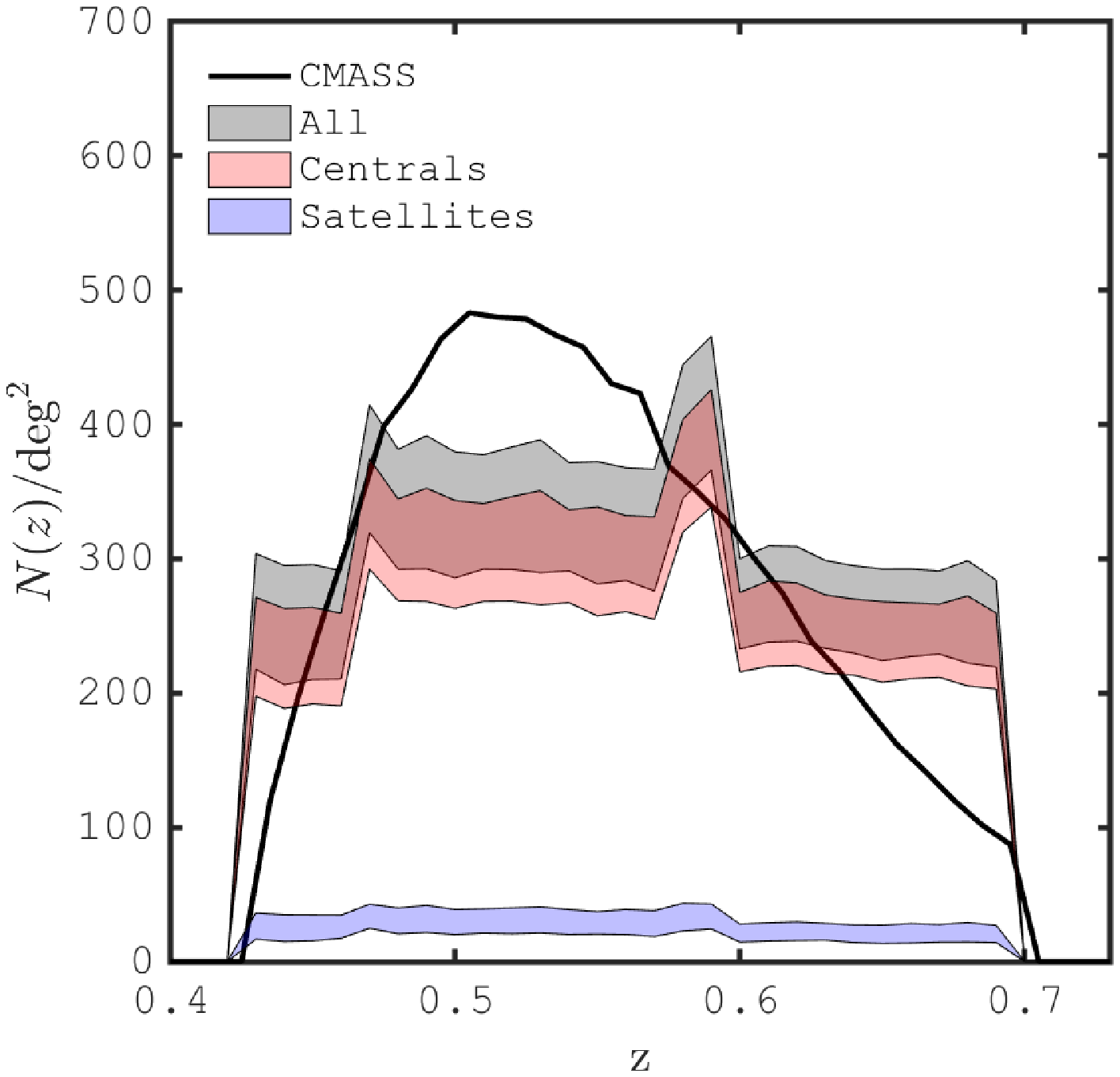}
\hspace{-6mm}
\includegraphics[width=2.5in]{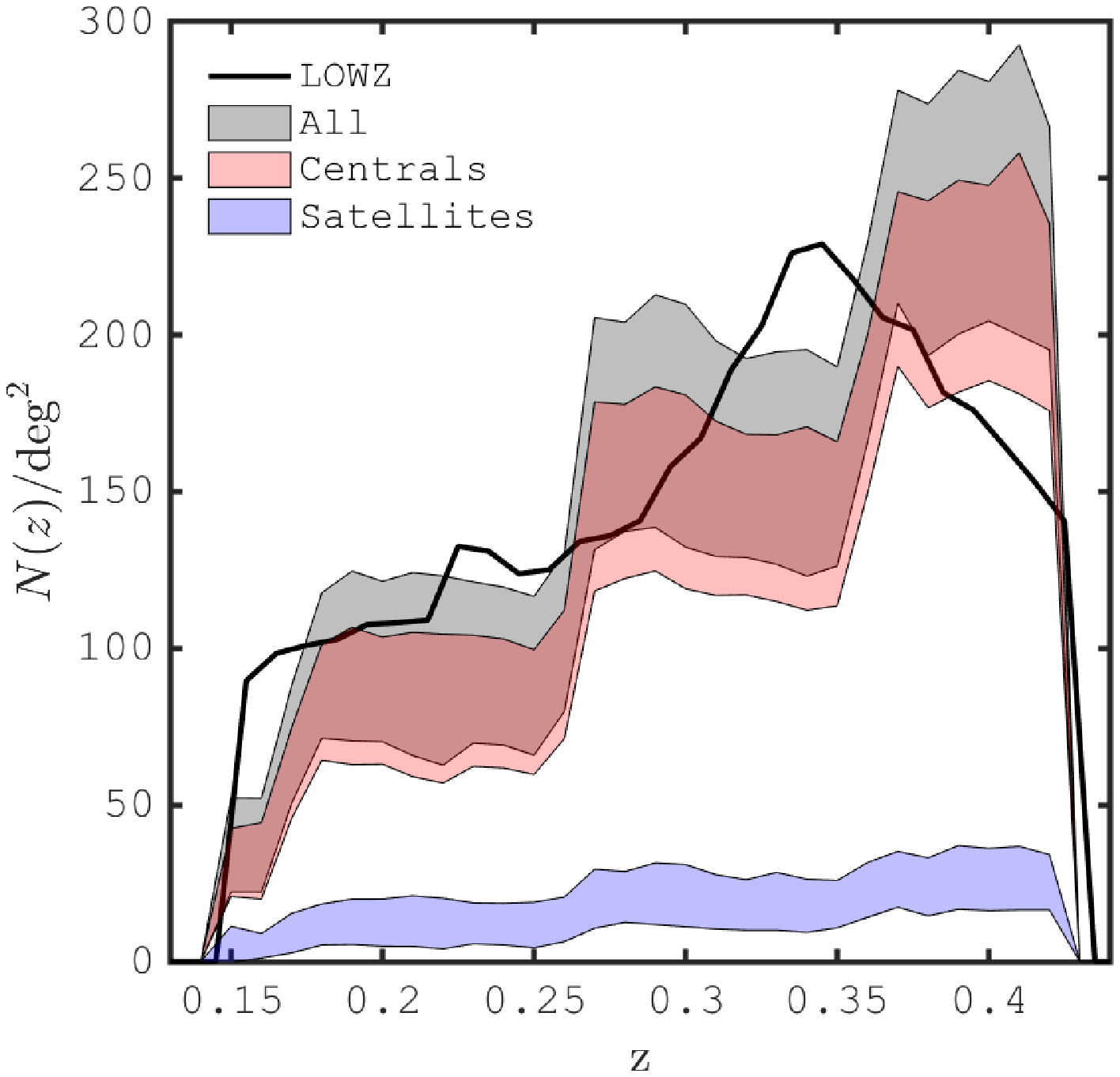}
\includegraphics[width=2.5in]{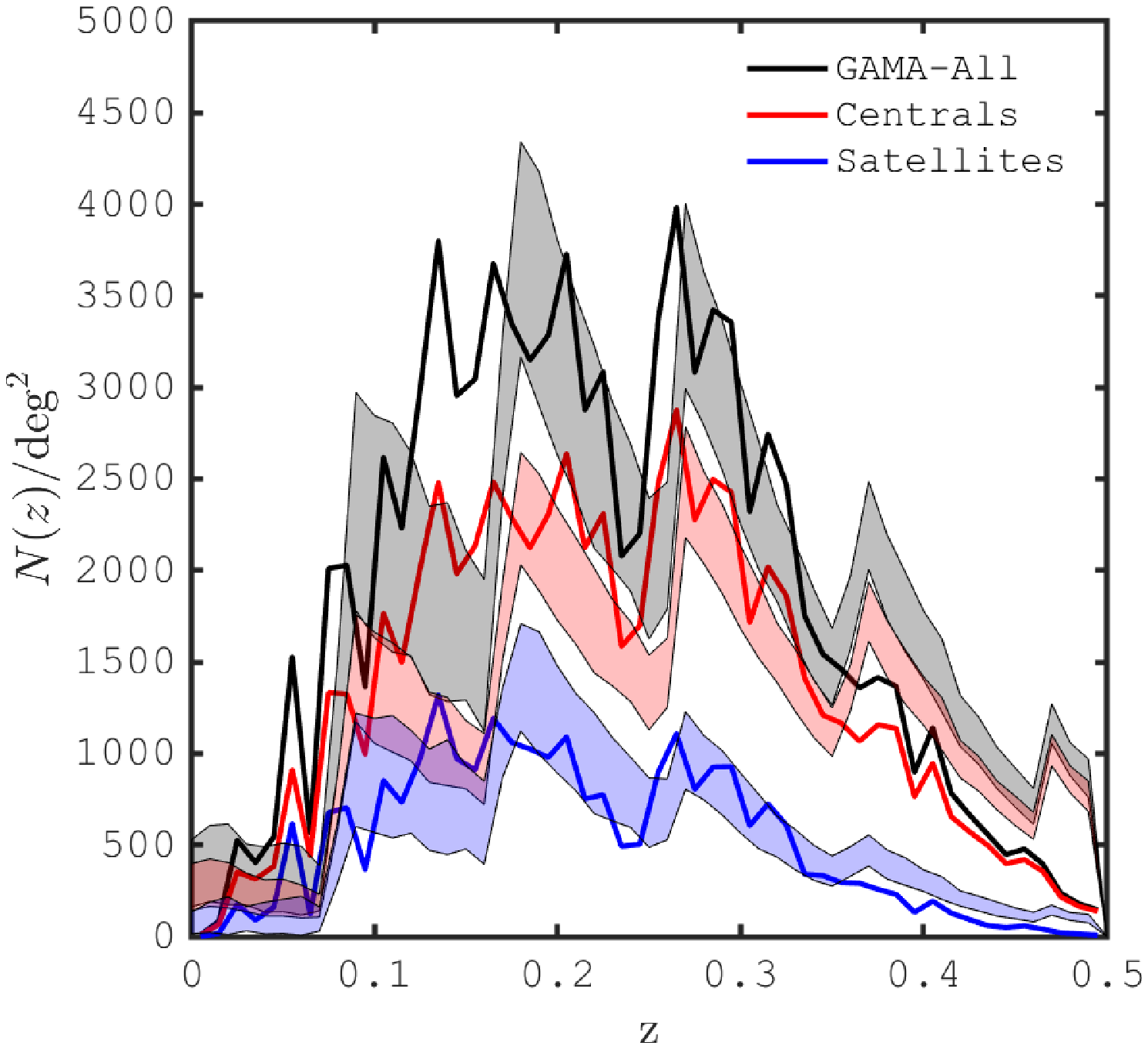}
\caption{Redshift distribution of the CMASS (left), LOWZ (centre), and GAMA (right) mock galaxies, 
for satellites (blue), centrals (red) and all combined (black). Solid lines are obtained from the data. 
Although the shape of the distributions differ between data and mocks, the mean redshifts and number densities are in good agreement, as discussed in the main text.}
\label{fig:GAMA_nz}
\end{center}
\end{figure*}

\begin{figure}
\begin{center}
\includegraphics[width=3.2in]{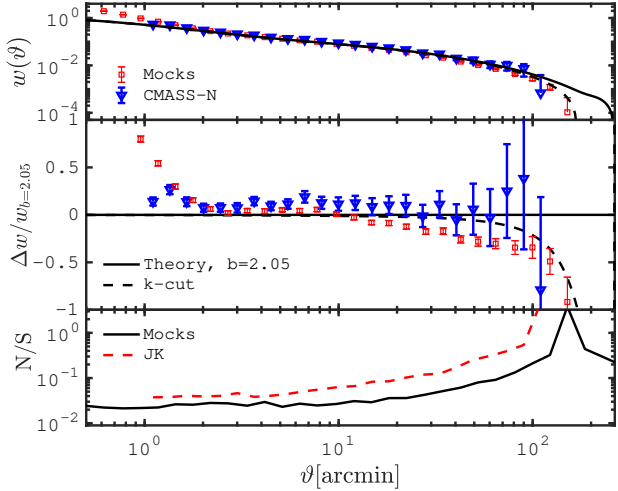}
\caption{({\it upper:}) Angular correlation function of the CMASS mocks (red squares), compared to the CMASS-NGC data (blue triangles). 
The mocks are averaged from 100 lines-of-sights, the error bars are on the mean;  the error on the data comes from jack-knife resampling.
The predictions shown in solid black assume the SLICS cosmology and the best fit bias of $b_{\rm CMASS}$ = 2.05.
The dashed black line illustrate the impact on theory  of excluding the $k$-modes larger than the simulation box.
({\it middle:}) Fractional difference between the measurements and the predictions. The clustering signal in the mocks is about 10\% lower than in the data.
({\it lower:}) Error over signal, for the mock and the jack-knife estimates of the covariance.}
\label{fig:w_theta_CMASS}
\end{center}
\end{figure}

The CMASS HOD prescription is largely inspired by  \citet[][equation 18 therein]{2017MNRAS.465.4853A}, with some adjustments made
to improve the match between our mocks and the data\footnote{We have also experimented with the implementation from \citet{Manera2013}, 
a HOD calibrated on the DR10 BOSS data release. This other calibration prefers higher number densities, but the resulting clustering amplitude  is too low compared to the DR12 data, hence we adopted the \citet{2017MNRAS.465.4853A} HOD model.   
}.
We approximate CMASS as a volume-limited sample and construct a volume-limited mock catalogue, avoiding the need to compute luminosity or stellar mass related quantities.
This means that the residual magnitude-related features seen at high redshift cannot be implemented with a magnitude cut from our  mocks.
To reproduce the decreasing number of high-redshift galaxies, we down-sample the high-redshift tail of the mock catalogues, as detailed below. Additionally, there are noticeable differences 
between the  target selection of the BOSS data in the north and south \textcolor{black}{Galactic} cap \citep{2016MNRAS.455.1553R}, therefore  we calibrate our CMASS and LOWZ HODs on the northern patches, which cover a larger area. Hereafter, when referring to CMASS and LOWZ data/area, we are using short notation for the `CMASS-NGC' and `LOWZ-NGC' subsamples
of the DR12 public data release\footnote{BOSS-DR12: https://data.sdss.org/sas/dr12/boss/lss/}. 

As a first step, we assign central and satellite galaxies to dark matter haloes over a broad redshift range, 
and find in a second step the selection in the mocks that best reproduces the density and mean $n(z)$ of the CMASS data.
For  dark matter haloes of mass $M_{\rm h}$, the average number of central galaxies  $\langle N_{\rm cen} (M_{\rm h}) \rangle$ varies
from one for massive haloes, to zero for light haloes. The full occupation distribution 
is well described  by \citep{2017MNRAS.465.4853A}:
\begin{eqnarray}
\langle N_{\rm cen} (M_{\rm h}) \rangle = \frac{1}{2} {\rm erfc} \bigg[ \frac{{\rm ln} (M_{\rm cut}/M_{\rm h})}{2\sigma}\bigg],
\label{eq:CMASS_Ncen}
\end{eqnarray}
where erfc$(x)$ is the complementary error function, $M_{\rm cut}$ controls the minimal 
halo mass that can host a central galaxy, and $\sigma$ introduces a  spread about this minimal mass.
 The average number of satellite galaxies $\langle N_{\rm sat} (M_{\rm h}) \rangle$ follows a power law, 
assigning more satellites to more massive systems: 
\begin{eqnarray}
\langle N_{\rm sat} (M_{\rm h}) \rangle = \langle N_{\rm cen} (M_{\rm h}) \rangle  \bigg[ \frac{M_{\rm h} - \kappa M_{\rm cut}}{M_1}\bigg]^\alpha. 
\label{eq:CMASS_Nsat}
\end{eqnarray}
Here $M_{1}$ corresponds to the average mass a halo must have to host a single satellite,  
$\kappa$ affects the minimal mass below which a halo has no satellite,  and $\alpha$ is the slope of the number of satellites as a function of halo mass. 
The values of the HOD parameters are taken from \citet{2017MNRAS.465.4853A} and  reported in Table \ref{table:HOD_para}.
Once computed,   $\langle N_{\rm cen} (M_{\rm h}) \rangle$ and $\langle N_{\rm sat} (M_{\rm h}) \rangle$
are used as the means of Poisson distributions, from which  we finally sample the actual number of objects.

\begin{table}
   \centering
       \caption{HOD parameters in the CMASS  and LOWZ  mocks, described by Eq. \ref{eq:CMASS_Ncen} and Eq. \ref{eq:CMASS_Nsat}.
       The parameters $M_{\rm cut}$ and $M_1$ are both in units of $h^{-1}M_{\odot}$.}
   \begin{tabular}{ccccccc} 
  \hline
   &$M_{\rm cut}$ & $\sigma$&  $M_{1}$  & $\kappa $& $\alpha$\\
\hline
 CMASS& $1.77 \times 10^{13} $& 0.897 & $1.51 \times 10^{14} $ & 0.137& 1.151\\
 LOWZ& $1.95 \times 10^{13}$ &  0.5509& $1.51 \times 10^{14} $&  0.137 & 1.551\\
\hline
    \end{tabular}
   \label{table:HOD_para}
\end{table}


The mass function of the mock CMASS galaxies is presented in the upper panel of Fig. \ref{fig:MassFunction}, where we see that the HOD
preferentially selects  haloes in the range $M_{\rm h} \in [10^{12} -10^{15}] M_\odot/h$, in accordance with the survey target selection strategy \citep{2016MNRAS.455.1553R}.
The number of satellite galaxies for haloes of different masses is shown in the lower panel of Fig. \ref{fig:MassFunction}. 
The dashed blue line shows the input HOD model (Eq. \ref{eq:CMASS_Nsat}), while the points show the measurement from one of the mock CMASS catalogues. 

The redshift distribution of the CMASS mocks is shown in the left-most panel of Fig. \ref{fig:GAMA_nz}, and compared with the 
distribution of the CMASS data. 
After selecting the redshift range [0.43-0.7], this public catalogue consists of about $579,000$ galaxies, with an effective area of 6,851 deg$^2$.
Note that the $n(z)$ shown here does not include the weights applied to the CMASS data, which only induce minor modifications to this histogram \citep[see][for more details about the data and the weights]{2016MNRAS.455.1553R}. 

We next implement in our volume-limited mocks the residual incompleteness seen in the data  at high redshift.
We first select all simulated CMASS galaxies in the range $0.43 < z_{\rm spec} < 0.7$, 
then randomly suppress a third of the galaxies in the range $0.6 < z_{\rm spec} < 0.7$.
The resulting $n(z)$ is not a perfect match to the data, however we achieve a 2\% agreement of the mean redshifts,
with  $\langle z \rangle = \sum n(z)\ z\ dz = 0.547$ in the data and 0.557 in the mocks.
The number densities match to within 2\%, with $n_{\rm gal} = 0.0225$ gal/arcmin$^2$ in the CMASS mocks and 0.0230 gal/arcmin$^2$ in the data.


\subsubsection{Clustering of the CMASS mocks}

We assess the accuracy of the mock lens catalogues by comparing the angular correlation function $w(\vartheta)$, described by Eq. \ref{eq:w_theta}, 
to measurements from the data and to predictions from {\small CosmoSIS}.
\textcolor{black}{Both data and mocks are obtained from {\small TREECORR}.} 
For the data measurement, we use random catalogues that are  50 times denser, 
and include the optimal `FKP' weights \citep{FKP} for both the D and R catalogues, and `systematic' weights in the D only \citep[see][for more details on these weights]{2016MNRAS.455.1553R}. 
As discussed therein, one cannot measure $w(\vartheta)$ below the fibre collisions radius of 62 arcsec.
We computed $w(\vartheta)$ in the mocks without any weights, using a set of random catalogues tailored for these simulations and described in Section \ref{subsubsec:randoms}. The results are presented in Fig. \ref{fig:w_theta_CMASS}, showing
that the amplitude of $w(\vartheta)$ is about 10-20\% lower in the mocks  than in the data  in the range \textcolor{black}{$2.0 < \vartheta < 60.0$ arcmin},
just under the $1\sigma$ error. 
Scaling up the  {\small CosmoSIS} $b=1.0$ predictions by a free linear bias parameter, we find that our CMASS mocks have a bias of $b_{\rm CMASS}=2.05$.

At the sub-arcminute scale, the non-linear bias in the mocks becomes important, as shown from the rising clustering amplitude in Fig. \ref{fig:w_theta_CMASS}. 
This should have no impact on current analyses since these scales must be excluded from the data due to fibre collisions. 
One could imagine, however,  to extrapolate the data signal in this region and infer new conclusions about the CMASS galaxies based on our mocks, however we strongly \textcolor{black}{advise}
against this. The reason is that the HOD and NFW parameters have been optimised to match the clustering only over these measured angles, 
and that the mocks could potentially be very wrong at smaller scales.
At large angles, the clustering amplitude in the mocks is again affected  by finite-box effects. 
The dashed black lines in the upper and middle panels of Fig. \ref{fig:w_theta_CMASS} show predictions excluding
these super survey modes, and the effect is relatively well modelled. 
This, along with other known issues, is summarised in Section \ref{subsec:issues}. 

We next compare the  sampling variance measured from the mocks to the jack-knife estimation technique.
Given a data vector $X^j=\{X_1, X_2...X_i \}$ measured $N_{\rm sim}$ times from the mocks ($j = 1, 2,  ... N_{\rm sim}$), 
the covariance between the data elements $X_1$ and $X_2$ is obtained from:
\begin{eqnarray}
{\rm Cov}(X_1,X_2) = \frac{1}{N_{\rm sim}-1}\sum_{j=1}^{N_{\rm sim}} \bigg(X_1^j - \overline{X_1}\bigg)\bigg(X_2^j - \overline{X_2}\bigg).
\label{eq:cov}
\end{eqnarray}
The over-bar denotes the average over the sample and the variance is simply given by the diagonal of the matrix.
 The jack-knife covariance matrix is obtained 
by splitting the CMASS galaxies in 158 sub-volumes, resampling the data 158 time removing one of the sub-volumes at every iteration,
and computing the covariance between these jack-knife samples. 
The mock covariance has been multiplied by $(100/6851)$
in order to area-rescale the results and thereby estimate the covariance of a  CMASS area survey.

We show in the lower panel of Fig. \ref{fig:w_theta_CMASS}  the noise-to-signal ratio, for both the mocks and the data.  
The two estimates converge to within 20\% below 10 arcmin, although 
    the JK estimate is significantly higher than the mock estimate at larger angles.   This result is consistent with previous findings  \citep[][who further compare mock errors with JK estimates in clustering measurements of the RCSLenS, WiggleZ and CMASS data]{Norberg2009, 2016MNRAS.456.2806B}.
The large cusp at $\vartheta \sim 150$ arcmin is caused by the signal crossing zero.


\subsection{Mock LOWZ lens galaxies}
\label{subsubsec:LOWZ}

\begin{figure}
\begin{center}
\includegraphics[width=3.2in]{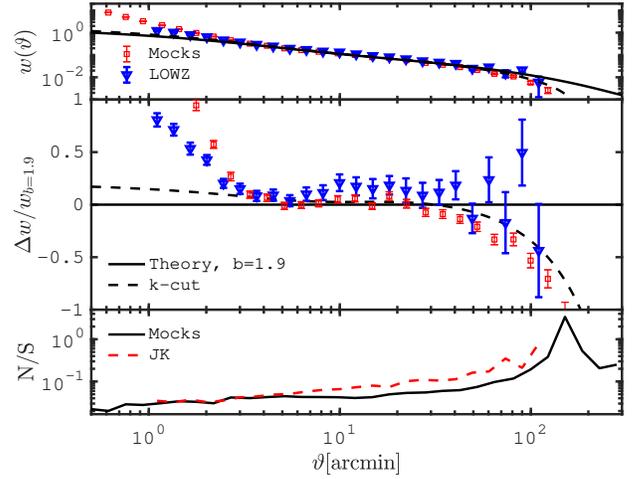}
\caption{Same as Fig. \ref{fig:w_theta_CMASS}, but for the LOWZ mocks, LOWZ-NGP data and predictions.
 The bias in the mocks is comparable to that in the data, 
with $b_{\rm LOWZ} = 1.9$.}
\label{fig:w_theta_LOWZ}
\end{center}
\end{figure}

We construct a suite of LOWZ mock galaxy catalogues that is  meant to reproduce the clustering, 
density and redshift distribution of the BOSS DR12 LOWZ data. 
The HOD follows the same prescription as the CMASS mocks (i.e. Section \ref{subsubsec:CMASS}, 
with Eq. \ref{eq:CMASS_Ncen} and Eq. \ref{eq:CMASS_Nsat}),
but with parameter values now given by the second row in Table \ref{table:HOD_para}.
The mass function ${\rm d}N / {\rm dlog}M_{\rm h}$ and satellite function  $\langle N_{\rm sat} (M_{\rm h}) \rangle$ 
are presented in Fig. \ref{fig:MassFunction}. They generally follow the CMASS mocks, but with noticeable 
differences at the high-mass end. 

The redshift distribution in the mocks is selected in the same range as the data, requiring $z \in [0.15-0.43]$ (see the central panel in Fig. \ref{fig:GAMA_nz}).
After this selection, we are left with a sample of 255,387 LOWZ galaxies from the BOSS NGC region, spread over an effective area of  5836 deg$^2$.
The mean values of the distributions are in good agreement, with $\langle z \rangle$ = 0.31 in the data and 0.32 in the mocks, a 3\% difference.
The effective number density of galaxies in the mocks is $n_{\rm gal} = 0.012  \mbox{ gal/arcmin}^2$,
which is  within 2\% agreement of the data.

\subsubsection{Clustering of the LOWZ mocks}

Our measurement of the angular correlation function from the LOWZ mocks is presented in Fig. \ref{fig:w_theta_LOWZ} and compared against data and predictions
assuming  our best fit galaxy bias of $b_{\rm LOWZ} = 1.9$. The measurement strategies for mock and data are identical to those used for CMASS (see  Section \ref{subsubsec:CMASS}).
We observe that the model agrees well with the mocks and the data for $ \vartheta >3$ arcmin,
and the amplitude of the clustering is about 10\% larger in the data than in the mocks.
The non-linear bias behaves differently in the mocks than in the data at smaller scales,  
such that there is a 10-20\% excess in clustering in the former. 
A similar effect was also observed in the CMASS mocks but for $\vartheta<1$ arcmin (see Fig. \ref{fig:w_theta_CMASS}), 
and  we note here again that these small angular scales are not well fitted 
by the HOD model and should therefore not be over-interpreted. At the largest scales, the finite-box effect is visible and well captured by our modelling 
that excludes the super-box $k$-modes.

As for the CMASS mocks, we see that the (area-rescaled) error estimated from the LOWZ mocks  reconnects with the jack-knife estimate for $\vartheta<10$ arcmin, and that the latter exceeds the former at larger angular separations. 



\subsection{Mock GAMA lens galaxies}
\label{subsubsec:GAMA}

\begin{figure}
\begin{center}
\includegraphics[width=3.2in]{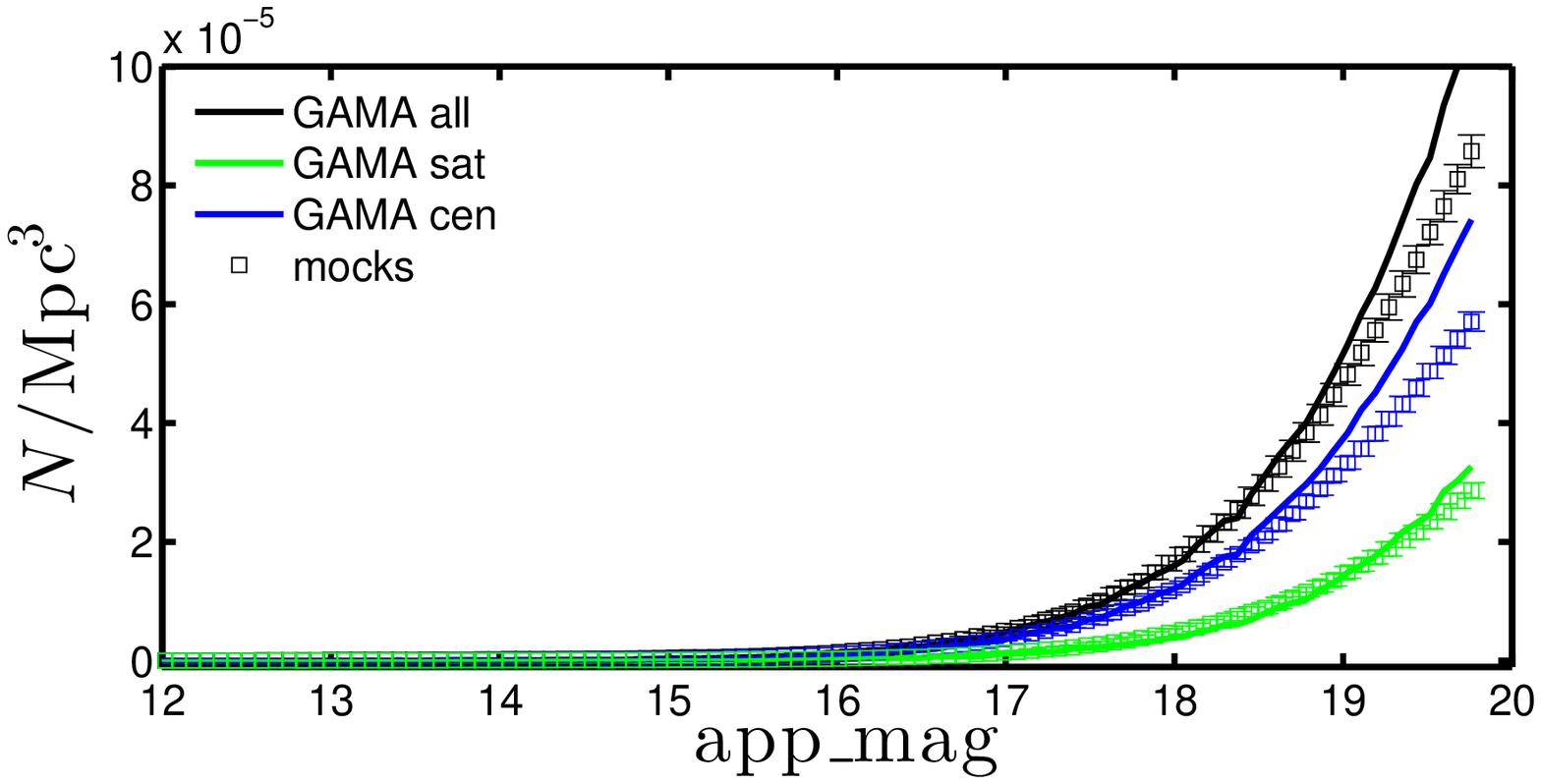}
\includegraphics[width=3.2in]{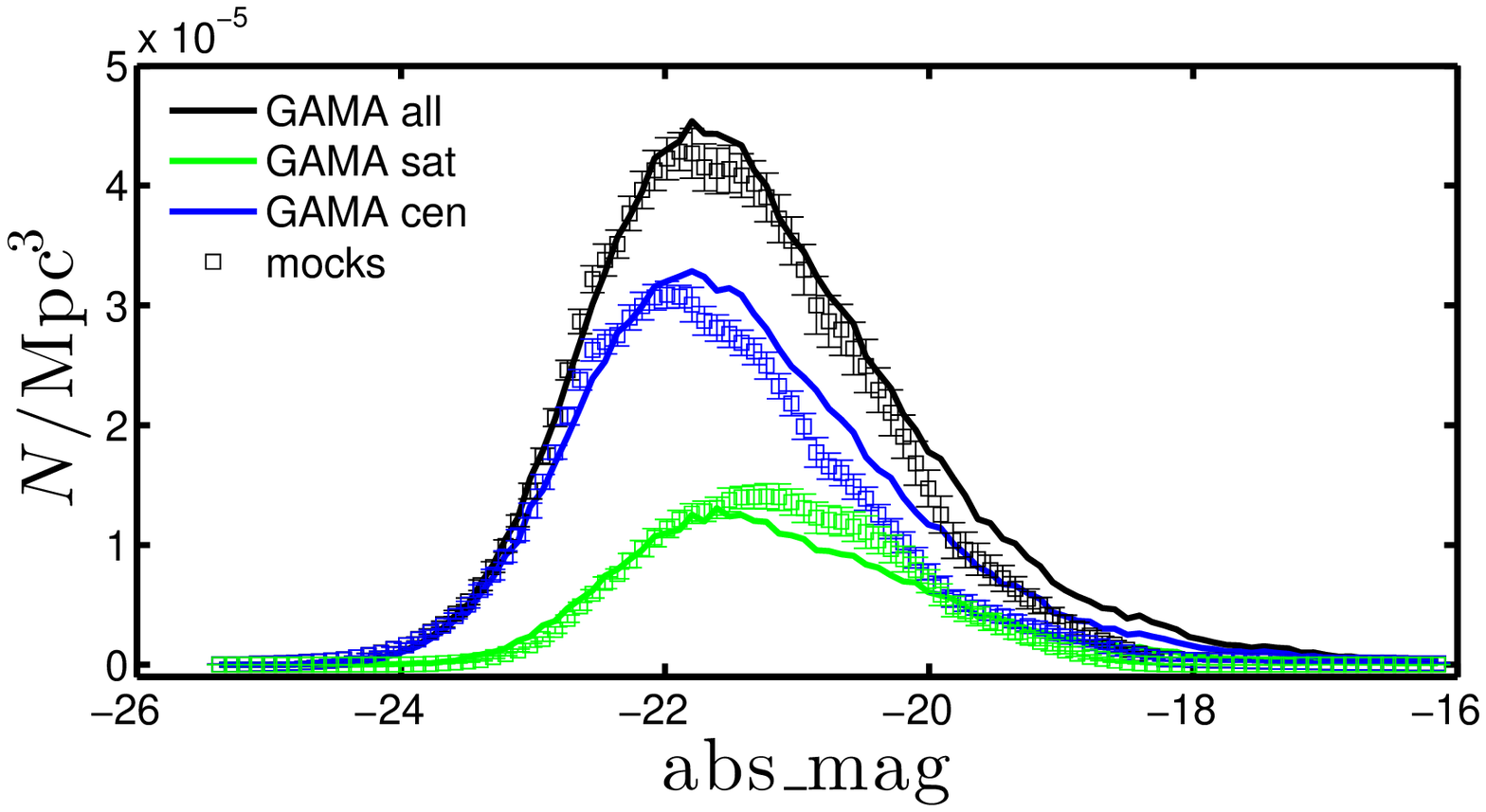}
\caption{ Apparent and absolute $r$-band magnitudes of the GAMA mocks, compared to the data.
There are missing faint objects in the mocks, as seen in the right part of these two panels.}
\label{fig:mag_GAMA}
\end{center}
\end{figure}

\begin{figure}
\begin{center}
\includegraphics[width=3.2in]{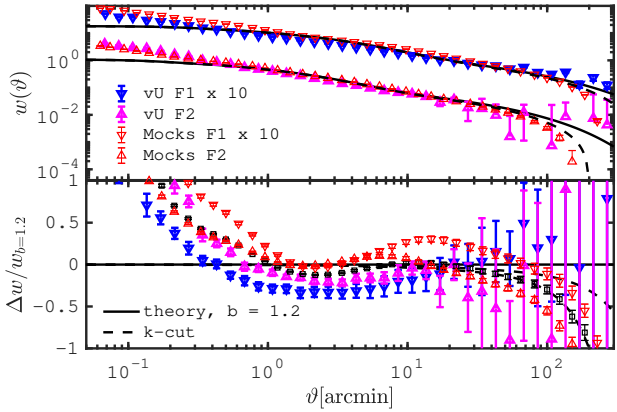}
\caption{({\it upper:}) Angular correlation function of the GAMA mocks, compared to the measurement presented in \citet{2017arXiv170605004V}.
Data and mocks are split in two redshift bins, F1 ($z < 0.2$) and F2 ($z>0.2$), shown in blue and magenta, respectively.
 The F1 data, mocks and theory lines have been multiplied by 10 for improved readability. 
The mocks are averaged from 100 lines-of-sight, the error bars are on the mean. 
Predictions assume a constant galaxy bias of $b_{\rm GAMA}=1.2$, which match well the mocks but is 10\% higher than the data. 
({\it lower:}) Fractional difference with respect to the theoretical predictions. }
\label{fig:w_theta_GAMA}
\end{center}
\end{figure}

The Kilo Degree Survey overlaps with the Galaxy And Mass Assembly survey \citep[][GAMA]{GAMA}, a spectroscopic survey designed  to resolve galaxy groups with
unprecedented completeness. With mean redshift of about $z=0.23$, GAMA probes lower redshifts compared to BOSS, and has been used in combination with KiDS
in a number of galaxy-galaxy lensing analyses that measure halo properties \citep[see][]{2015MNRAS.452.3529V, 2015MNRAS.454.3938S, 2016MNRAS.459.3251V},
scaling relations in groups \citep{2017arXiv171205463J} or  combined-probe cosmological analysis \citep{2017arXiv170605004V}. 
Of particular interest, the GAMA galaxies are marked as satellites, centrals of field galaxies in a group catalogue \citep{GAMA_groups}, which enables astrophysical investigations 
based on these properties.
The additional complication in modelling mock catalogues here is that  GAMA is a magnitude-limited survey, 
which means that in order to match the redshift and clustering of the spectroscopic data, we must first reproduce its apparent magnitude.
 This also means that the volume-limited CMASS and LOWZ HODs that we used in the last sections are not suitable here.

The GAMA HOD prescription follows the model of \citet{2017arXiv170106581S}, 
which is based on a conditional luminosity function (CLF). 
In this approach,  the mean numbers of satellites and centrals depend on the mass of the host halo and on the luminosity range, 
which in absolute magnitude we set to [$-26.7 < M_r<-18.0$].
The number of central galaxies is obtained by integrating the central CLF over that luminosity range. 
Given a halo mass $M_{\rm h}$ and minimum luminosity threshold $L_{\rm min}$, the number of central and satellite galaxies are given by Eq. \ref{eq:CMASS_Ncen} and Eq.\ref{eq:CMASS_Nsat}, provided that we include a luminosity dependence in the following quantities\footnote{Note that the HOD parameters in Eq. \ref{eq:CMASS_Ncen} and Eq. \ref{eq:CMASS_Nsat} are named differently in the papers where they are first introduced. 
There is nevertheless a one-to-one correspondence between our notation $(M_{\rm cut}, \sigma, M_{1}, \kappa, \alpha)$ and that used in \citet{2017arXiv170106581S}:  $(M_{\rm min}, \sigma_{{\rm log} M}, M_0, M_{1}', \alpha)$.} :
\begin{eqnarray}
 N_{\rm cen} (M_{\rm h})& \rightarrow &N_{\rm cen} (>L_{\rm min} | M_{\rm h})\nonumber \\
M_{\rm cut}& \rightarrow& M_{\rm cut}(L_{\rm min})\nonumber \\
\sigma &\rightarrow &\sigma(L_{\rm min})
\end{eqnarray}
and
\begin{eqnarray}
N_{\rm sat} (M_{\rm h})& \rightarrow &N_{\rm sat} (>L_{\rm min} | M_{\rm h})\nonumber \\
M_1 &\rightarrow &M_1(L_{\rm min})\nonumber \\
\kappa &\rightarrow &\kappa(L_{\rm min})\nonumber \\
\alpha &\rightarrow &\alpha(L_{\rm min})
\end{eqnarray}
Therefore, most of the GAMA HOD parameters depend on the host halo mass, on the redshift and on the luminosity limit of the mock survey.
To ease the reading, we report the calculation of these dependences in Appendix \ref{sec:GAMA_HOD}, and skip ahead  
to describe how the luminosity is assigned in the first place.
The luminosity-mass relation of the central galaxies is constructed from a mean function $\langle L_{\rm cen}(M_{\rm h}, z) \rangle$
that is then multiplied in ${\rm log}_{10}$-space by a scatter function implemented from a Gaussian with $\sigma = 0.314$.
This scatter has been chosen such as to introduce stochasticity in the luminosity-mass relation that closely matches the spread in luminosity of the GAMA data.
 We use the modelling and parameter values of  \citet{2017arXiv170106581S} for the mean  luminosity-mass function, taken from \citet{2011ApJ...736...59Z}:
\begin{eqnarray}
\langle L_{\rm cen}(M_{\rm h}, z) \rangle= L_\star \bigg[A_{\rm t}(M_{\rm h}/M_{\rm t})^{\alpha_{\rm M}}  {\rm exp}\bigg(\frac{-M_{\rm t}}{M_{\rm h}} +1.0\bigg)\bigg] \times  10^{0.4Q(z - 0.1)}.
\label{eq:GAMA_L_M}
\end{eqnarray}
 It behaves as a power law with index $\alpha_{\rm M}= 0.264$ at the high-mass end,
that is exponentially suppressed at the low-mass end. The transition occurs around  $M_{\rm t} = 3.08\times 10^{11} ~h^{-1}M_{\odot}$, and is modulated by an amplitude
parameter $A_{\rm t}$ = 0.32 in units of $L_\star = 1.20\times10^{10} ~h^{-2}L_\odot$. The redshift evolution is captured by the parameter $Q = 0.7$, which 
can be turned off by setting $Q=0$. 

The  CLF-based HOD described above provides a luminosity-mass relation and a number of satellites as a function of a luminosity range.
The  luminosity-mass relation is used to assign luminosity to the central galaxies, but this relation does not apply to the satellites,
hence we need a different approach.  We first split the wide [$-26.7 < M_r<-18.0$]  absolute magnitude range into 30 finer bins, then 
use the CLF (Eq. \ref{eq:CMASS_Nsat}) to compute the number of satellites {\it per fine bin}:
\begin{eqnarray}
 \langle N_{\rm sat}^{\rm bin} \rangle = \langle N_{\rm sat} (>L_{\rm max} | M_{\rm h})\rangle  -  \langle N_{\rm sat} (>L_{\rm min} | M_{\rm h}) \rangle,
\end{eqnarray}
 where $L_{\rm min}$ and $L_{\rm max}$ are the fine bin boundaries. 
 These detected satellite objects are then written to the catalogue, and their luminosities are randomly drawn from the luminosity range of the fine bin under study.
At this stage, every object has been assigned a luminosity, which we then convert into absolute and apparent magnitudes (the apparent magnitudes have been $K$-corrected\footnote{The $K$-correction that is discussed here enters in the conversion between absolute and apparent magnitude. It is not to be confused with the $k$-mode correction mentioned previously,  which has to do with missing Fourier modes in a finite volume simulation box.} in the data and in the mocks to $z=0.1$, see details in Appendix  \ref{sec:GAMA_HOD}).
The GAMA mock data are then selected with $z<0.5$ and $m_r <19.8$.

\textcolor{black}{In this section and the next, these GAMA mocks are} compared with the DR3 release\footnote{GAMA:www.gama-survey.org} of the  GAMA data \citep{GAMA_DR3},
 for which the central/satellite status and stellar mass assignments have been estimated \citep{GAMA_groups, GAMA_Mstar}.
Note that the distinction between centrals and  satellites is not as accurate in the data as in the mocks;  
GAMA data assigns 3 classes of galaxies: centrals, satellites and `other', of which the last is interpreted as a field galaxy, or a central with no \textcolor{black}{observed} satellites.
Apparent and absolute magnitudes are extracted from the `{\tt Rpetro}' and `{\tt absmag\_r}' catalogue entries respectively, and 
the same $z<0.5$ and $m_r <19.8$ cuts are applied here as well.


The $r$-band magnitude distributions from the mocks and from the data are both plotted in Fig.~\ref{fig:mag_GAMA}, showing a good overall agreement, even though the details are not exactly reproduced. 
For instance, there is an excess of faint centrals in the mocks (blue line and symbols), but a deficit of faint satellites (green), and these do not perfectly cancel out,
as the deficit is also seen in the combined sample (black). Nevertheless, this disagreement only has a  minor impact on the covariance estimates.
The galaxy mass function and HOD prescription  are presented in Fig.~\ref{fig:MassFunction}, where we see that GAMA galaxies can be 
hosted by dark matter haloes down to $10^{11} M_{\odot}/h$, explaining the higher number density  relative to BOSS galaxies. 
The resulting $n(z)$ is shown  in  the right panel of Fig.~\ref{fig:GAMA_nz}, where the mean redshift of the GAMA data ($\langle z \rangle = 0.227$) and mocks ($\langle z \rangle = 0.253$)
differ by 0.025, or 11\%. 
The number densities match to better than 6\%, with $n_{\rm gal} = 0.244$ (0.260) gal/arcmin$^2$ in the mocks (data). 

\subsubsection{Clustering of the GAMA mocks}

The clustering in the GAMA mocks is presented in Fig. \ref{fig:w_theta_GAMA}, which shows results for all mock galaxies in black,
and for two subsets: F1 selects the $z<0.2$ objects shown with downward-pointing triangles, while F2 selects $0.2 < z < 0.5$, shown with upward-pointing triangles.
These are compared to predictions (in black) and to the measurements from  \citet{2017arXiv170605004V}  (in blue and magenta).
The clustering measurements in the mocks are generally 20\% higher than those from the data (bias is 10\% higher).
We note some deviations from the theory at large scales in the F1 mock data, where the clustering from the mocks overshoots the model by up to 20\% at $\vartheta = 20$ arcmin.
 Scaling the predictions by a free amplitude parameter, we conclude that our mock GAMA sample has a galaxy bias of  $b_{\rm GAMA} = 1.2$.
 Interestingly, the non-linear bias seen at small scales in the mocks is similar to that observed in the data.
 The area of the GAMA survey is too small to allow for  jack-knife resampling,
hence we do not show a comparison between the mocks and the jack-knife error estimates.

\subsubsection{Stellar mass in the GAMA mocks}

\begin{figure}
\begin{center}
\includegraphics[width=3.2in]{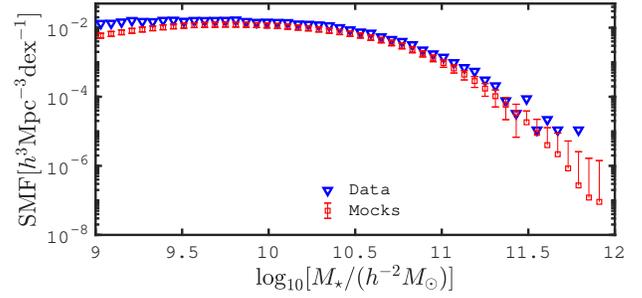}
\caption{
Stellar mass function observed in the GAMA survey (blue), compared to that in the GAMA mocks (red).
Error bars are the  $1\sigma$ scatter, scaled to the survey area.
In both data and mocks, we applied a cut on redshift, requiring $0.01 < z_{\rm spec} < 0.15$.
 The SMF from the mocks significantly undershoots the data  below $10^{9.5}  h^{-2} M_{\odot}$.
}
\vspace{-5mm}
\label{fig:GAMA_SMF}
\end{center}
\end{figure}

 We show in this section how each GAMA galaxy is assigned a stellar mass, thereby opening the possibility of further expanding the data vector in combined-probe analyses. 
The central galaxies are assigned a stellar mass based on the conditional stellar mass function  described in \citet{2016MNRAS.459.3251V} and in \citet{Dvornik18}, with its parameters derived directly from fitting the model to the GAMA data \citep{2016MNRAS.459.3251V}.
The stellar masses for the satellites are assigned with a different method, due to the difficulty in dealing with the sparsity at the low-mass end of the conditional stellar mass function.  
Instead, we take advantage of the linear relation between the absolute $r$-band magnitude and mean stellar mass $\langle M^{\rm sat}_\star \rangle$,
which for the GAMA satellites  in the data can be well fitted by:
\begin{eqnarray}
\log_{10}(\langle M^{\rm sat}_\star \rangle/h^{-2} M_\odot) = -0.47 M_r + 0.56.
\end{eqnarray}
To this linear relation, a magnitude-dependent scatter is added to obtain the satellite stellar mass, with:
\begin{eqnarray}
M^{\rm sat}_\star = \langle M^{\rm sat}_\star \rangle + \sigma_{M^{\rm sat}_\star }.
\end{eqnarray}
The scatter $ \sigma_{M^{\rm sat}_\star }$ is extracted from the data and increases monotonically as the luminosity becomes fainter. 
The typical scatter of $\log_{10}(M^{\rm sat}_\star)$ at the low $M_r$ end ($M_r\gtrsim-19$) is constant at $\sim0.25$, but narrows to $0.14$ at the brighter end ($M_r\sim-22$), where the data starts to become sparse.

We note that for the purpose of generating mock covariance estimates involving galaxy-galaxy lensing in stellar mass bins, assigning the correct stellar mass to the centrals is more important than for the satellites.  
This is because (a) the centrals tend to be more massive, dominating the signal at the high-mass end, (b) centrals in the mocks are directly correlated to the halo centre, hence to the peak of the lensing signal, and c) there are far fewer satellites than centrals in the data and in the mocks.  
We also note that the mock satellites are not correlated to any sub-halo mass concentrations, yielding less lensing signal than would be expected in true data at the small scales (i.e. within a halo).
Instead, the lensing signal from mock satellites is on average close to the expected signal at large separations.



The combined centrals+satellites stellar mass function is shown in Fig.~\ref{fig:GAMA_SMF}, for galaxies with $0.01 < z_{\rm spec} < 0.15$.
This redshift cut is imposed in order to construct a volume-limited sample from the GAMA data, which is necessary for the stellar mass / absolute magnitude relation to stay linear \citep{2016MNRAS.459.3251V}.
We see a deficiency in the overall galaxy counts in the mock, which only comes from the difference in number densities at low redshifts
(see central-right panel in Fig. \ref{fig:GAMA_nz}).  
These mass function data points are nearly fully covariant, and since the mock agrees with the data within a little over $1\sigma$, 
we can expect the error bars derived from the mocks to be representative of the true covariance.
We also see that the mock galaxy counts start to drop significantly relative to the true GAMA counts at stellar masses lower than $10^{9.5}h^{-2}M_\odot$.  
Therefore we recommend that  the covariance estimate from the GAMA mocks should  be limited to stellar masses above this value.


\subsection{KiDS-HOD mocks}
\label{subsubsec:KiDS}

\begin{figure}
\begin{center}
\includegraphics[width=2.95in]{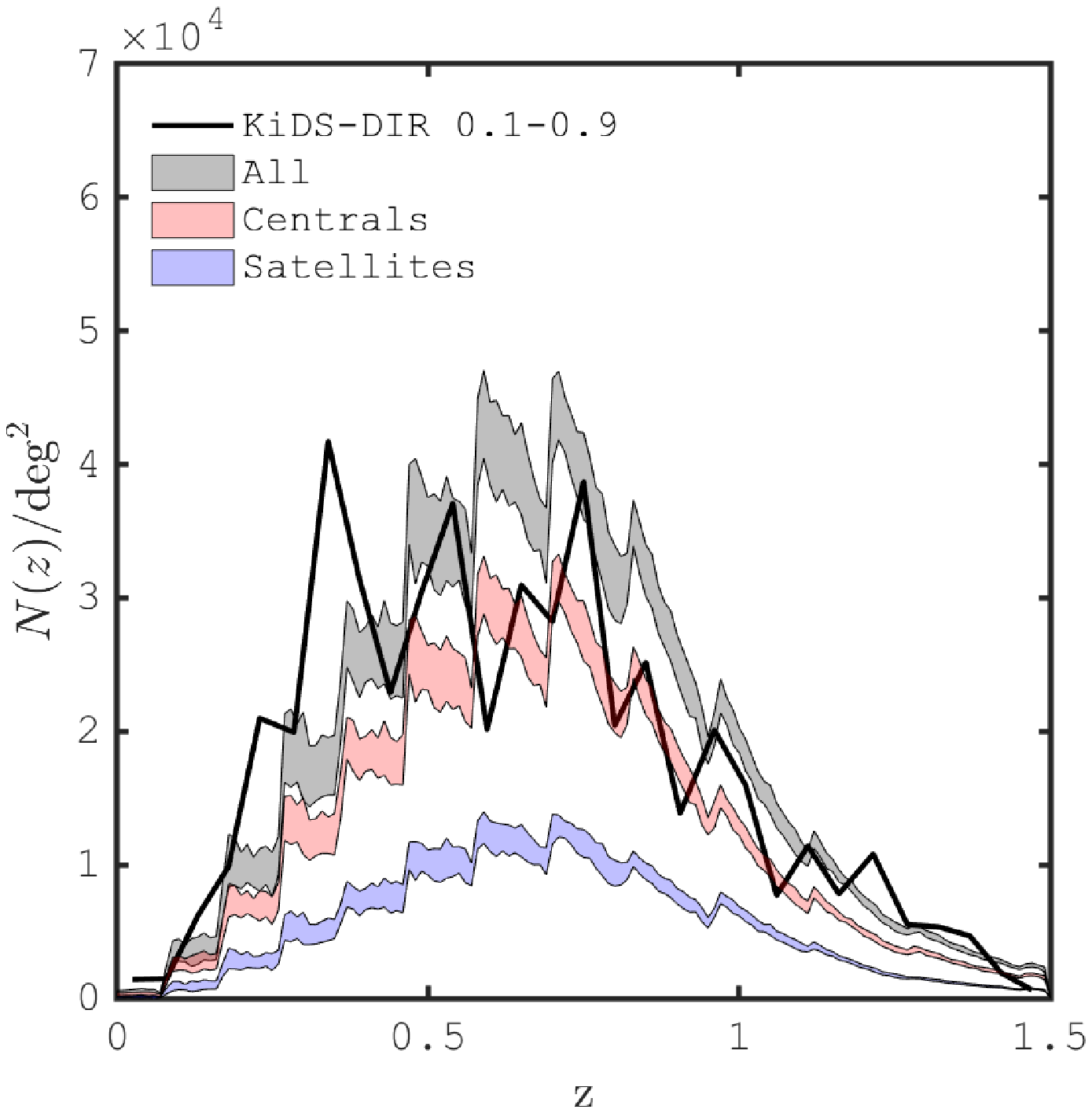}
\includegraphics[width=2.95in]{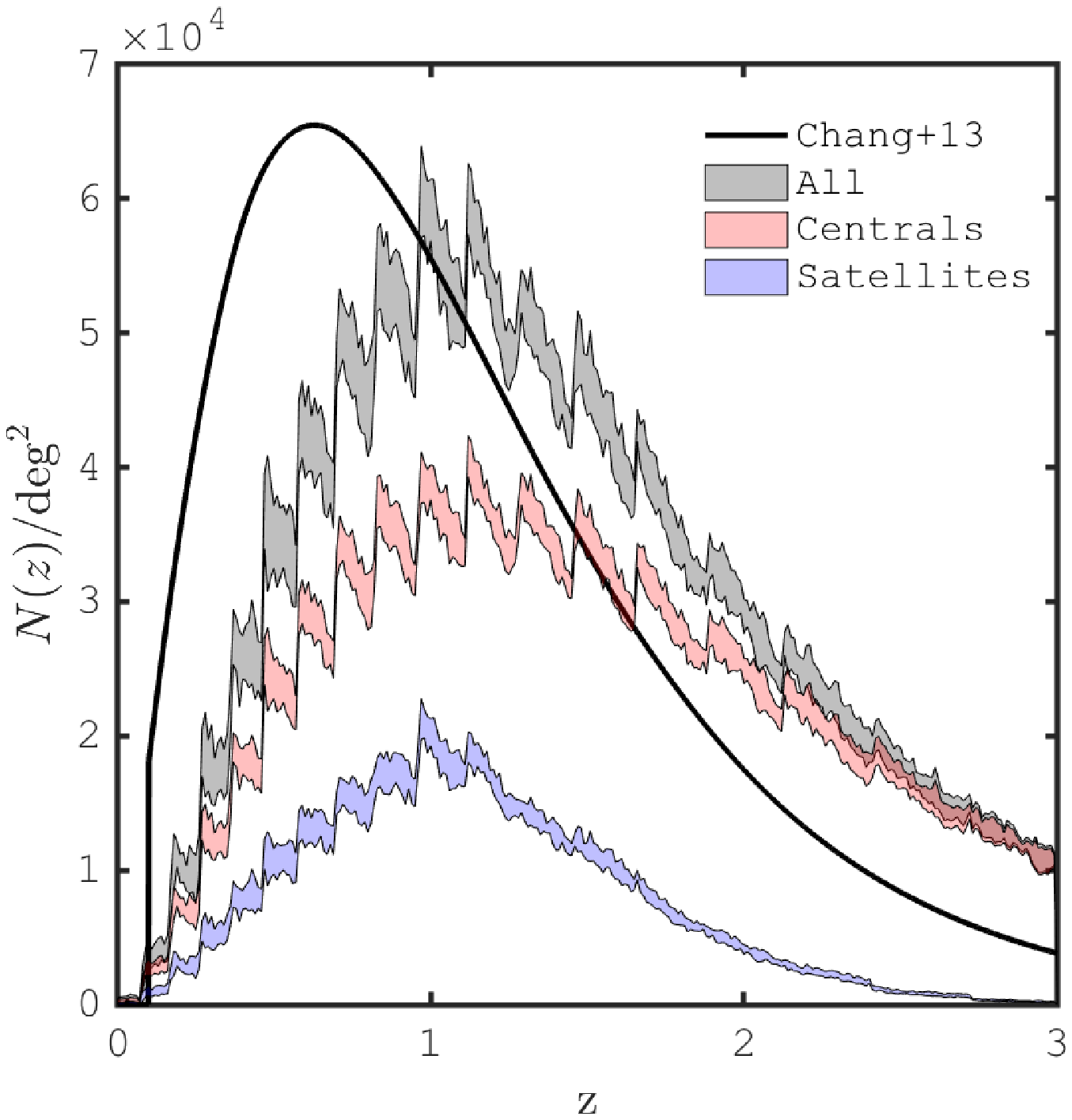}
\caption{Redshift distributions of the KiDS-HOD (upper) and LSST-like HOD (lower) mock catalogues, both based on the GAMA HOD prescription
described in Section \ref{subsubsec:GAMA}. The solid black line in the upper panel is from the KiDS DIR estimate of the distribution after requiring $0.1 < Z_{\rm B} < 0.9$; 
in the lower panel, we show the forecast by \citet{LSSTgal}. \textcolor{black}{The saw-tooth distributions are caused by the multiple-plane tiling algorithm that introduces
step functions in the co-moving volume as a function of redshift. This occurs  at every boundary redshifts listed as $z_{\rm s}$ in Table \ref{table:redshifts}.}}
\label{fig:lsst_nz_hod}
\end{center}
\end{figure}

\begin{figure}
\begin{center}
\includegraphics[width=3.2in]{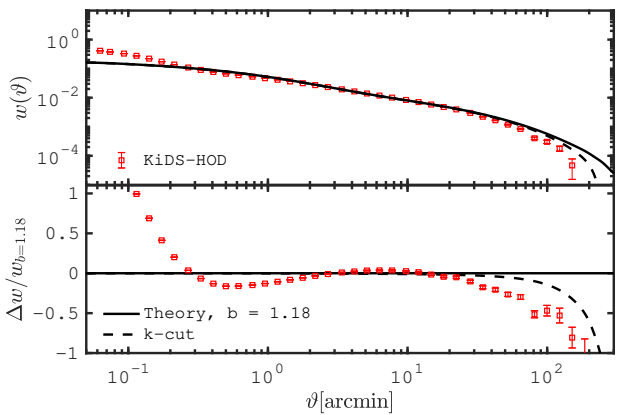}
\caption{({\it upper:}) Angular correlation function of the KiDS-HOD mocks compared to the predictions, assuming a galaxy bias of $b_{\rm KiDS} = 1.18$. 
The mocks are averaged from 100 lines-of-sight, their error bars are on the mean. This measurement has not yet been carried out in the KiDS-450 data.
({\it lower:}) Fractional difference with respect to the predictions. }
\label{fig:w_theta_KiDS}
\end{center}
\end{figure}

 We describe in this section a distinct simulation product  in which galaxies are assigned via a HOD up to $z_{\rm spec} = 2.0$, 
each containing spectroscopic and photometric redshifts, as well as lensing information.
 These galaxies can therefore be used both as sources and lenses,
which can help to explore systematics effects related to weak lensing in a realistic environment. 
Given the large size of these catalogues \textcolor{black}{and their specific application}, we generated these mocks only for a subset of the full SLICS,  providing 120 lines-of-sight.


These catalogues are a straightforward extension of the GAMA HOD model, which is representative of the KiDS data for  apparent $r$-band magnitudes down to 19.8 and $z<0.5$
by construction, and provides empirically motivated mock catalogues at fainter magnitudes and higher redshifts. 
Photometric redshift estimates $Z_{\rm B}$ are based on the joint PDF presented in Fig. \ref{fig:nz_w} and
the lensing quantities $\gamma_{1,2}$ and $\epsilon^{\rm obs}_{1,2}$ are computed from the shear maps, the latter assuming $\epsilon^{\rm n}=0.29$ per component.

Important features of these HOD galaxies relevant for weak lensing measurements are:
\begin{itemize}
\item the spectroscopic $n(z)$ and number density naturally emerge from the HOD,
\item all objects are clustered in a realistic manner,
\item the CLF based calculation allows for selection strategies based on apparent or absolute magnitude,
\item by construction, the different light cones have  different numbers of haloes, hence different numbers of galaxies.
\end{itemize}
This mock can be used, for example,  to validate redshift recovery methods based on cross-correlations \citep{TheWiZZ, 2017arXiv171002517D}, 
to verify the residual impact of source-lens coupling \citep{2007MNRAS.379.1507F}, or to study detailed selection effects caused by close neighbours (see Section \ref{subsec:close_pairs}).

The construction of the KiDS-HOD mocks starts with the same steps as the GAMA mocks (same HOD parameters, same luminosity function, see Section \ref{subsubsec:GAMA}).
Instead of applying a $K$-correction followed by a redshift and magnitude cut however,
we  match the KiDS-450 DIR redshift distribution (assuming a cut in photometric redshift of $Z_{\rm B} \in [0.1 - 0.9]$) 
by down-sampling the volume-limited mock catalogue.
 In particular, we want to preserve the shape of the $n(z)$ for $z<0.4$, but at the same time we need to suppress higher-redshift galaxies in a manner that reproduces the tail seen in the data. 
After exploring a few different methods, we find a good match by filtering the galaxy sample with a down-sampling function $f(z_{\rm spec})$ 
defined as:
\begin{eqnarray}
    f(z_{\rm spec}) = \begin{cases} \frac{0.95}{17.0(z_{\rm spec}-0.4)^4 + 1.0} & \mbox{ for \hspace{5mm}}  z_{\rm spec}> 0.4 \\
                0.95 & \mbox{ for \hspace{5mm}}  z_{\rm spec}< 0.4 \end{cases}
\end{eqnarray}
In other words, we randomly select a fraction $f(z_{\rm spec})$ of all galaxies with spectroscopic redshift $z_{\rm spec}$.
This empirical function suppresses the high redshift objects by the right amount up to  $z_{\rm spec}=2.0$.
The resulting $n(z)$ is shown in the upper panel of Fig. \ref{fig:lsst_nz_hod}, 
which highlights the match between the KiDS mocks and the KiDS-450 data.
The mean redshift in the mocks is $ \langle z \rangle = 0.69$ whereas it is 4\% higher in the data with $Z_{\rm B} \in [0.1-0.9]$.
The number density is $n_{\rm gal} = 7.55$ gal/arcmin$^2$ in the mocks and matches the data to better than a percent, 
where $n_{\rm gal} =$ 7.53 gal/arcmin$^2$.
Alternatively, we could have down-sampled the mocks to match  the KiDS DIR $n(z)$ bin-by-bin,  however this distribution
is relatively noisy, and we opted instead for a strategy that did not introduce more features.

We measure the clustering  $w(\vartheta)$ from these mocks,  shown in Fig. \ref{fig:w_theta_KiDS}, where we compare the results  
to a theoretical calculation with the same $n(z)$ and scale by a free linear bias parameter. 
We see that the mocks and predictions agree over a range 
of scales, from which we deduce that the bias in our mock data is  $b_{\rm KiDS} = 1.18$.  Departure from the linear bias model  apparent for $\vartheta < 2.0$ arcmin,
and significant for $\vartheta < 0.2$ arcmin. 


Since the number density of galaxies fluctuates between lines-of-sights,  the distributions  of the sources and of the lenses
would both contribute to the covariance in a weak lensing measurement. This would introduce an additional variance compared to a suite of mock catalogues  all constructed with 
a fixed $n(z)$, such as the KiDS-450 source catalogue presented in Section \ref{subsec:V0}.
Depending on the error analysis strategy, this additional variance might already have been included elsewhere, such 
that there is a risk of double counting that component to the uncertainty. 
This is why we advocate against using these KiDS-HOD mocks for cosmic shear covariance estimation. 


We also want to stress that there is no guarantee that the (low-redshift) GAMA luminosity function is accurate once extrapolated to higher redshifts.
This could have an impact on some science applications, but not if the  requirements on the mocks are only to be realistic and representative, such
as  for the study of the neighbour-exclusion bias (see Section \ref{subsec:close_pairs}).


\subsection{LSST-like HOD mocks}
\label{subsubsec:LSST-HOD}

Although the KiDS-HOD mock presented in the previous section is designed to emulate current weak lensing surveys, its galaxy number density is  lower than 
the forecasted values of future surveys. 
Following the same procedure, we describe here a separate mock that can be used for upcoming experiments:
we extend the GAMA HOD up to $z=3$ and produce an LSST-like mock\footnote{Note that this mock product differs from the other LSST mock presented in Appendix \ref{subsubsec:LSST}, in which the $n(z)$ is imposed
and galaxy positions are placed at random in the light cones.}
with the redshift distribution presented in the lower panel of Fig. \ref{fig:lsst_nz_hod}. This corresponds to a magnitude-limited survey with 
a cut at $m_r = 26.8$, and has a number density of $n_{\rm gal} = 25.8$ gal/arcmin$^2$. 

We observe that the redshift distribution is shifted to higher redshifts
compared to the  \citet{LSSTgal} forecast, due to the difficulty to produce as many low redshift galaxies as required by the 
forecasted $n(z)$. This would require the SLICS to resolve  lower mass haloes, or the HOD  to populate 
each halo with more satellites, or even to include these missing objects as `field galaxies', placed at random in the light cones.
It is not clear which of the above-mentioned methods would provide the most realistic mock data, hence we decided to simply extrapolate the GAMA HOD to larger 
redshifts and find the apparent magnitude cut that best reproduces the object density, at the cost of biasing the mean redshift towards higher values. 
Overlooking this difference, these LSST-like mocks are representative  of what future lensing  data might look like, and can be used to test 
different aspects of the weak lensing analyses that require a HOD back-bone construction (source-lens coupling, close neighbours studies, etc.). 
In particular, we use them in our analysis of the  neighbour-exclusion bias in Section \ref{subsec:close_pairs}.

\subsection{Preparing mocks for other surveys}
\label{subsubsec:2dFLenS}

HOD prescriptions similar to those presented in the preceding sections can be used in conjunction with the halo catalogues to generate mock galaxy catalogues that emulate  other surveys. This task can be  made easier when the data selection strategy resembles that of a surveys for which mocks are already available.
 For example, the galaxy selection of the 2-degree Field Lensing Survey LRG sample \citep[][2dFLenS]{2016MNRAS.462.4240B} is very close to the BOSS CMASS and LOWZ targets,
with the main difference being a lower redshift completeness  \citep{2016MNRAS.462.4240B}.
We hence do not need to construct a separate 2dFLenS mock sample, as it is possible to match the density of the  data simply 
by randomly down-sampling the BOSS mocks by 50\%. This approach has been used in \citet{2016MNRAS.462.4240B} and in \citet{Amon2017}.

 Our HOD method could be used to construct mock data that resemble the Dark Energy Survey lens sample \citep[the redMaGiC sample, see][]{redmagic}, 
the WiggleZ spectroscopic galaxy sample \citep{WiggleZ_data} or upcoming data
from LSST\footnote{LSST: https://www.lsst.org} or DESI\footnote{DESI: desi.lbl.gov}, as they become available.

 \subsection{Random Catalogues}
\label{subsubsec:randoms}

When measuring clustering in configuration space (with i.e. the Landy-Szalay estimator described in Eq. \ref{eq:w_theta}), 
 a `random' catalogue must be provided. Extra care must be taken to ensure that the random catalogue reproduces the $n(z)$ and the two-dimensional geometry of the 
data (or mocks), otherwise the estimator is no longer unbiased, and can contain significant systematic features. 
The density of the randoms is typically increased compared to the data, while the mask and survey boundaries
are preserved. It has become common for public releases of clustering data to also provide a set of random catalogues tailored for the survey,
and we describe in this section how we construct a similar set of randoms to be used with our simulated data products.

This is not as straightforward as it seems, owing to the fact that  the SLICS simulations are produced from the multiple plane approximation (see Section \ref{subsec:planes}).
The three-dimensional volume that ends up in the light cone {\it is not} a cone or a pyramid, but a sequence of steps.
 It is essential that the randoms follow this three-dimensional selection function
inherent in the mocks.
Also, since the randoms must be tailored to the mock data for which we wish to measure $w(\vartheta)$, 
they follow the $n(z)$ from the  mock surveys (and not the $n(z)$ from the data).

We populate the randoms with ten times the density of the mock data, 
and  distribute the galaxies randomly within the pixels of the 100 deg$^2$  light cone (what we call the `ray-tracing' coordinate frame).
We finally transform these positions into `clustering coordinates', a procedure that imparts the three-dimensional geometry of the SLICS light cones (see Appendix \ref{subsec:coordinates} for details on these two coordinate frames). 
We produce randoms for the CMASS, LOWZ, GAMA and  KiDS-HOD mocks, as well as for the $z=0.2$ halo sample used in Fig. \ref{fig:w_theta_halo}. 
These catalogues contain three quantities per object: $(x_{\rm clustering},y_{\rm clustering}, z_{\rm spec})$, and are used in all $w(\vartheta)$ measurements presented in this paper.

\subsubsection{Data products: galaxy catalogues}
\label{subsubsec:data:galaxies}

We provide the following galaxy catalogues:
\begin{itemize}
\item{KiDS-450 and KiDS-450-dense source galaxies, whose positions are placed at random in the light cone (see Section \ref{subsec:V0});}
\item{LSST-like source galaxies, whose positions are placed at random in the light cone (see Appendix \ref{subsubsec:LSST});}
\item{CMASS, LOWZ and GAMA spectroscopic lens galaxies, whose positions emerge from the HODs (see Sections \ref{subsubsec:CMASS} - \ref{subsubsec:GAMA});}
\item{KiDS-HOD and LSST-like HOD galaxies, whose positions emerge from the HODs (see Sections \ref{subsubsec:KiDS} -\ref{subsubsec:LSST-HOD}).}
\item{Random catalogues for clustering measurements with the CMASS, LOWZ, GAMA and KiDS-HOD catalogues.}
\end{itemize}

Additionally, we provide mock KiDS-450 observations covering the full mosaics,  with mock galaxies placed at the exact same location as in the data.  
This additional mock is meant to be used primarily for peak statistics \citep[as in][]{2017arXiv170907678M} or other measurements sensitive to variations in source  number density,  and is described in Appendix \ref{subsec:V1}.

\subsection{Summary of known limitations}
\label{subsec:issues}

Numerical simulations, \textcolor{black}{including all those listed in Table \ref{table:Nbody}}, always have built-in limitations that must be documented and acknowledged, especially when choosing the regime 
where the mocks are accurate and suitable for their science case.
It is sometimes possible to forward-model these limitations in a comparison between mock measurements and predictions, 
as for the case of mass resolution or finite-box effect.   When this is possible, the observed mis-matches are significantly reduced and can be ignored, 
especially when using the mocks for the calibration of estimators. 
However,  fully accounting for these systematic effects is generally less obvious, for example for the estimation of covariance matrices, as discussed in HvW15.
It is advisable then to exclude the elements of the data vectors  for which the contamination level is important.
 
We list in this section all the known limitations from the SLICS mock catalogues that might or might not affect the analyses they are used for.
These were previously discussed in the main text, and we strongly recommend that the users carefully read them in order to make precise statements about their measurements from the SLICS simulations.

\begin{enumerate}
\item{There are no neutrino nor baryon feedback mechanisms: these mocks emulate a post-recombination universe in which all matter behaves as collisionless dark matter, with the imprint from the baryonic acoustic oscillations.}
\item{The particle mass resolution is $2.88\times 10^9 ~h^{-1}M_{\odot}$, and haloes made of less than 100 particles are not fully resolved. 
This incompleteness is visible from the halo mass function, in Fig. \ref{fig:dndm}, and could be inconsistent with some data samples 
that have a significant fraction of these low-mass haloes.}
\item{Finite resolution affects small angles (i.e. $\vartheta\lesssim$ 1 arcmin in $\xi_+$ at $z\sim0.5$, and $\vartheta\lesssim$ 5 arcmin in $\xi_-$). 
For the $w(\vartheta)$ measurement, this effect is degenerate with the non-linear halo bias that occurs at small scales. Generally, $k$-modes smaller that $2.0 ~h{\rm Mpc}^{-1}$
are well resolved.}
\item{Finite box effects affect large angles (i.e. $\vartheta\gtrsim$ 1 deg in $w(\vartheta)$ and  $\vartheta\gtrsim$ 0.5 deg in $\xi_+(\vartheta)$ ).
These can be identified and modelled from predictions in which $k$-modes larger than $2\pi/(505~h^{-1}{\rm Mpc})$ have been removed. 
The sampling variance extracted from the SLICS mocks should also be scaled using this modelling, as shown in HvW15.}
\item{The correlation across mass sheets has been explicitly broken, hence any 3D measurement should be performed only inside
individual lens sub-volumes. We refer the reader to the values of $z_{\rm s}$ in Table \ref{table:redshifts} in order to split the mock data
in a manner that is insensitive to this. The data should then be split in the same way for consistency.}
\item{Although the $n(z)$ and $n_{\rm gal}$ of the KiDS-450 mocks match the data without the $Z_{\rm B}$ cuts, discrepancies are observed in tomographic analyses (see Table \ref{table:n_eff_comp}). In that case, the $n(z)$ still matches the data by construction, but  $n_{\rm gal}$ does not. 
Since this can be critical to many analyses, we recommend to use the KiDS-450-dense mock  instead,
then down-sample the catalogues to recover the  $n_{\rm gal}$ from the data in whatever $Z_{\rm B}$ slice is being analysed.}
\item{We have only measured the angular correlation function in broad redshift bins. Finer tomographic binning may reveal larger discrepancies. }
\item{Clustering measurements in our mocks are generally in close agreement with the data, but the linear bias sometimes differs by about 10\%.
 This is partly caused by differences in cosmology, which affects the clustering. We nevertheless recommend to consider and propagate these differences in data analyses, possibly by rescaling the mock measurements.}
\item{In the GAMA mocks, the $K$-corrections are degenerate with the redshift evolution of the luminosity function. 
We calibrate these together to empirically reproduce the $n(z)$ given an apparent magnitude cut. 
However, the underlying luminosity function in the mocks might no longer be a good match to that of the data without the $K$-correction. }
\item{Satellite galaxies are placed according to spherical NFW profiles. We have decided not to use the triaxial profiles 
as there is no strong consensus that sub-haloes necessarily trace the dark matter. 
Additionally, the HOD prescriptions are calibrated assuming spherical NFW, which could make the interpretation less accurate.
However, this means that the galaxy-galaxy lensing signal from the satellites is weaker than in the data, in which many satellite galaxies are believed to reside in sub-haloes/cores.}
\item{The concentration parameter is allowed to vary in order to maximise the agreement with the data in clustering measurements.
This means that a detailed study of the one halo term -- i.e. precise reconstruction of the halo profiles -- might differ between the data and the mocks.}
\item{The inertia matrix provided by our halofinder is not very accurate since no phase space cleaning has been applied before measuring this quantity.
Certainly, the shapes are not reliable for haloes made of less than 400 particles, possibly more.}
\end{enumerate}
\textcolor{black}{Despite these limitations, the SLICS mocks stand out as a particularly useful tool for combined probe studies involving weak lensing 
and remains accurate within the dynamical range listed above.}

\section{Combined-Probe Analyses}
\label{sec:combined_probe}

 Different cosmological probes are sensitive to different redshifts and/or dynamical ranges of the underlying large-scale structure formation. 
Differences in instruments and measurement strategies also \textcolor{black}{mean} that the systematic effects are typically distinct and uncorrelated.
Combinations of probes at the data vector level exploit these advantages and offer complementary cosmological information \textcolor{black}{and opportunities for self-calibration} 
 \citep[see][for recent combined-probes analyses]{2017arXiv170605004V, 2017arXiv170706627J, 2017arXiv170801530D}.
Control samples such as the SLICS are  critical for the estimation of  the correlation between the elements of the combined-probe data vector.

In the next sections we first carry out a galaxy-galaxy lensing measurement in the mocks by combining our KiDS-450 source catalogues with different spectroscopic  lens catalogues, comparing our results with measurements from the data.
We then construct a larger data vector by  adding  1) the clustering of the lenses and 2) the cosmic shear  of the sources. 
 We present 
the full covariance matrix of this combined data vector as a demonstration of what can be achieved with the SLICS.

\subsection{Galaxy-galaxy lensing}
\label{subsec:GGL}

\textcolor{black}{In a galaxy-galaxy lensing measurement, foreground galaxies serve as tracers of the foreground mass concentrations
around which the shapes of background sources are analysed. Even though the full matter distribution is responsible for the lensing signal, 
we hereafter refer to the foreground tracers as `the lenses'.}
This is usually performed with a $\gamma_{\rm t}(\vartheta)$  measurement, obtained by stacking the tangential component of the 
source ellipticities $\epsilon_{\rm t}^{jk}$ for all  pairs  of lenses  and sources (labelled $k$ and $j$, respectively) separated by an angular distance
$\vartheta$. Lenses and sources are \textcolor{black}{generally} assigned weights, $w_k$ and $w_j$ respectively, and the estimated $\widehat{\gamma_{\rm t}}$
is given by: 
\begin{eqnarray}
\widehat{\gamma_{\rm t}}(\vartheta) = \frac{\sum^{N_{\rm pairs}}_{j,k} \epsilon_{\rm t}^j w_j w_k}{\sum^{N_{\rm pairs}}_{j,k}  w_j w_k }.
\label{eq:gamma_t_est}
\end{eqnarray}
The sums are over all pairs for which $\vartheta_{jk}$ falls within predetermined bins. 
 
 Although $\gamma_{\rm t}$ is straightforward to implement in cosmological analyses, it is not necessarily the most optimal  choice.
Instead, one can extract instead the differential surface mass density $\Delta \Sigma(R)$, defined as\footnote{We use the galaxy-galaxy lensing notation from \citet{Dvornik18}:
$\Delta \Sigma_{\rm com}(R)$ and $\Sigma_{\rm cr, com}$ are sometimes labelled $\Delta \Sigma(R)$ and $\Sigma_{\rm crit}$, respectively. This is to be distinguished from measurements in `proper' distance, which we do not use in this section.}:
\begin{eqnarray}
 \Delta \Sigma(R_{\rm com}) = \gamma_{\rm t}(\vartheta) \Sigma_{\rm cr,com},
\end{eqnarray}
where $R \equiv R_{\rm com} = \vartheta \chi(z_{\rm l})$ is the  comoving distance perpendicular to the line-of-sight, and 
\begin{eqnarray}
\Sigma_{\rm crit} = \frac{c^2}{4 \pi G} \frac{D(z_{\rm s})}{D(z_{\rm l})D(z_{\rm l},z_{\rm s})}\frac{1}{(1+z_{\rm l})^2},
\end{eqnarray}
 is the  comoving critical surface mass density. In the above expression, $c$ is the speed of light in vacuum, $G$ is Newton's constant, while 
 $D(z_{\rm s})$, $D(z_{\rm l})$ and $D(z_{\rm l},z_{\rm s})$ are the angular diameter distances to the sources, to the lenses, and between the sources and the lenses.
This estimator is more optimal than $\gamma_{\rm t}$ since the geometrical term downweights source-lens pairs that are close in redshift and that hence carry only little signal 
\citep{2005MNRAS.361.1287M}. In the case where the source redshift is not known for individual objects but estimated for a population,  we measure  instead
\begin{eqnarray}
 \Delta \Sigma(R) = \gamma_{\rm t} /\overline{\Sigma_{\rm cr, com}^{-1}},
\end{eqnarray}
where now the  comoving critical surface mass density is measured for a given lens redshift $z_{\rm l}$: 
\begin{eqnarray}
\overline{\Sigma_{\rm cr, com}^{-1}}[z_{\rm l}] = \frac{4 \pi G}{c^2} (1+z_{\rm l})^2 D(z_{\rm l}) \int_{z_{\rm l}}^{\infty} n(z') \bigg[1 -  \frac{D(z_{\rm l})}{D(z')}\bigg] {\rm d}z'.
\label{eq:sigma_crit_inv}
\end{eqnarray}
\textcolor{black}{We then compute $\gamma_{\rm t}$, $\Sigma_{\rm crit}^{-1}$ and $\Delta \Sigma(R)$} in thin lens slices of width $\Delta z_{\rm l} = 0.01$ 
and stack the signals, weighted by the number of lens per slice.
The angular scales are converted to  comoving scales with the relation $\vartheta = R/\chi(z_{\rm l})$.
\textcolor{black}{To reduce the contamination from foreground galaxies we only consider source galaxies whose photometric redshift satisfy $Z_B > z_l + 0.1$  \citep[see][for full details about the measurement]{Amon_iband}. }

\begin{figure}
\begin{center}
\includegraphics[width=3.2in]{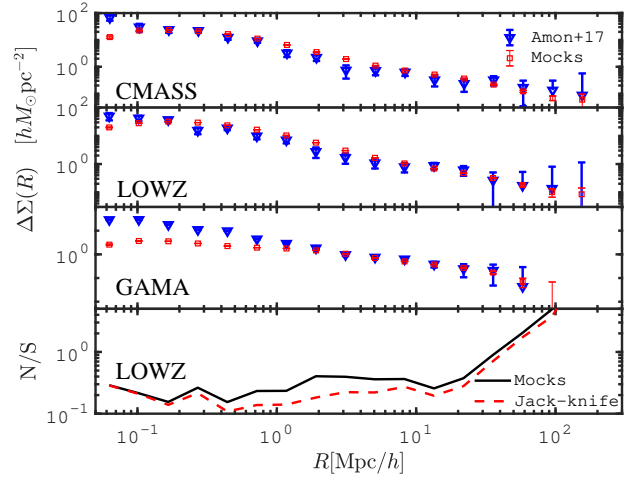}
\caption{({\it upper three panels:}) Differential surface mass density, $\Delta \Sigma$, as measured in the  KiDS-450 and CMASS/LOWZ/GAMA mocks, 
 and compared to the measurement from the data by \citet{Amon2017}. The error bars on the mocks are on the mean, 
 while that on data are from the mocks, scaled to the overlapping survey areas.
 ({\it lowest panel:}) Comparison between the error obtained from the mock covariance about the LOWZ $\times$ KiDS-450 measurement, 
 and the jack-knife estimate from the data.}
\label{fig:gt_theta_CMASS}
\end{center}
\end{figure}

We compare in Fig. \ref{fig:gt_theta_CMASS} the  $\Delta \Sigma(R)$ signal extracted from the KiDS-450 mock sources around the CMASS/LOWZ/GAMA targets, with the measurements from data presented in \citet[]{Amon2017}.  \textcolor{black}{To further ease the comparison with the measurements  of $w(\vartheta)$ for these three mock surveys presented in Figs. \ref{fig:w_theta_CMASS}, \ref{fig:w_theta_LOWZ} and \ref{fig:w_theta_GAMA}, 
it is convenient to note that at their mean redshift ($\langle z \rangle=$ 0.58, 0.32, 0.25), the angles subtended by the comoving size $R=1.0 ~h^{-1}$Mpc  are respectively 2.4, 3.9 and 4.8 arcmin.}

The mocks and data agree within $1\sigma$ over a range of scales,
 however some discrepancies are observed. There is a noticeable difference in the signal at small angular scales for the GAMA survey,
 which is sourced by  the implementation of the satellites in the mocks. Whereas the satellite galaxies in the data are believed to be highly correlated with sub-haloes \citep{2017MNRAS.471.2856V}, 
 the satellites in the mocks are placed at random positions within a NFW profile (see Section \ref{subsec:V2}), 
 which destroys the satellite contribution to the galaxy-galaxy lensing signal. The LOWZ and CMASS mocks are less affected by this missing signal 
because of the smaller satellite fraction, compared to GAMA. 
 
  In absence of an ensemble of mock data, the errors on galaxy-galaxy lensing measurements are often estimated from analytical calculations 
 that neglect the sampling variance, or from bootstrap or jack-knife resampling of the data, which are generally accurate at small scales but perform less well 
 at larger angles  \citep[see figure 5 in][for a comparison between these methods]{2015MNRAS.452.3529V}. 
 Galaxy-galaxy lensing analyses interested in intra-halo properties need not to worry about this, 
 but the same cannot be said about measurements that target the two-halo term, e.g.  to constrain the galaxy bias.

 The SLICS simulations are ideal to test the accuracy of these assumptions, since the error estimated from them 
 contains both the shape noise and the sampling variance. 
   A comparison between the SLICS and an analytical covariance is presented in \citet{2018arXiv180500562B}.
 We show here, in the lowest panel of Fig. \ref{fig:gt_theta_CMASS}, a comparison between the error on $\Delta\Sigma(R)$ obtained from the mocks in a LOWZ $\times$ KiDS-450   analysis,  versus a jack-knife estimate from the data.  
Both error estimates are normalized by the data signal to improve the readability, and their measurements of the noise-to-signal ratio agrees to 
within 10\% at the smallest scales shown here,  but differ by up to a factor of two for $R > 0.7 h^{-1}$Mpc.  

 \textcolor{black}{A clear asset of the SLICS mocks is that they can provide error estimates even for surveys of smaller area (e.g. GAMA), where
internal resampling is not reliable. One can also inspect with these mocks  the relative contributions to the covariance from the shape noise and the sample variance,  
and/or combine the data vectors with other cosmological probes, as shown in the next section.}

\subsection{Covariance for 3$\times$2 point data vectors}
\label{subsec:3x2}

\begin{figure}
\begin{center}
\includegraphics[width=3.2in]{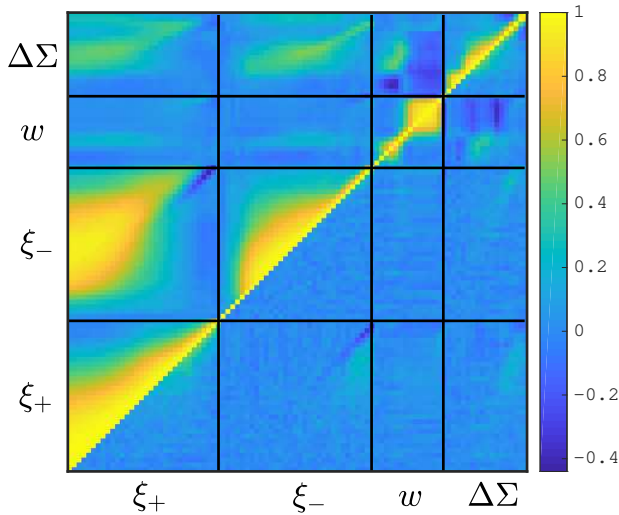}
\caption{Cross-correlation coefficient, $r_{ij} = \mbox{Cov}_{ij}/\sqrt{\mbox{Cov}_{ii}\mbox{Cov}_{jj}}$, for a combined-probe measurement
  involving cosmic shear $\xi_\pm(\vartheta)$ from the KiDS-450 mocks, galaxy clustering $w(\vartheta)$ from the LOWZ mocks, and galaxy-galaxy lensing $\Delta \Sigma(R)$
from the combination of both. 
The cosmic shear segments  of the data vector represent a single tomographic bin of the KiDS-450 mock data, selected 
with $Z_{\rm B} \in [0.5 - 0.7]$. The sources are not binned in the galaxy-galaxy lensing signal. 
Shape noise is included in the lower triangle part of this matrix, which explains why the off-diagonal components of the lensing data are suppressed
compared to the upper triangle part.}
\label{fig:3x2cov}
\end{center}
\end{figure}


 We present here the covariance matrix of  a mock measurement similar in nature to that presented in \citet{2017arXiv170605004V} and \citet{2017arXiv170801530D},
which  combined three measurements of 2-point correlation functions related to the foreground matter field.
The data vector we analyse here consists of the $\xi_{\pm}({\vartheta})$ cosmic shear data points measured from the KiDS-450 mock sources
selected with $0.5 < Z_{\rm B} < 0.7$ (presented in Section \ref{subsec:V0}),
the angular correlation function $w(\vartheta)$ measured from the LOWZ mock  spectroscopic survey (Section \ref{subsubsec:LOWZ}) and the
galaxy-galaxy lensing signal measured from their combination (Section \ref{subsec:GGL}). 
In the latter case, the source galaxies are not binned in $Z_{\rm B}$, and the signal is promoted from $\gamma_{\rm t}(\vartheta)$ to $\Delta \Sigma(R)$
when compared to the two data analyses  mentioned above.

We merge the mock data vector from each  line-of-sight as $X = [\xi_+(\vartheta), \xi_-(\vartheta), w(\vartheta), \Delta \Sigma(R)]$ and compute the full covariance matrix from \textcolor{black}{844} lines-of-sight with Eq. \ref{eq:cov}.
 We present the results in Fig. \ref{fig:3x2cov}, normalized such that the diagonal is equal to one.
The different blocks represent distinct components of the combined data vector, separated by thick black lines.
We use a large number of bins in order to highlight the structure of the matrix, however most  of the points are highly correlated;
far fewer points are required to capture the same information content.

The lensing data used in this calculation include shape noise as an option, 
which down-weights the off-diagonal components of the normalized matrix when turned on  (see the lower triangle part of  Fig. \ref{fig:3x2cov}). Even in this case, it is possible to observe some structured correlation in most blocks, although the noise level on individual elements is significant. The noise free case is shown in the upper triangle part of the matrix in  Fig. \ref{fig:3x2cov},
where we can distinguish significant amounts of cross-correlation between most blocks.

 The covariance matrix presented here  is only one example of a  3$\times$2 point data vector that can be formed from the SLICS, 
and it is straightforward to expand on this and include other data types such as redshift-space distortions, CMB lensing, void lensing or lensing peak count, just to name a few.
In some cases, a combined-probe covariance matrix can be estimated analytically,
which provides an opportunity to validate the two approaches.  
Indeed, the halo model offers a prescription to compute this quantity via the trispectrum \citep{Takada2009a, cosmolike}.
Some measurements however are  harder or currently impossible to integrate in this framework (redshift-space distortions, void lensing, peak counts, non-linear transforms, etc.),
but are fully accessible with the SLICS mocks. The caveat, of course, is that the covariance estimated from the mocks will be at a fixed cosmology,
and subject to a limited precision due to the finite number of mocks.
Lastly, as mentioned in the introduction, the estimate will be biased low due to the missing Super Sample Covariance term. In practice, this term can be evaluated with response functions
from `separate universe' simulations \citep{SSC}, by comparing the results to simulations with larger volumes (e.g. the HSC mocks presented in Table \ref{table:Nbody}) or from
Gaussian realisations. In \citet{2017arXiv170605004V}, it was shown that the  SLICS mocks contain some contribution to the SSC term from the simulation volume outside of the light-cone.
The missing contribution would inflate the cosmic shear error by 10-70\% depending on the angular scale, but has no effect on the galaxy-galaxy lensing error. This is indeed an important ingredient that must supplement the covariance estimate extracted from the SLICS.

\section{Neighbour-Exclusion Bias on Cosmic Shear}
\label{subsec:close_pairs}

{In this second part of the paper, we make use of the 120 KiDS-HOD and LSST-like HOD mocks described in Sections \ref{subsubsec:KiDS} and \ref{subsubsec:LSST-HOD} to revisit a selection effect related to close neighbours that was first identified in \citet{2011A&A...528A..51H}. 
The general idea is that isophote overlap makes the positions  and shape measurement generally difficult, inaccurate or biased for these objects.
For this reason,  they are either fully removed or downweighted, depending on the analysis strategy,
and this choice introduces a selection effect that is not random on the sky due to clustering. 
Indeed, galaxy clusters have the highest density of objects, hence they have higher chances to contain close neighbours and blended objects.
This effect exists even in smaller systems, since any background galaxy that exactly aligns with a foreground massive dark matter halo is obscured by the 
 central galaxy.
 This translates into an effective down-sampling of the foreground over-densities compared to the rest of the sky, a bias that affects the cosmic shear signal. 
As a corollary, voids in the foreground have lesser chances of containing close neighbours and are thus given more weight.
}


\citet{2011A&A...528A..51H} studied this effect using the Millennium Simulations, and reported an impact of a few percent to tens of percent 
on  cosmic shear measurements, depending on the scales, redshifts and definition of `close galaxy pairs' that {are excluded}.
 \citet{DES-SV_MacCrann} studied this effect -- which they referred to as the {\it blend-exclusion bias} --  in a  cosmic shear analysis of the DES-Science Verification data
 targeted at small-scales. 
They estimated its impact from image simulations and provided a physically motivated model of this bias, and finally investigated different degeneracy directions, notably  with the sum of neutrino mass and with baryon feedback parameters. {In this paper, we term this effect the {\it neighbour-exclusion bias} on the weak lensing signal, following the 
terminology of \citet{DESY1_Samuroff}\footnote{The {\it neighbour-exclusion bias} is the exact same phenomenon that was coined the {\it blend-exclusion bias} in  \citet{DES-SV_MacCrann}, but this latter name can lead to a confusion since by definition,  `blended' galaxies refer to nearly complete overlap of two objects that makes them nearly undistinguishable. 
These normally appear as a single catalogue entry with high shape noise.
Our naming captures the fact that this  selection effect operates mainly on pairs of galaxies that are close but distinguishable.}}, 
in which a number of neighbour-induced biases have also been studied in the context of cosmic shear measurements with the first year of DES data. 
Their strategy was different as they used pairs of simulations with and without clustering, and merged multiple shear biases into a scale-dependent multiplicative correction term.
This method is accurate, but its calibration also depends on the survey depth, redshift distribution, and on the exact galaxy sample used. 
In the end, mock data such as the SLICS are required for validating the framework.

 We build on these preceding results by exploring 
the impact on different cosmic shear analysis pipelines with shallow
and deep mock data, with an eye on the signal in cross-tomographic bin combinations and on the residual effect at large angles.  
We focus at first on the (shallower) KiDS-450 survey, and discuss the (deeper) LSST-like survey afterward.
%

\subsection{Measurement from mocks}
\label{subsec:blend_mocks}

We start by identifying close pairs in the full  KiDS-HOD mocks with a KDTree \citep{KDTree} algorithm\footnote{We used the python module 
 {\it scipy.spatial.KDTree}.}. During this stage, we use different definitions of close pairs: objects separated by less than: 1.0, 2.0, 3.7, and 5.0 arcsec
 on the sky.    These {\it \textcolor{black}{exclusion angles}} are meant to represent the variable shape measurement quality confronted to realistic seeing conditions,
and the largest three of these are taken from \citet{2011A&A...528A..51H} for validation and cross-reference;  the 1.0 arcsec separation targets future surveys.   
 We next adopt two strategies to deal with the close pairs we have found. 
 Either we reject the faintest of the two galaxies in a pair -- we refer to this technique as `FAINT' --
 or we remove both galaxies from the catalogues -- we refer to this as  `BOTH'.
These two cases emulate different pipelines currently used in weak lensing  analyses.

Removing close pairs from the catalogues has two effects: 1) it modifies the mean redshift distribution,   preferentially removing high-redshift galaxies since their mean angular separation is smaller due to their larger distance from us, and 2) it preferentially reduces the number of galaxy pairs aligned with foreground structures, which is the anisotropic selection effect at the core of this study.
The first effect does no harm to a data analysis as it is effectively a down-sampling of the data: as long as the estimated final $n(z)$ is accurate, the inferred cosmology 
will not be affected by this.
The second effect is more problematic however as it correlates with the foreground matter distribution.
This is similar to the selection effect described in  \citet{2015MNRAS.449.1259S}, who examined the impact of  cluster obscuration
on cluster mass reconstruction with stacked shear and magnification signals \textcolor{black}{\citep[see also][]{2015MNRAS.449..685H}}. 

To isolate the neighbour-exclusion bias,
we need to factor out the first effect. Following the `FIX' criterion \citet{2011A&A...528A..51H} we proceed as \textcolor{black}{follows}:
\begin{enumerate}
\item{Find and remove close pairs from the KiDS-HOD mocks.
The outcome of this are `filtered catalogues'. We repeat this for the range of \textcolor{black}{exclusion} angles and for the two selection criteria, BOTH and FAINT.}
\item{Split the original and filtered catalogues in four tomographic bins.  For this work, we reduce the complication arising from photometric error and split our data 
according to their true redshift: $z_{\rm spec} \in [0.1 - 0.3], [0.3 - 0.5], [0.5 - 0.7]$ and $[0.7 - 0.9]$.}
\item{Measure the $\widehat{\xi^{ij}}_{\pm}(\vartheta)$ signal from the original and filtered catalogues, in all pairs of tomographic bins $(i,j)$. }
\item{Measure the original and filtered $n(z)$, and compute the associated theoretical  predictions $\xi^{ij}_{\pm}(\vartheta)$.}
\end{enumerate}
The neighbour-exclusion bias can then be quantified as:
\begin{eqnarray}
\beta_{\pm}^{ij}(\vartheta) = \bigg(\frac{\widehat{\xi^{ij}}_{\pm}(\vartheta) \big|_{\rm filtered}}{\widehat{\xi^{ij}}_{\pm}(\vartheta)} \bigg) \times\bigg( \frac{\xi^{ij}_{\pm}(\vartheta)}{\xi^{ij}_{\pm}(\vartheta) \big|_{\rm filtered}} \bigg)
\label{eq:beta}
\end{eqnarray}
with $i,j = 1...4$. The second factor on the right hand side of  Eq. \ref{eq:beta} removes the effects of the modified $n(z)$ after filtering, and leaves 
$\beta^{ij}_{\pm}$ sourced only by the neighbour-exclusion effect. 
In the absence of selection effects, both brackets would cancel out exactly, resulting in $\beta^{ij}_{\pm} = 1.0$.

\begin{figure}
\begin{center}
\includegraphics[width=3.2in]{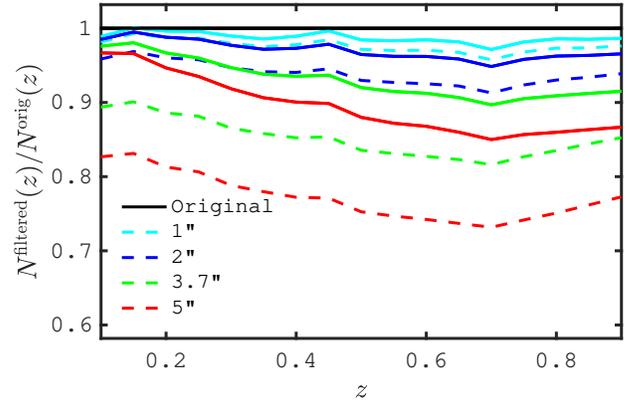}
\caption{ Effect of the close pair selection on the number density of objects in the 120 KiDS-HOD mocks galaxy catalogues,
presented as the ratio between the filtered and original $N(z)$.
Different colours represent different opening angles in the close pairs selection criteria,
solid lines represent the FAINT rejection scheme, dashed lines show BOTH. 
 Approximately twice as many objects are rejected in the latter case.}
\label{fig:nz_cp_both}
\end{center}
\end{figure}

\begin{figure*}
\begin{center}
\includegraphics[width=2.9in]{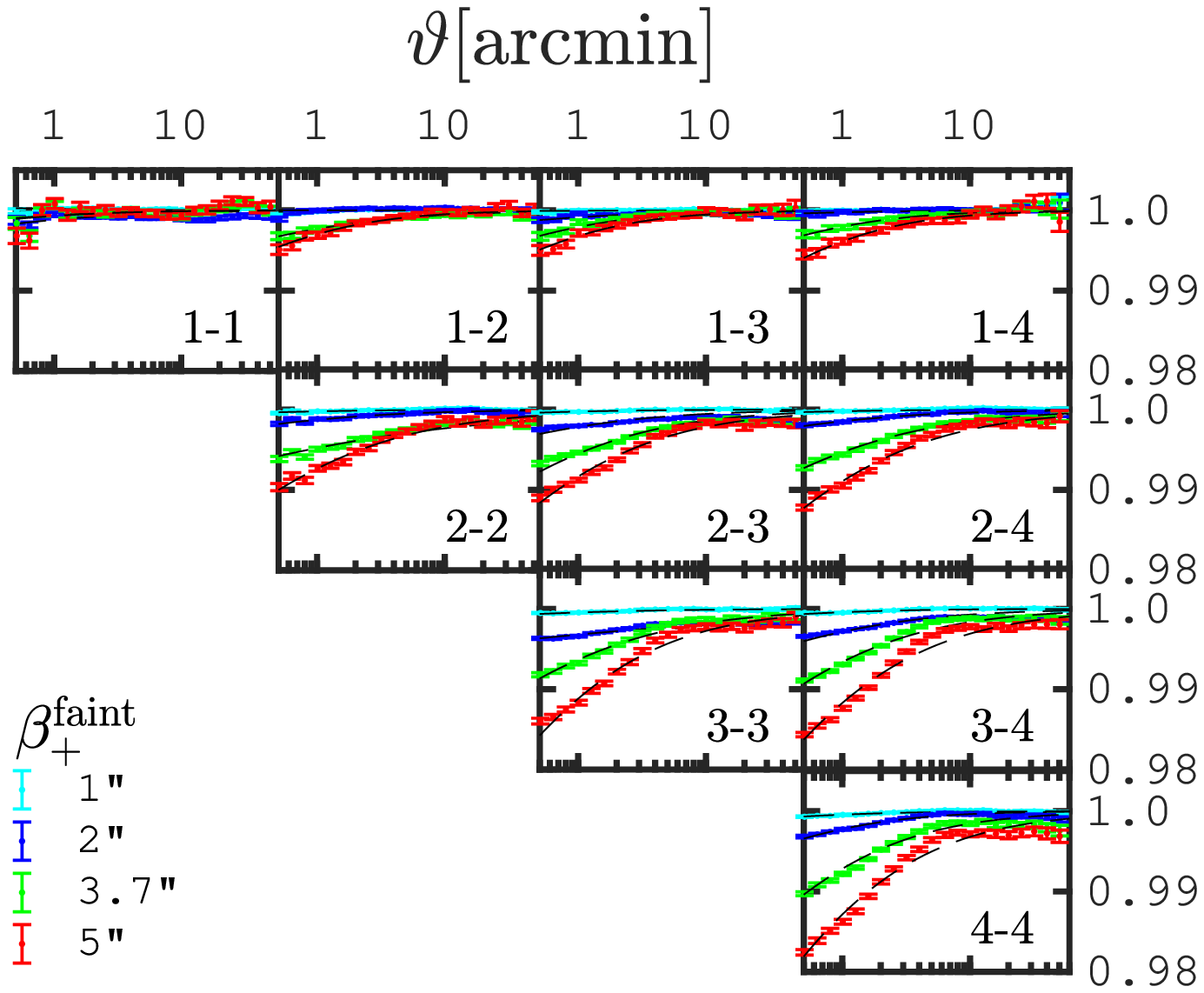}
\hspace{20mm}
\includegraphics[width=2.9in]{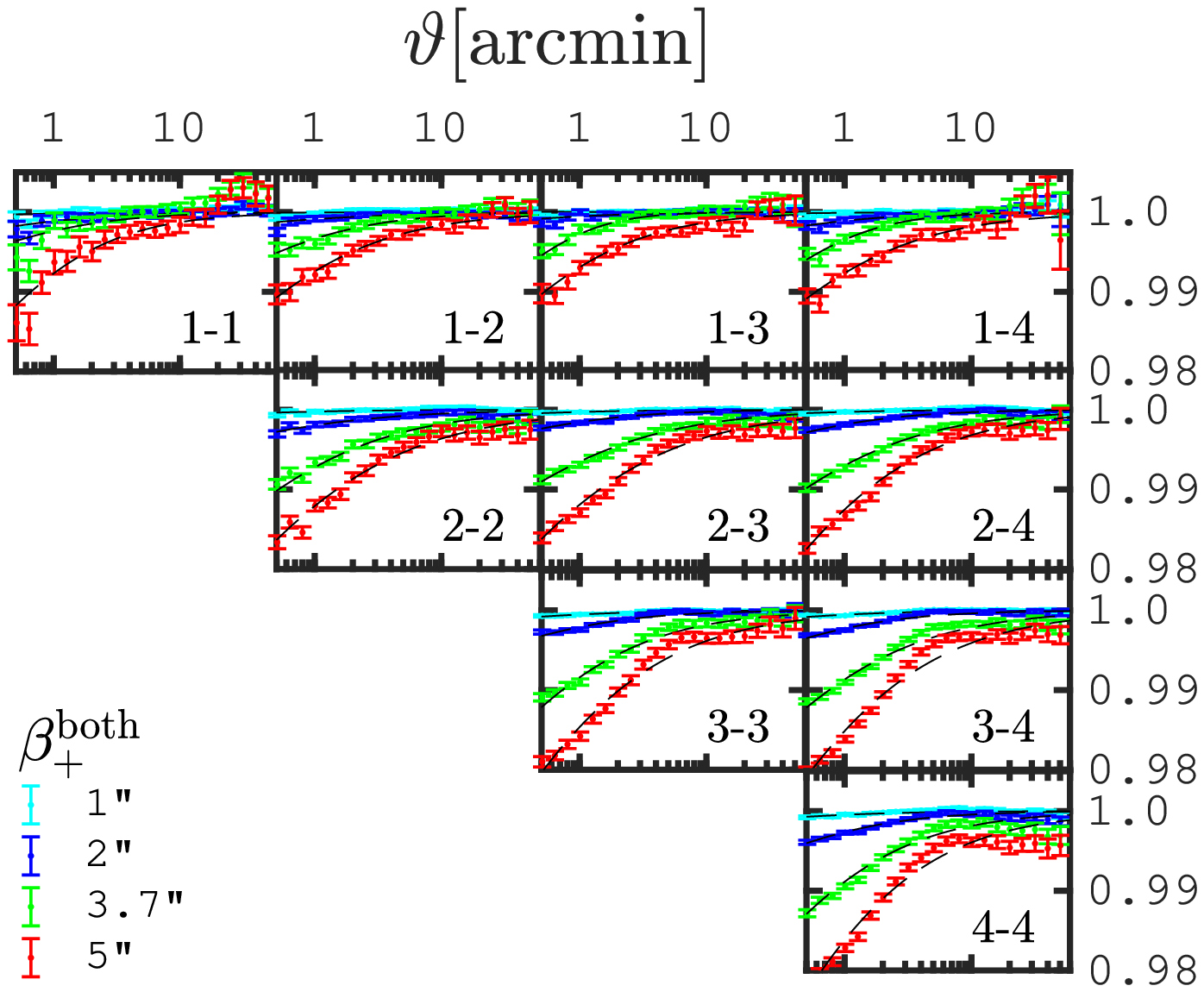}
\vspace{15mm}
\caption{Effect of the close-pair selection on $\xi_+$ signal in the KiDS-HOD mocks galaxy catalogues. Left and right panels show the FAINT and BOTH
prescriptions, respectively. $y$-axes show $\beta_{+}$, defined in Eq. \ref{eq:beta}, while $x$-axes show the separation angle in arcmin.
Different colours represent different opening angles used in the definition of close pairs. 
Upper left to lower right show the results  for tomographic bins with increasing redshift. 
Dashed black lines represent the best fit from Eq. \ref{eq:beta_fit}.
(right panels:) Same as left, but for the BOTH prescription. The KiDS-450 cosmic shear analyses included angular scales down to 0.5 arcmin in their analysis \citep{KiDS450}, while the smallest angles included in the DES cosmic shear measurement went from 7 arcmin at lower redshifts to  3.5 arcmin at higher redshifts  \citep{DES1_Troxel}. }
\label{fig:beta_p}
\end{center}
\end{figure*}

\begin{figure*}
\begin{center}
\includegraphics[width=2.9in]{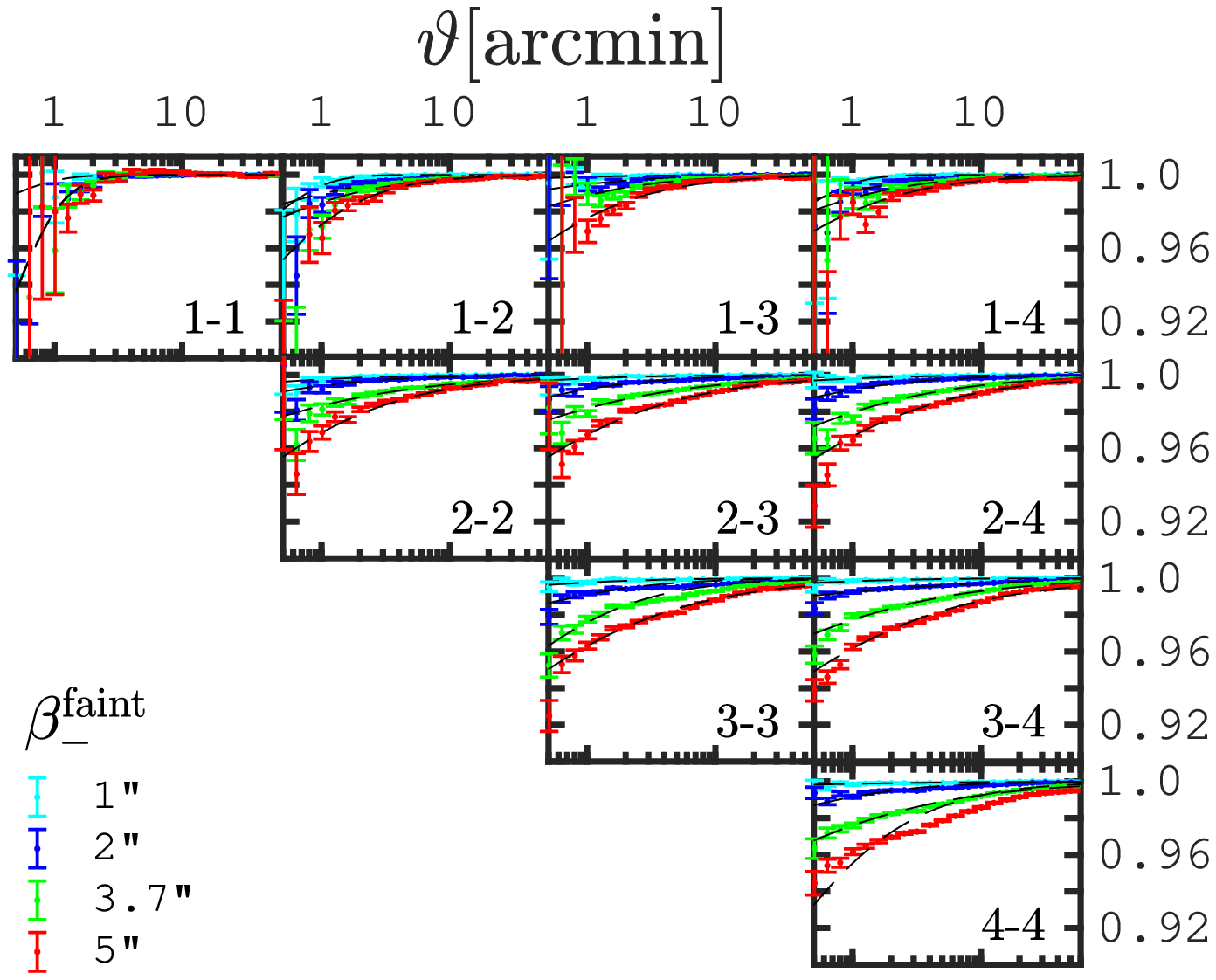}
\hspace{20mm}
\includegraphics[width=2.9in]{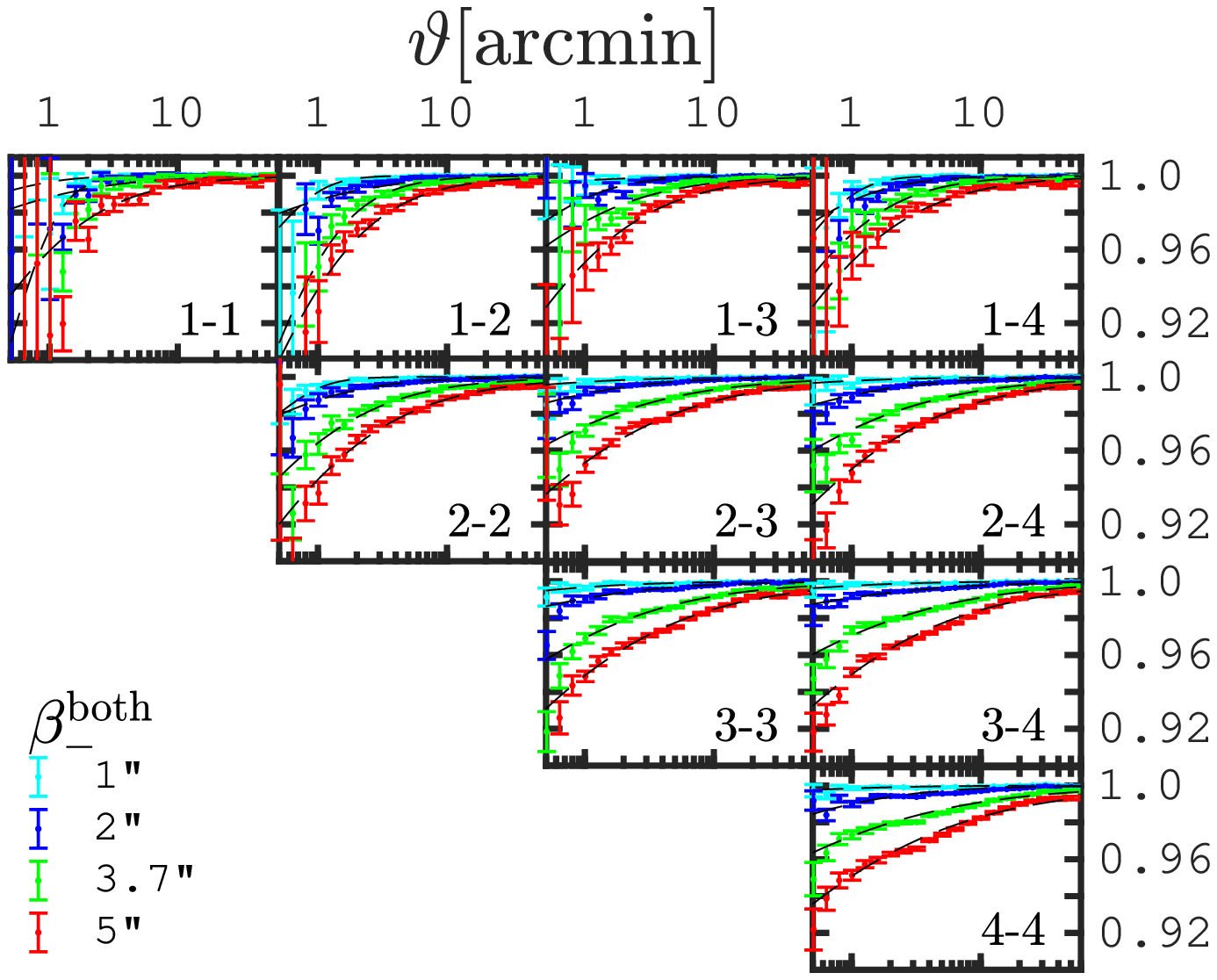}
\vspace{15mm}
\caption{Same as Fig. \ref{fig:beta_p}, but this time for $\beta_-$. Note the change in $y$-scaling.
The KiDS-450 cosmic shear analyses included angular scales down to 4.2' in their analysis, 
while the DES cosmic shear measurement excluded scales smaller than 70' at low redshift and 35' at high redshift. }
\label{fig:beta_m}
\end{center}
\end{figure*}

We show in Fig. \ref{fig:nz_cp_both} the ratio between the redshift distributions  of the full sample, with and without filtering, for the the FAINT and BOTH techniques
and for the different \textcolor{black}{exclusion} angles. 
As redshift increases, all curves first show that the filtering removes an increasing number of galaxies,
which simply reflects the fact that more galaxies are contained within the same solid angle. 
This trend starts to reverse beyond redshift $z = 0.7$, where the $n(z)$ of the KiDS-HOD mocks begins to fall off (see the top panel of  Fig. \ref{fig:lsst_nz_hod}). 
The BOTH filter  rejects  approximately twice as many objects as the FAINT as expected; the FAINT filter preferentially rejects objects at higher redshift that appear dimmer,
and preserves almost all low-redshift objects. 

Our measurements of $\beta_{+}^{ij}(\vartheta)$ and  $\beta_{-}^{ij}(\vartheta)$ are presented in Figs. \ref{fig:beta_p} and \ref{fig:beta_m},
respectively. We see that larger \textcolor{black}{exclusion} angles exhibit larger effects, as expected from the larger fraction of objects with close neighbours. 
These results are in excellent agreement with the equivalent results  from \citet[][see their `FIX' method]{2011A&A...528A..51H}.
We additionally find that higher redshift measurements are more affected by this selection bias due to the higher fraction of close pairs, 
and that cross-tomographic signals are impacted as well.  {Because  the neighbour-exclusion bias mainly occurs at small scales, 
the measurement of $\widehat{\xi}_-$ suffers from a stronger bias than $\widehat{\xi}_+$, at fixed angle.} 

\begin{figure}
\begin{center}
\includegraphics[width=2.9in]{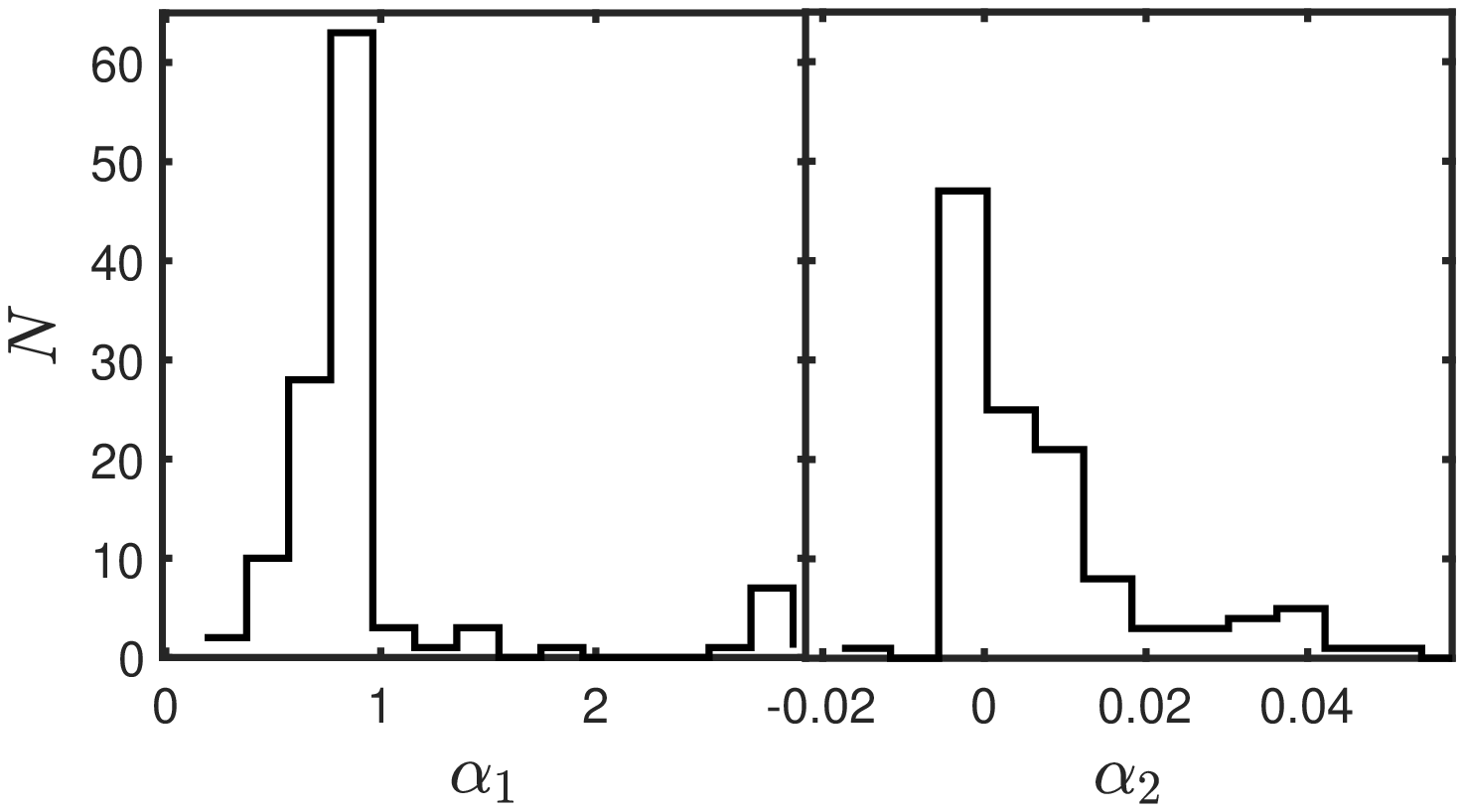}
\caption{Distributions of the best fit parameters $\alpha_{1,2}$, which model the neighbour-exclusion bias according to Eq. \ref{eq:beta_fit}.
 These histograms could be used as informative priors on these two parameters in a  procedure that marginalises over this selection effect.}
\label{fig:alpha_prior}
\end{center}
\end{figure}

\citet{DES-SV_MacCrann} have developed two models to describe this effect, the first one calculated from a third-order correction to the shear-shear correlation,
the second one as a toy model based on the luminosity of the neighbouring cells. These two models were compared to simulations and to the DES-Science Verification catalogue,
and were shown to reproduce most of the features of the neighbour-exclusion bias, but not all. 
Given the relative size of this effect and the complexity to model and measure it with high accuracy, we instead propose here a simple parametric description that can be included in an MCMC with two extra nuisance parameter.  
We find that the shape of both $\beta_{+}$ and $\beta_-$ is well modelled by:
\begin{eqnarray}
\beta_{\rm fit}(\vartheta) = \frac{1}{(1+\vartheta^{-\alpha_1})^{\alpha_2}},
\label{eq:beta_fit}
\end{eqnarray}
with $\alpha_1 > 0$, and $\vartheta$  in arcmin. At large angles,  $\vartheta^{-\alpha_1}$ tends to zero, hence $\beta_{\rm fit}(\vartheta)$
approaches unity. We fit our tomographic measurements of $\beta_+$ in the range $0.5 < \vartheta < 317$ arcmin, whereas we restrict  $\beta_-$ at small angles to $\vartheta>1.6'$
in order to minimise the impact of the noise seen in some panels.
The best fits are shown as dashed lines in Figs. \ref{fig:beta_p} and \ref{fig:beta_m},
with parameter values spanning the range $\alpha_1 \in [0.0,3.0]$ and  $\alpha_2 \in [-0.02,0.05]$,
shown in Fig. \ref{fig:alpha_prior}. This fit was carried out on the mean measurement of $\beta_{\pm}^{ij}(\vartheta)$, 
averaged over all lines-of-sight. The scatter per realisation would be larger, but we are not interested in that noisy quantity. 

Additionally, given that the shapes of $\beta_{\pm}$ are similar to those arising from the impact of baryon feedback on the matter density field \citep{Semboloni11},
this contribution must be included in the interpretation of the measured baryon feedback parameters, something  omitted in 
\textcolor{black}{previous} analyses \citep[H17]{HWVH15} and forecasts \citep{Foreman16}, but first pointed out in  \citet{DES-SV_MacCrann}. 
Alternatively, marginalizing over the baryon  feedback parameters \textcolor{black}{and/or over the reduced-shear model}
should, at the same time, mitigate this selection effect, whose amplitude is lower than some of the most extreme feedback models.

We now investigate the importance of this effect on two current weak lensing measurement strategies.

\subsection{\textcolor{black}{Neighbour-exclusion bias in a {\it lens}fit-like pipeline}}
\label{subsubsec:cp_KiDS}

As a first example, we examine the KiDS-450 analysis pipeline of  H17. It first uses {\sc SExtractor} \citep{SExtractor} to provide a catalogue of deblended objects.
These catalogue entries are then passed to  {\it lens}fit \citep{2007MNRAS.382..315M, 2013MNRAS.429.2858M},
which performs the galaxy shape measurement on the images \textcolor{black}{with as few cuts as possible on the selected objects,
as described in H17.}
{\it lens}fit masks neighbouring objects when measuring each target object, but as this process does not fully correct for light leaking outside the neighbours' masked regions, a `contamination radius' statistic, to test for the presence of close neighbours, is also measured, calculating the distance to the nearest detected neighbour \citep{2013MNRAS.429.2858M}. 
If the contamination radius is less than 4 pixels, the object is flagged and excluded from the analysis.
The flagging system is described in \citet{2013MNRAS.429.2858M}, and captured by the  {\tt FITCLASS} flag in the KiDS-450 data and image simulation catalogues. 
For the KiDS-450 cosmic shear analysis, a stricter criterion of 4.25 pixels was employed to minimise additive bias (see Appendix D in H17).  
In this method, the effect of blending at a given centroid-to-centroid separation strongly depends on the galaxy sizes and is found to preferentially remove fainter galaxies. 
With a pixel size of 0.214 arcsec, this means that the blending strategy here corresponds closely to the FAINT technique, with full blending occurring 
approximately at 0.9 arcsec.

A second selection is at play in this shape measurement strategy:
\textcolor{black}{the presence of a close neighbour may affect the weight assigned to that galaxy: close neighbours tend to be measured as being more elliptical and to have higher weights than they would if they had been measured in isolation.}
Although these objects are not excluded from the analysis, their detection rate and weight are affected by the presence of neighbours, 
which can be related to an `effective' exclusion angle. 
In order to study this, we make use of
image simulations similar to those on which {\it lens}fit was calibrated  for the KiDS-450 cosmic shear analysis \citep{2016arXiv160605337F}.
 These improved simulations  are augmented with realistic input galaxy properties which are inferred from the HST COSMOS data. 
 This current study, however, uses only a subset of the full suite designed for shear calibration; we investigate the effect of bad and good seeing conditions,
 but do not include variations in the lensing shear or galaxy rotations. These are required for a full shear calibration, but have minimal impact on close pairs selection.
We run the {\it lens}fit shape measurement tool on these simulations and constructed object catalogues based on the input  (the `true' objects) 
and output (the `{\it lens}fit \textcolor{black}{measurement}' of these objects). 
For each input object, the catalogues contain the input and detected positions and magnitudes,  a shape weight, a `source-type' flag  ({\tt FITCLASS}) that identifies stars, galaxies, blends, badly measured objects, etc., and a flag for objects that  were not matched to the simulation input.  Our matching condition requires that the centroid of an observed object resides within a 3 pixel radius of an input centroid.

We construct our baseline catalogue by first removing all the \textcolor{black}{input} stars, then applying a $m_r <24.5$ cut such as to mimic the observed data.
Ignoring this step would overestimate the effect by artificially boosting the depth. 
We next construct the {\it lens}fit \textcolor{black}{measurement} catalogue by requiring {\tt FITCLASS=0} and by  rejecting unmatched galaxies.
We then count the close pairs that are present in the `true'  and `filtered' catalogues (optionally summing the  {\it lens}fit weights, in the second case) as a function of separation, 
and finally take the ratio between the two measurements.  We normalise the ratio to unity  at 6\,arcsec, where {filtering} should be minimal. 
This ratio is shown in the lower panel of Fig. \ref{fig:ratio_des}, where we see that close pairs are in fact unaffected by {close-neighbours selection} for angular scales larger than 3 arcsec, 
but  \textcolor{black}{reliable shape measurements for more than half the close pairs are not produced by the pipeline} below 1.8 arcsec. This means that the KiDS-450 measurement strategy can be representatively identified 
as the dark blue lines in Fig.  \ref{fig:beta_p} and \ref{fig:beta_m}, in the left panels describing the FAINT technique.

\textcolor{black}{In the cosmic shear analysis of H17, the $\widehat{\xi}_+$ and $\widehat{\xi}_-$ measurements extend from $\vartheta>0.5$ and $4.2$, respectively.   
From Figs.   \ref{fig:beta_p} and \ref{fig:beta_m} we can therefore expect the KiDS cosmic shear signal to be affected by the neighbour-exclusion bias by less than a percent.}

\subsection{\textcolor{black}{Neighbour-exclusion bias in a {\small NGMIX}-like pipeline}}
\label{subsubsec:cp_DES}

\begin{figure}
\begin{center}
\includegraphics[width=2.9in]{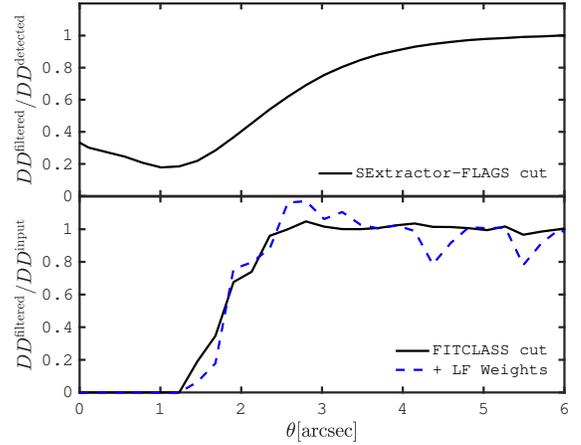}
\caption{(upper:) Ratio between the number of pairs in the DES-Science Verification catalogue with and without applying the SExtractor flag.   
(lower:) Ratio between the number of pairs in the image simulations that mimic the KiDS-450 data, with and without \textcolor{black}{including the effect of the {\it lens}fit measurement}.
 }
\label{fig:ratio_des}
\end{center}
\end{figure}

As a second example, we examine one of the  pipelines used by the Dark Energy Survey Collaboration {on their Science Verification (SV) data, 
presented in \citet{2016PhRvD..94b2002B} and re-examined in \citet{DES-SV_MacCrann}.}
In  their  strategy, the {\sc ngmix} shape measurement pipeline with meta-calibration \citep{METACAL} was run on all objects that passed the 
{\sc SExtractor} {\tt FLAG\_i<2} cut, which rejected both blended objects identified by {\sc SExtractor}. Once that was done however, all shapes were given the same weight. This is similar in nature to the filter BOTH presented in Section \ref{subsec:blend_mocks}. The question now is to find the effective \textcolor{black}{exclusion} angle at which the filter operates.

We measure this by running  our close-pairs finder algorithm on the public Science Verification catalogue\footnote{DES-SV data: https://des.ncsa.illinois.edu/releases/sva1/doc/gold}, with and without applying the  {\tt FLAG\_i<2} filter.  We then compare the number of close pairs in the upper panel of Fig. \ref{fig:ratio_des},
again normalizing the ratio to unity at the largest angle.
The effect of the selection becomes apparent already at  4 arcsec, and by 2.5 arcsec almost half of the pairs are filtered out.
Note that this measurement differs in nature from that carried out on the KiDS image simulation, since we do not know the `input' here,
but  only the objects detected by {\sc SExtractor}. This explains why the ratio does not converge to zero at zero lag: many pairs separated by less than one arc second were not even detected to start with due to obscuration by the foreground \textcolor{black}{member}.

According to this figure, this  measurement strategy has a close pair definition bracketed between 2 and  3.7 arcsec, plotted as blue and green symbols on Figs. \ref{fig:beta_p} and \ref{fig:beta_m}. 
 {This is in excellent agreement with the results on the impact of close pairs reported in  \citet{DES-SV_MacCrann} for the same DES-SV data,
 which provides robust validation of both approaches. }

We emphasise that the results quoted in this section cannot be directly applied to the DES data, since this survey has a different depth and density than the KiDS-HOD mocks analysed here. Instead, one should think of our results as the outcome of a DES-like analysis (notably the shape measurement method) performed on KiDS-like data.  The technique is general though, and hence some conclusions can be reached for the DES-year1 {cosmic shear analysis of \citet{DES1_Troxel}}.
While the neighbour-exclusion bias  significantly deviates from unity at small scales, their cosmic shear analysis is protected against this effect for three reasons:
1) their upgraded meta-calibration strategy no longer requires the {\tt FLAG\_i<2}  cut, reducing even more the size of the effect, 
2) they have folded this effect into their new shear calibration  \citep{DESY1_Samuroff}, 
and 3) they applied aggressive angular cuts  on the measurements: in order to minimize the contamination from baryon feedback, 
they excluded angular scales smaller than 3.5 (7.0) arcmin in the highest (lowest) tomographic bin for $\xi_+$, 
and  35 (70) arcmin in the highest (lowest) tomographic bin for $\xi_-$. As seen in the right panels of Figs. \ref{fig:beta_p} and \ref{fig:beta_m},
{the amplitude of the neighbour-exclusion bias on} these angular scales is less than a percent. 



\subsection{Future surveys}
\label{subsec:cp_future}

\begin{figure}
\begin{center}
\vspace{-2mm}
\includegraphics[width=4.0in]{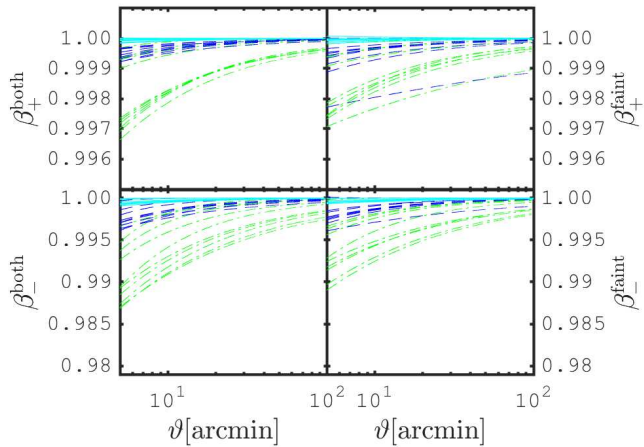}
\vspace{-15mm}
\caption{Zoom-in on  the neighbour-exclusion bias {model} at large angles in {mock data} at KiDS-depth, estimated from the fit function (Eq. \ref{eq:beta}). 
Upper and lower panels show $\beta_+$ and $\beta_-$, respectively; 
right and left panels show FAINT and BOTH methods, respectively.   The different colours  (cyan, blue and green) match the different separation angles presented in Fig. \ref{fig:beta_p}
 (1, 2 and 3.7 arcsec, respectively), and the different lines  of the same colour show the fits from the 10 panels in that figure.}
\label{fig:cp_large_angle}
\end{center}
\end{figure}

The upcoming weak lensing experiments such as LSST and {\it Euclid} are expected to achieve sub-percent precision  on cosmological parameters from cosmic shear measurements.
The neighbour-exclusion bias must therefore be accurately captured in order to interpret the measurement correctly. 
At KiDS depth, this effect is mostly significant on smaller angular scales, but a residual effect propagates to all scales.
We illustrate this in Fig.  \ref{fig:cp_large_angle}, which zooms in on the fit function described by Eq. \ref{eq:beta}, over the range  $5 < \vartheta < 100$ arcmin.
The different colours match the separation angles presented in Fig. \ref{fig:beta_p}, and the four panels show $\beta_{\pm}$ for the FAINT and BOTH
methods. All tomographic bins are over-plotted. \textcolor{black}{Even at large angular separations, these models are mostly consistent with the measurements. The agreement is not perfect in all tomographic bins, but the trends are captured with enough accuracy to support our result:} the effect on $\xi_+$ is below 0.3\% at all scales, but  $\xi_-$ can be affected by 0.5\% at 20 arcmin.

\begin{figure}
\begin{center}
\vspace{-2mm}
\includegraphics[width=3.5in]{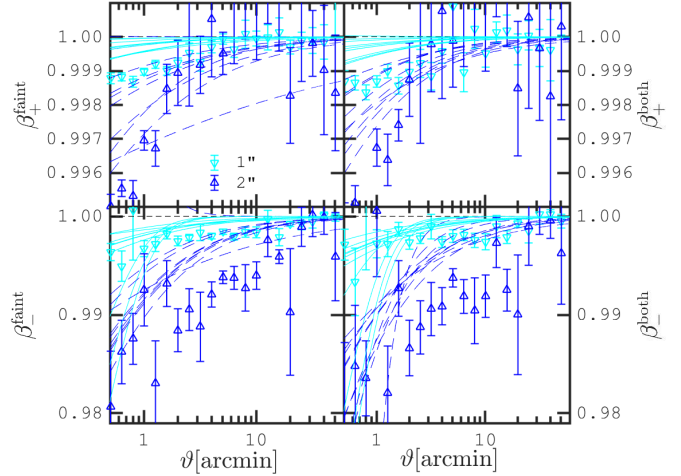}
\caption{Neighbour-exclusion bias measured from the LSST-like HOD mocks with 1 and 2 arcsec \textcolor{black}{exclusion} angles,
\textcolor{black}{compared to the best fit models estimated from the KiDS-HOD reported in Fig. \ref{fig:cp_large_angle}}.
Upper and lower panels show $\beta_+$ and $\beta_-$, respectively; 
left and right panels show FAINT and BOTH methods, respectively.  }
\vspace{-5mm}
\label{fig:beta_lsst}
\end{center}
\end{figure}

We investigate this further with the LSST-like HOD mocks presented in Section \ref{subsubsec:LSST-HOD}, 
where the number density is almost four times higher than the KiDS-HOD mocks, with a redshift distribution that now extends to $z=3$.
We carry out a single two-dimensional cosmic shear analysis over 20 of these mocks, and extract the neighbour-exclusion bias 
for the BOTH and FAINT cases, assuming that {rejection of close neighbours} occurs at 1.0 or 2.0 arcsec separation. The results, presented in Fig. \ref{fig:beta_lsst},
indicate that the $\xi_+$ and $\xi_-$ measurements are affected by half a percent and up to two percent, respectively,
{when the exclusion angle is set to 2 arcsec. If we reduce the exclusion angle down to one arcsec separation,}
then the $\xi_+$ and $\xi_-$ measurements are affected by 0.2 and 0.5 percent, respectively. 

Also shown in Fig. \ref{fig:beta_lsst} are the best fit models estimated from the shallower KiDS-HOD mocks, previously reported in Fig. \ref{fig:cp_large_angle}.
\textcolor{black}{We clearly see that the LSST-like data points are systematically lower than the \textcolor{black}{KiDS-HOD} best-fit lines.
We can further read off from this figure that, at fixed close-pairs model and exclusion angle, the neighbour-exclusion bias affects more severely our LSST-like mock data compared to the KiDS-HOD mock data, by about a factor two, which means that  if we are to marginalize over the neighbour-exclusion bias using $\alpha_{1,2}$ as nuisance parameters, 
then their priors need to be revisited. }


\textcolor{black}{There is a key caveat in our analysis which stems from our choice to populate the mocks with unit weight sources that match the effective number density of the KiDS and LSST surveys (8 and 26 gal per arcmin$^2$, respectively), rather than matching the raw number density with non-unit weights. 
The number of close pairs in the raw data is larger, hence the neighbour-exclusion bias is expected to be larger.
Furthermore our study does not include any dependence on the size distribution that slowly varies with redshift and magnitude.   The technique presented here will therefore be extended in future analyses to extend the complexity of the mock source sample in order to determine a more accurate amplitude for the neighbour-exclusion bias.  For future high-precision surveys we would advocate marginalising over a model given by Eq. \ref{eq:beta_fit} using informative priors on the two nuisance parameters from a mock galaxy analysis.}

\section{Conclusions}
\label{sec:conclusion}

We describe a suite of numerical simulation products tailored for the estimation of covariance matrices 
in combined-probe analyses involving weak lensing data from the Kilo Degree Survey. 
Many of these have already been used to date, hence the first part of this paper serves as the 
main reference for the description of the methodology and performance of the mock data used in these analyses.
More specifically, we generate 844 fully independent realisations of  mock lensing data that emulate the KiDS-450 and an LSST-like survey described in \citet{LSSTgal},
in individual patches of 100 deg$^2$ each. 
In the same simulated light cones we also  include  mock catalogues that emulate spectroscopic galaxy surveys such as GAMA, CMASS, LOWZ and 2dFLenS,
as well as CMB lensing convergence maps. Used in conjunction with the lensing mocks, 
these different simulation products can serve for pipeline validation and uncertainty estimation in combined-probe analyses involving e.g. cosmic shear, galaxy-galaxy lensing estimators, galaxy clustering, redshift-space distortions, and their cross-correlation with the CMB lensing data.
We quantify the accuracy of the galaxy catalogues by comparing the redshift distributions and clustering
with the data they are meant to emulate;  we reach 20\% agreement or better on the two-point correlation function $w(\vartheta)$ over a range of dynamical scales,
 with residual differences partly caused by our choice of cosmological parameters.
 At small angular scales, the variance obtained from the mock clustering  and galaxy-galaxy lensing measurements
are consistent with jack-knife estimations of the error; we identify from the mocks scales where the latter becomes  unreliable. 
We generate a 3 $\times$ 2-point function data vector that includes cosmic shear, lens clustering and galaxy-galaxy lensing measurements, 
and present an estimation of the covariance matrix for these combined-probes.

In the second part of the paper, we demonstrate how these mocks can be used to estimate the neighbour-exclusion bias at KiDS and LSST depth, 
inspired by the early work of \citet{2011A&A...528A..51H}. For this particular science case, we produce two additional 
suites of mock data, in which both the lenses and the sources catalogues are extracted from an HOD prescription.
These are meant to be representative of the KiDS-450 and LSST surveys, and include realistic levels of source-lens coupling, 
photometric uncertainty, galaxy clustering and redshift distributions.
We identify   {galaxies with close neighbours} in our mock lensing data with four different {exclusion angles},
and investigate two methods to cope with them, representative of the shape measurement techniques used in the Dark Energy Survey and in the KiDS-450 data.
We compare the cosmic shear signal with and without the filtering of these close pairs, 
in the context of a four-bin tomographic analysis.  We find a redshift dependence in the selection effect: the neighbour-exclusion bias
is larger at higher redshift due to the increase in number of  objects at fixed solid angle. At KiDS-depth and assuming poor seeing conditions blurring objects 
separated by less than 5 arc seconds, the impact 
on the $\xi_{+}$ measurement is of the order of a few percent, while it reaches up to 10\% \textcolor{black}{for the same angular scales} for $\xi_-$ (see Figs. \ref{fig:beta_p} and \ref{fig:beta_m}). 
In all cases, the angular dependence of this effect has a simple shape that we model with a two-parameter function (see Eq. \ref{eq:beta_fit}).
We measure the distribution of these two parameters over all tomographic bins, which could serve as a  prior in a MCMC marginalization pipeline
for current surveys. 
\textcolor{black}{This prior will however need to be revisited for future deeper surveys using the methodology outlined in this paper.}

We investigate the sensitivity of current cosmic shear analyses to this selection bias by identifying the 
filtering technique that best matches the data measurement procedure.
The {\sc ngmix } pipeline uses {\small SExtractor} flags to reject blended objects, which effectively suppresses most pairs
separated by less than 2.5 arcsec, as verified on the DES-Science Verification data. 
Given the conservative cuts that were applied on the angular scales, 
we find that this $\xi_{\pm}$ measurement is affected by less than a percent. 
The DES year1 results are further protected since the updated meta-calibration method does not require the cut of {\small SExtractor} flags.
The KiDS-450 pipeline uses the {\it lens}fit shape measurement tool, which returns a shape weight that is affected by the proximity
of close neighbours. We measure the effective {close-pairs exclusion} radius from KiDS-like image simulations
and find that more than  half the close pairs are rejected when separated by less than 1.8 arcsec. 
The KiDS-450 cosmic shear analysis extended to 0.5 arcmin in $\xi_+$ and 4.2 arcmin in $\xi_-$, 
at which scales the amplitude of the neighbour-exclusion bias is always less  than a percent.  

We next measure this bias in deeper and denser mock data in a non-tomographic setup, and find that the amplitude of the effect is about twice the size measured from the shallower KiDS-HOD mocks.
For future lensing surveys like LSST, the neighbour-exclusion bias needs to be understood with high accuracy since it is degenerate with 
baryon feedback parameters \citep{DES-SV_MacCrann} and can be mostly addressed with an angle-dependent shape calibration technique \citep{DESY1_Samuroff}.
In any case, these future measurements will need to be calibrated against numerical simulations such as those presented in this paper,
possibly upgraded with actual images for each \textcolor{black}{object}.


The SLICS mocks can find a number of applications in data analyses, for estimator validation and calibration, in the data processing,
for estimation of covariance matrices in combined-probe measurements, for studies of statistical properties of covariance and likelihood functions, 
or for the investigation of systematic effects. Many of these applications  are relatively new and would require further exploration
in order to reach the level of accuracy and control required for future lensing surveys.
To encourage and accelerate this progress, we make all simulation products publicly available at {\tt slics.roe.ac.uk}.

\section*{Acknowledgements}

The HOD calculations used in the paper inherits from the code written by Marcello Cacciato, who also provided 
many advices on general HOD strategies. Shadab Alam and Chris Blake also contributed to these discussions, 
which helped us in deciding what strategy best suited our needs. 
We would like to thank Ian Fenech Conti, and Ricardo Herbonnet for their help with the image simulations,
 Alexander Smith for sharing the details of his GAMA HOD prescription,  Joe Zuntz for providing the LSST $n(z)$ and for his insights on the Dark Energy Survey shape measurement strategy,
  and the anonymous referee for their useful comments.
 This work also benefited from discussions about blending with Javier Sanchez and Joe Zuntz.
We would like to acknowledge the input of many users that have tested the different simulation products and provided invaluable feedback that helped us finding bugs and making the mock products easier to use, notably Chris Blake, Massimo Viola, Edo van Uitert, Marika Asgari, India Rose Friswell, Harry Johnston, Nicolas Martinet and Axel Buddendiek, Elena Sellentin and Chien-Hao Lin.
We thank Martin Kilbinger for help with the {\sc athena} correlation function measurement software and the {\sc nicaea} theoretical modelling software,
Mike Jarvis for maintaining {\sc treecorr} and Joe Zuntz for help with CosmoSIS.

JHD is supported by the European Commission under a Marie-Sk{\l}odowska-Curie European Fellowship (EU project 656869).
\textcolor{black}{CH and AA acknowledge support from the European Research Council under grant number 647112; 
AA is further supported by a LSSTC Data Science Fellowship.}
VD acknowledges the Higgs Centre Nimmo Scholarship and the Edinburgh Global Research Scholarship.
\textcolor{black}{AK and HHo acknowledge support from the Netherlands Organisation for Scientific Research (NWO) Vici grant 639.043.512}.
RN acknowledges support from the German Federal Ministry for Economic Affairs and Energy (BMWi) provided via DLR under project no. 50QE1103.
LvW is supported by the NSERC  of Canada.
HHi is supported by an Emmy Noether grant (No. Hi 1495/2-1) of the Deutsche Forschungsgemeinschaft.
\textcolor{black}{LM acknowledges support from STFC grant ST/N000919/1.}

Computations for the $N$-body simulations were performed in part on the Orcinus supercomputer at the WestGrid HPC consortium (www.westgrid.ca),
in part on the GPC supercomputer at the SciNet HPC Consortium.
SciNet is funded by: the Canada Foundation for Innovation under the auspices of Compute Canada;
the Government of Ontario; Ontario Research Fund - Research Excellence; and the University of Toronto.
The post-processing calculations were mainly carried out on the Cuillin cluster at the Royal Observatory of Edinburgh, 
which is run by Eric Tittley. 

The mock data presented in this paper are calibrated against observations from KiDS, GAMA and BOSS.
This KiDS data are based on data products from observations made with ESO Telescopes at the La Silla Paranal 
Observatory under programme IDs 177.A-3016, 177.A-3017 and 177.A-3018.

GAMA is a joint European-Australasian project based around a spectroscopic campaign using the Anglo-Australian Telescope. The GAMA input catalogue is based on data taken from the Sloan Digital Sky Survey and the UKIRT Infrared Deep Sky Survey. Complementary imaging of the GAMA regions is being obtained by a number of independent survey programmes including GALEX MIS, VST KiDS, VISTA VIKING, WISE, Herschel-ATLAS, GMRT and ASKAP providing UV to radio coverage. GAMA is funded by the STFC (UK), the ARC (Australia), the AAO, and the participating institutions. The GAMA website is http://www.gama-survey.org/.
Also based on observations made with ESO Telescopes under programme ID 177.A-3016.

 Funding for SDSS-III has been provided by the Alfred P. Sloan Foundation, the Participating Institutions, the National Science Foundation, and the U.S. Department of Energy Office of Science. The SDSS-III web site is http://www.sdss3.org/. SDSS-III is managed by the Astrophysical Research Consortium for the Participating Institutions of the SDSS-III Collaboration including the University of Arizona, the Brazilian Participation Group, Brookhaven National Laboratory, Carnegie Mellon University, University of Florida, the French Participation Group, the German Participation Group, Harvard University, the Instituto de Astrofisica de Canarias, the Michigan State/Notre Dame/JINA Participation Group, Johns Hopkins University, Lawrence Berkeley National Laboratory, Max Planck Institute for Astrophysics, Max Planck Institute for Extraterrestrial Physics, New Mexico State University, New York University, Ohio State University, Pennsylvania State University, University of Portsmouth, Princeton University, the Spanish Participation Group, University of Tokyo, University of Utah, Vanderbilt University, University of Virginia, University of Washington, and Yale University.

We would finally like to thank McGill University for its hospitality,  where 
an important part of the HOD code development was made.

{\footnotesize All authors contributed to the development and writing of this paper. The authorship list is given in three groups:
the lead author (JHD), followed by two alphabetical groups.
Members of the first alphabetical group carried out key infrastructure work specifically for this paper. 
Members of the second alphabetical  group provided proprietary data central to this work, or contributed to the analysis.}

\bibliographystyle{hapj}
\bibliography{mock}
\bsp

\appendix

\section{Additional source galaxy catalogues}
\label{sec:more_sources}

\subsection{Mock LSST-like source galaxies}
\label{subsubsec:LSST}

\begin{figure}
\begin{center}
\includegraphics[width=2.75in]{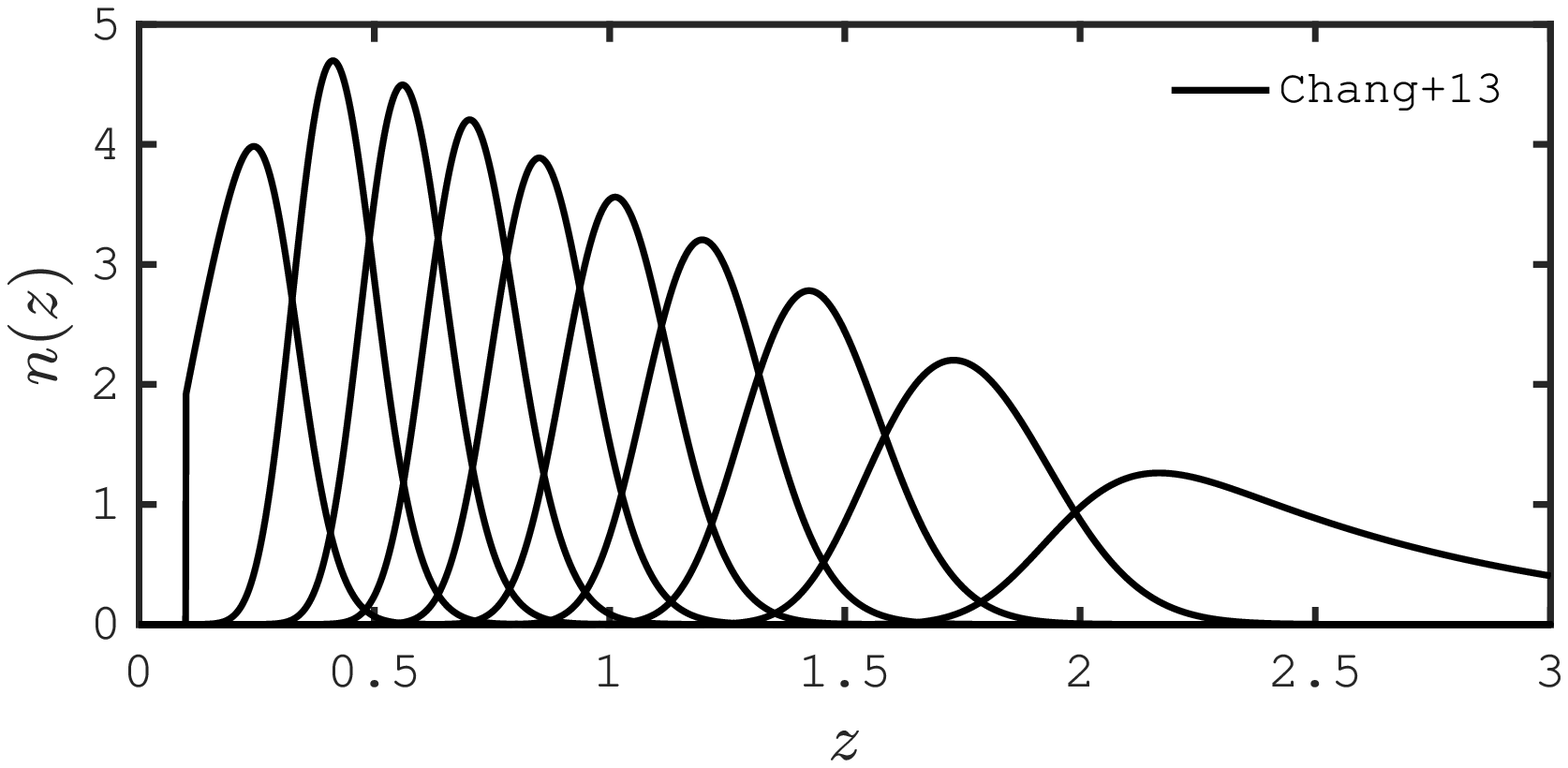}
\caption{Redshift distributions used for the 10 tomographic bins of the LSST-like source catalogues, 
assuming the survey specifications presented in  \citet{LSSTgal} with $\alpha  = 1.25$, $\beta = 1.0$ and $z_0 = 0.5$.}
\label{fig:lsst_nz}
\end{center}
\end{figure}

\begin{figure*}
\begin{center}
\includegraphics[width=3.5in]{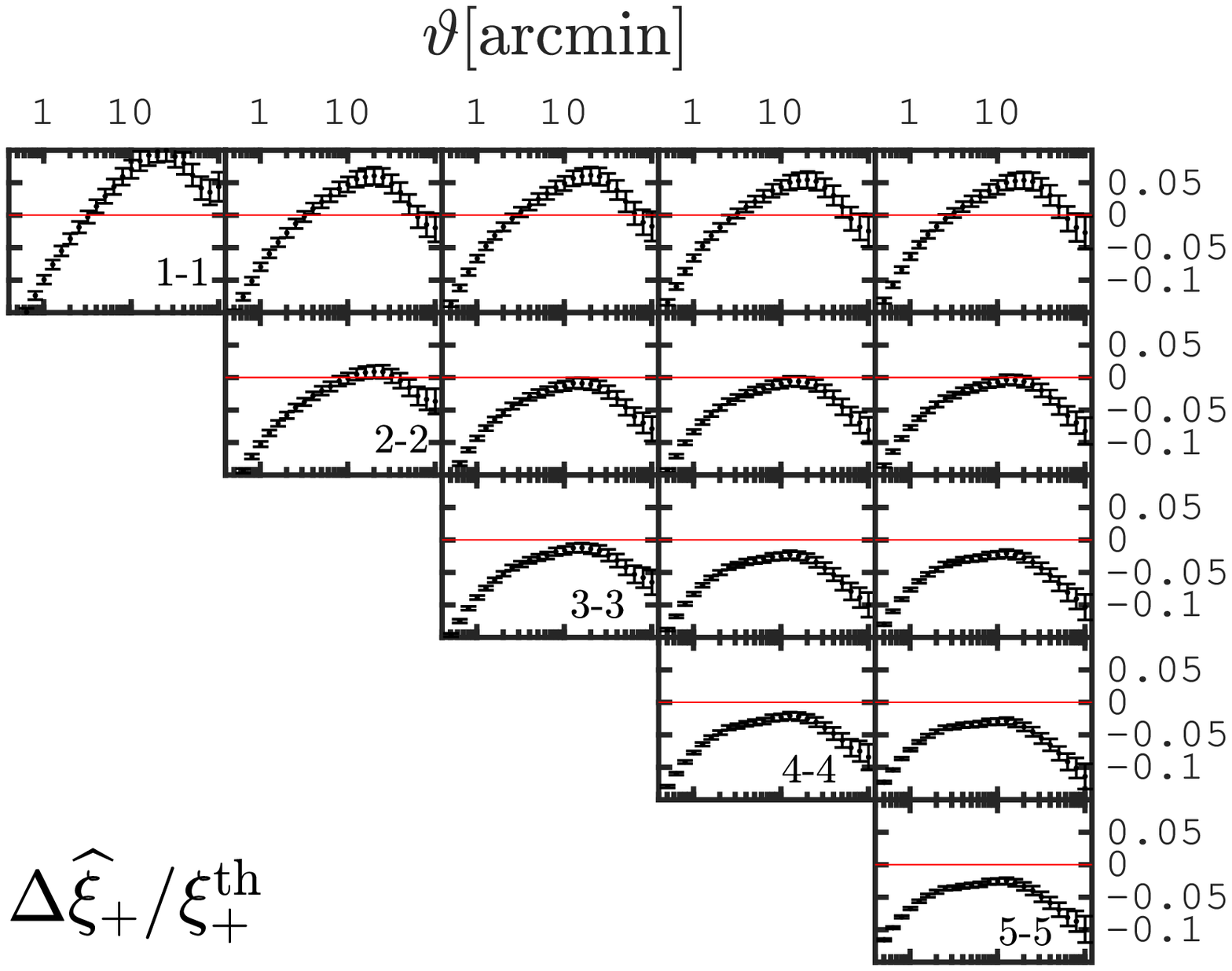}
\includegraphics[width=3.5in]{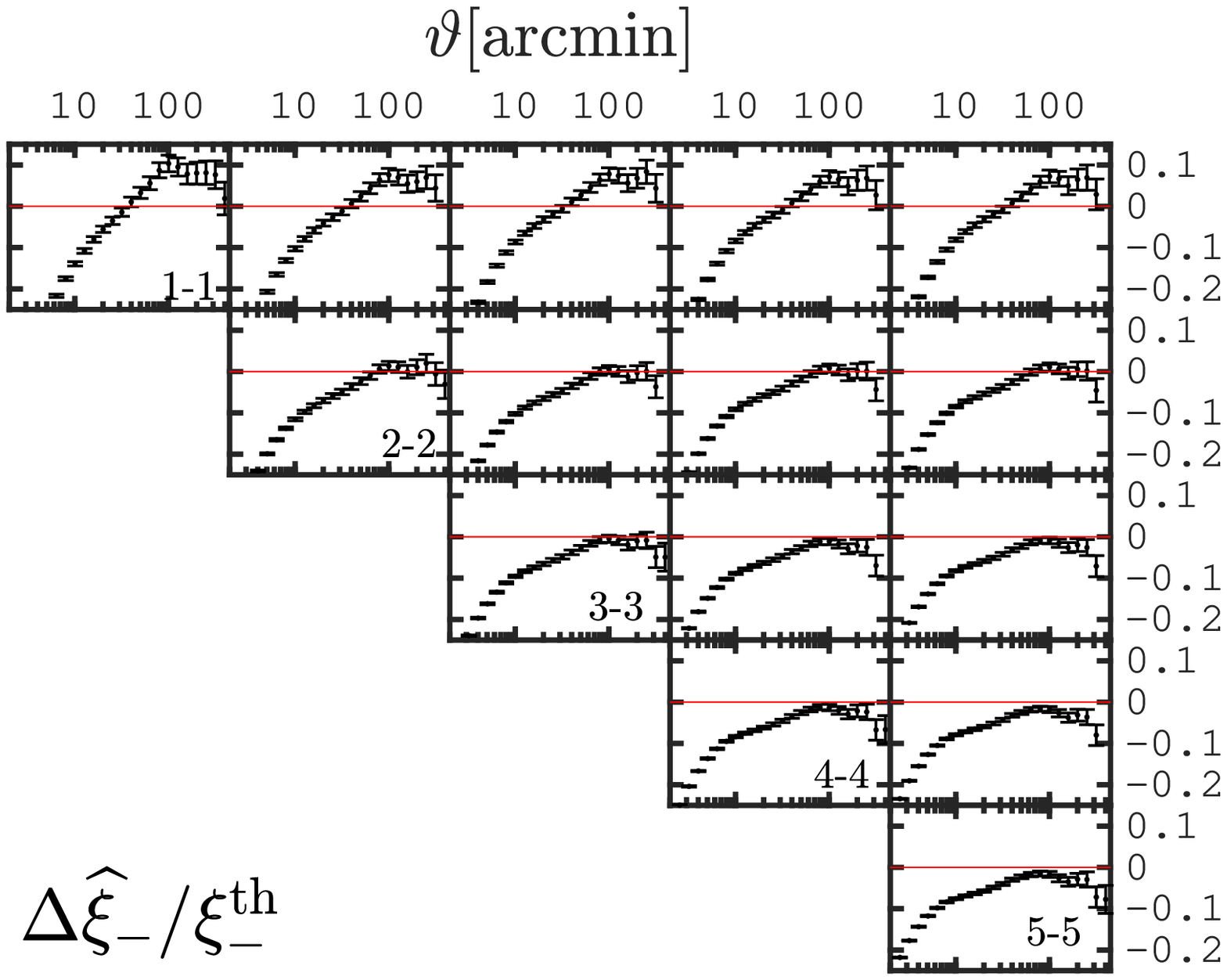}
\caption{Cosmic shear measured from the first 5 tomographic bins of the LSST-like source mocks, ignoring shape noise.
The $y$-axis shows the fractional difference between the measurements of $\xi_+$ (left) and $\xi_-$ (right) from the mocks and the predictions 
obtained from {\small NICAEA} with the input cosmology and $n(z)$.
The $x$-axis shows the angular separation $\vartheta$ in arcmin.
Error bars show the error about the mean, and the tomographic bins are labeled on the sub-panels. 
}
\label{fig:lsst_xip}
\end{center}
\end{figure*}

Following the same procedure as for the mock KiDS-450 source galaxies described in Section \ref{subsec:V0}, we produce mock galaxy catalogues with LSST-like specifications, based on forecasted survey specification from \citet{LSSTgal}:
\begin{eqnarray}
n_{\rm lsst}(z) = z^{\alpha} {\rm exp}\bigg[- \bigg(\frac{z}{z_0}\bigg)^{\beta} \bigg]
\end{eqnarray}
with $\alpha  = 1.25$, $\beta = 1.0$ and $z_0 = 0.5$, and assuming a galaxy number density of 26 gal/arcmin$^2$. 
We split this distribution in ten tomographic bins of equal number density, we convolve each of these with a Gaussian function that varies with redshift, 
i.e. $\sigma = \sigma_z(1+z)$ where $\sigma_z = 0.02$, and we finally truncate these distributions such that data lies in the range $z \in [0.1 - 3.0]$.
 The resulting tomographic distributions are shown in Fig. \ref{fig:lsst_nz}.

We compute the shear two-point correlation function from these mocks using Eq. \ref{eq:xi_estimator}, and the results are shown in Fig. \ref{fig:lsst_xip}
for all combinations involving the first 5 tomographic bins, and without shape noise.
We recover the results presented in Section \ref{subsec:V0} and in HvW15, namely that the angular scales comprised in the range [1-50] arcmin
in $\xi_+$ are generally modelled to better than 5\%, however smaller scales suffer from limits in particle mass resolution, while 
large scales are affected by the finite simulation box size.


\subsection{Source galaxies with clustering at fixed bias}
\label{subsec:V04}


 
 \textcolor{black}{In addition to the random position and HOD approaches we have produced mock galaxy catalogues in which the position of the galaxies trace the underlying dark matter with a controlled bias.  We do this by sampling the projected 2D density mass sheets $\delta_{2D}(\chi_l, \bf{\theta})$ at random such that the density distribution of galaxies in each redshift slice is proportional to the mass distribution projected within the slice.  This has the advantage that it contains lens clustering, but the bias is a controlled parameter, as opposed to being redshift, scale and mass dependent.  This is helpful when comparing measurements to theoretical models that assume linear bias 
 \citep[see H17 and ][for two applications of these mock data]{2017arXiv170605004V}.}

\subsection{Source galaxies with positions set by data }
\label{subsec:V1}

We have developed another type of mock catalogues also based on the SLICS light cones, 
and in which the position of the galaxies exactly match those of the KiDS-450 data.
The prime application of this approach is to reproduce the observed variation in source density,
which modulates the local noise properties and affect statistics such as weak lensing peak counts
\citep[for a detailed discussion on the importance of this, see][]{2017arXiv170907678M}.

\begin{figure}
\begin{center}
\includegraphics[width=3.2in]{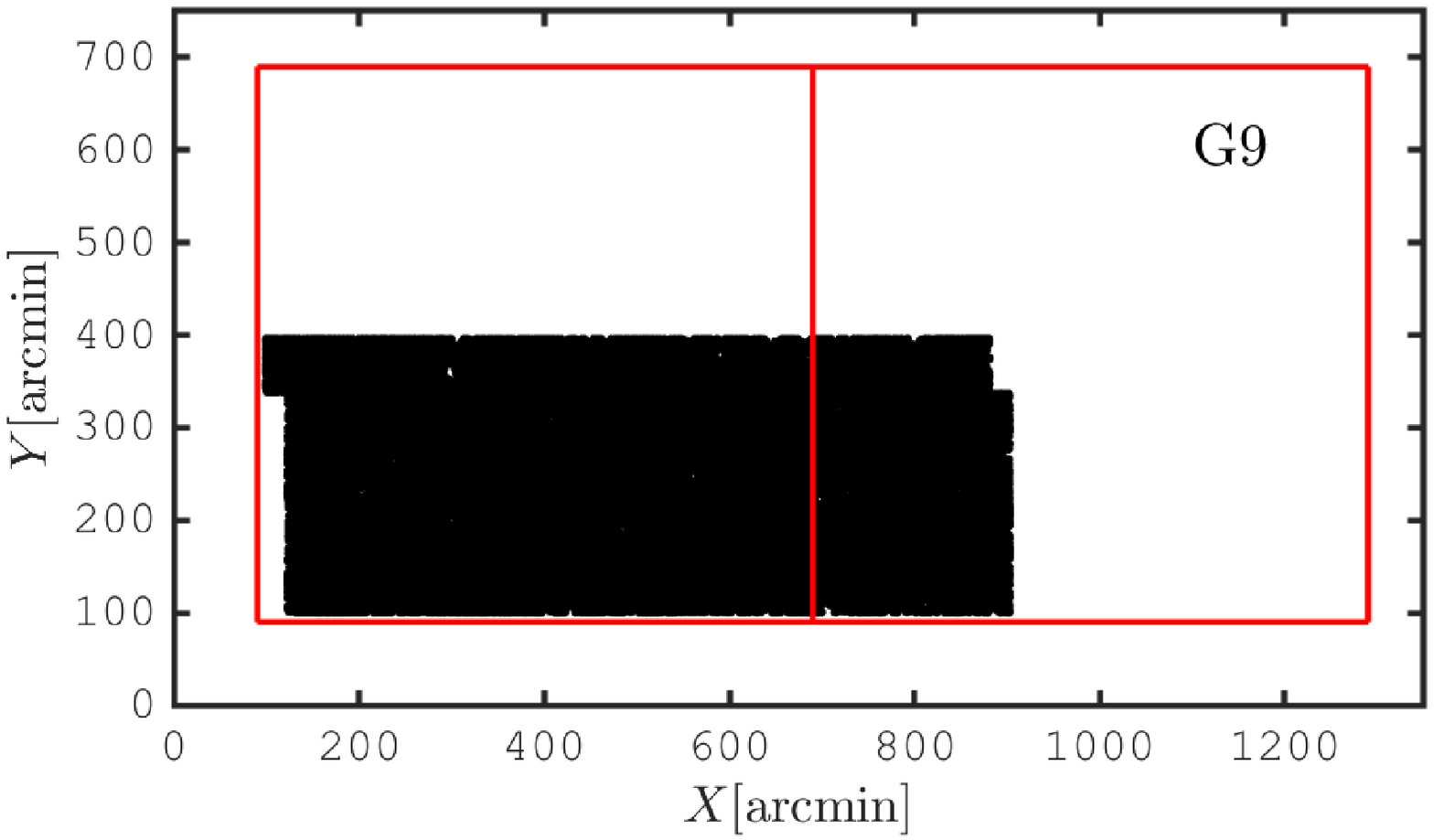}
\includegraphics[width=3.2in]{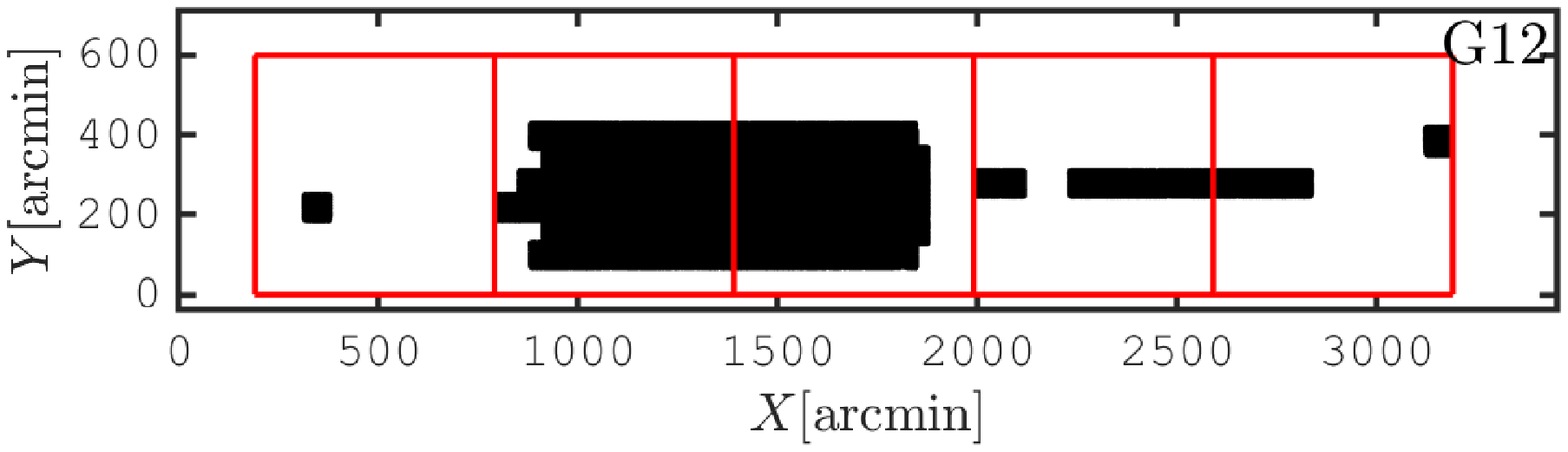}
\includegraphics[width=3.2in]{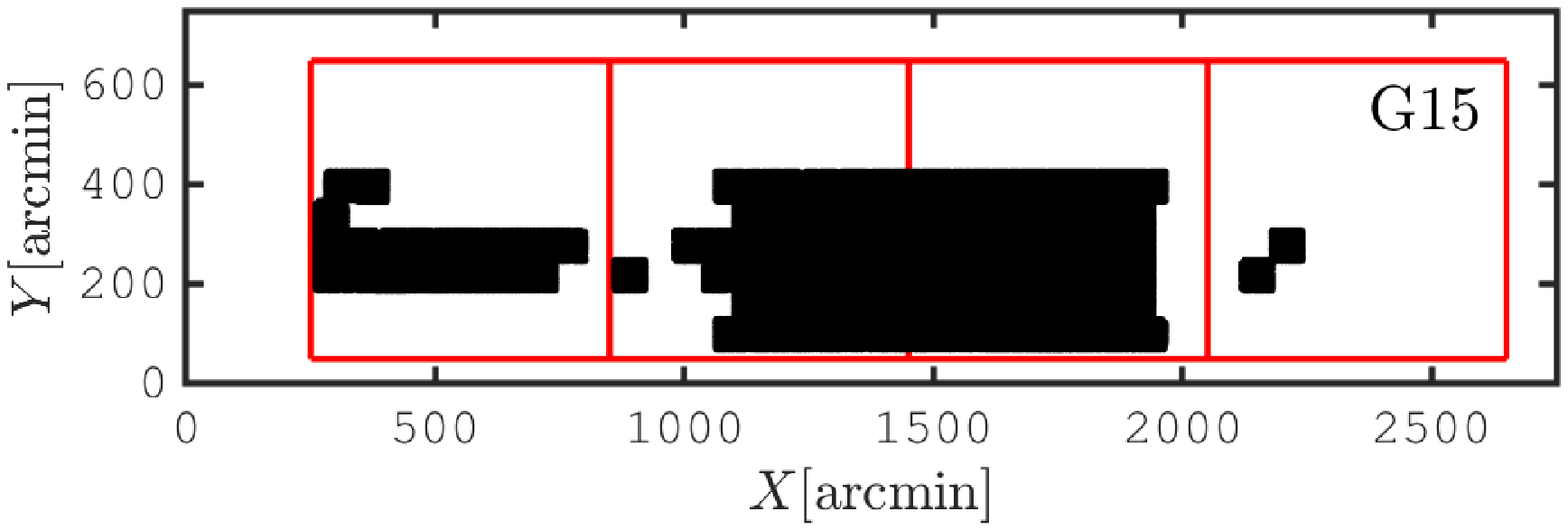}
\includegraphics[width=3.2in]{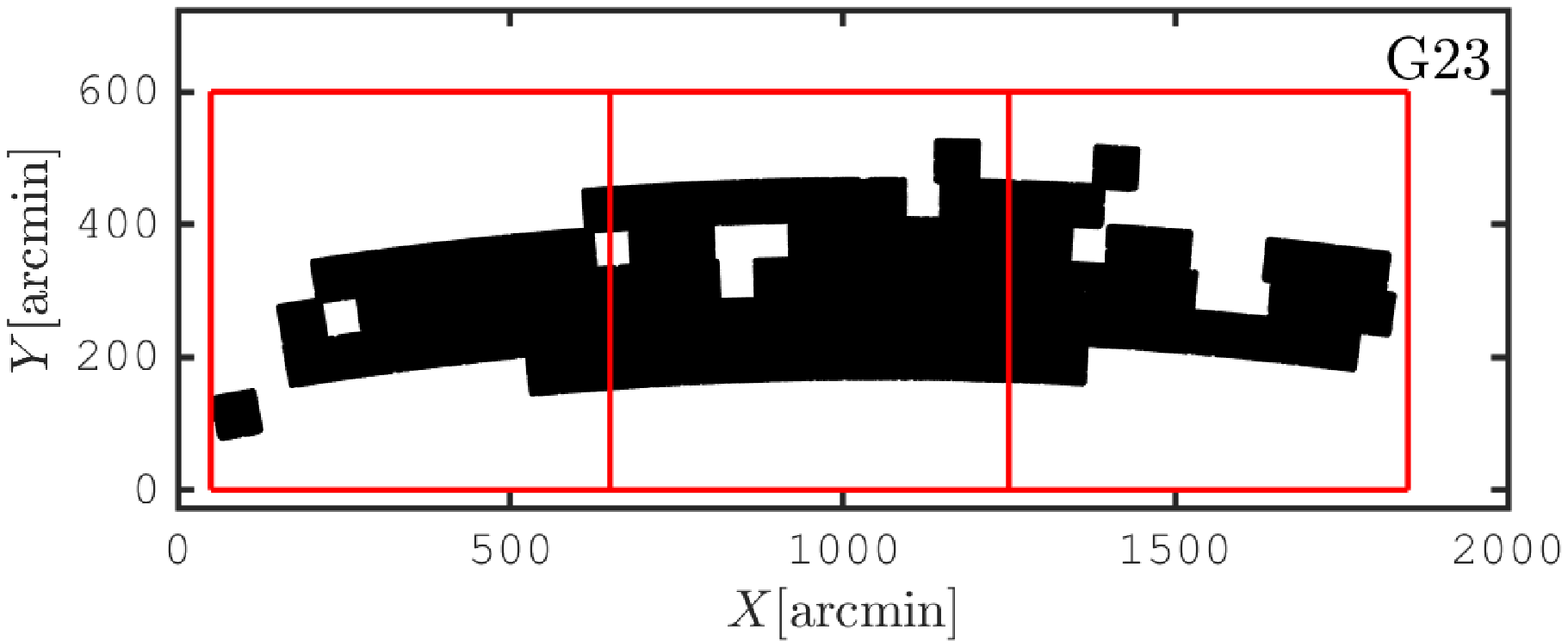}
\includegraphics[width=3.2in]{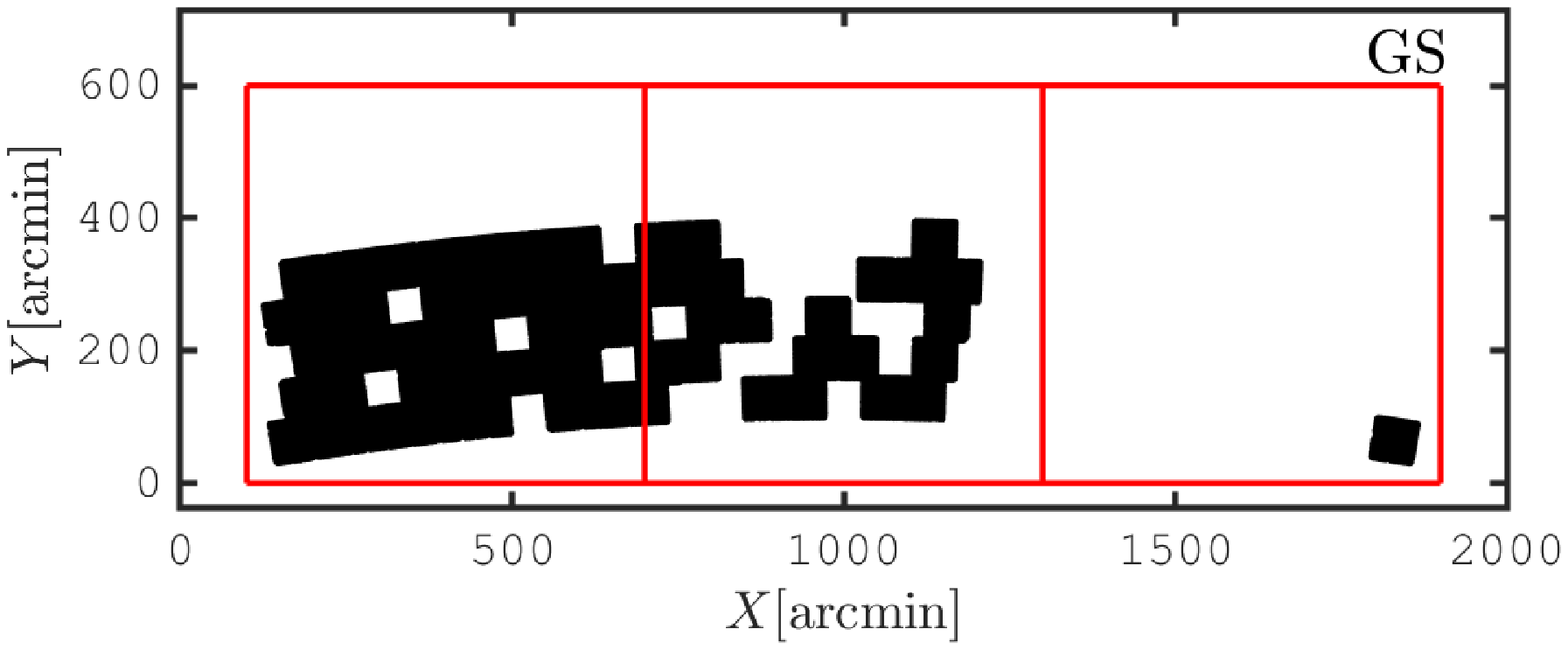}
\caption{Tiling configuration of the 17 SLICS simulations onto the 5 KiDS patches. The red squares represent the area of individual light cone. 
The axes are in the $xy$ coordinate frame of the masks, in units of arcmin. Each black point corresponds to a galaxy in the KiDS catalogue. }
\label{fig:tiling}
\end{center}
\end{figure}

Since we cannot capture all the data in one light cone, we break the observed sky coverage into 100 square degree patches,
and tile the mocks into a mosaic, as illustrated in Fig. \ref{fig:tiling}. 
The five KiDS-450 fields are decomposed into 17 `mock regions', shown as red boxes.
Unfortunately, the KiDS mosaic is not efficiently decomposed into $10\times10$ regions, 
which is why these 450 deg$^2$ of data take significantly more than 4.5 mock light cones to be covered.
However, many mock regions contain very little data and we could recycle some unused coverage,
keeping this to a minimum in order to avoid unphysical correlations. 
After this tiling technique, the position of every galaxy in the KiDS-450 data passing a $0.1 <Z_{\rm B} <0.9$ cut is matched to a pixel in one of 13 SLICS light cones,
organized into 17 regions. \textcolor{black}{To be clear, there is no correlation between the location of these mock galaxies and the large scale structure from the mocks.}

The next step is to assign a shear to these objects, which requires knowledge of their redshift in the simulation.
To achieve this, we draw a $z_{\rm spec}$ value from the DIR $n(z)$ and use this redshift to interpolate the two shear components from the shear planes
described in Section \ref{subsec:planes}, at the pixel location.
We  include in the mock the original coordinates of the galaxy, the coordinate in the mock light cone, 
the original observed ellipticity, the shear extracted from the SLICS, the $Z_{\rm B}$ and $z_{\rm spec}$ redshifts, as well as the shape weight
and the Field ID. These quantities, summarised in Table \ref{table:cat_V1}, are all required by peak statistic analyses such as the one carried out in \citet{2017arXiv170907678M}.
\textcolor{black}{We generated with that method a total of 67 independent mock replicas of the KiDS mosaic, based on tiling 871 SLICS light cones.
Note that the $n(z)$ is similar but not exactly identical to that of the `main' KiDS-450 sample (described in Section \ref{subsec:V0}), 
causing variations of order 10 percent on the cosmic shear signal.}

It is important to note that the simulated data contained in different mock regions (the red boxes in Fig. \ref{fig:tiling}) are not correlated, as they originate 
from different light cones. In contrast, these correlations exist in the data, which means that care must be taken to avoid 
being affected by this difference. For example, one should not compute correlation functions on the full mock mosaic, otherwise
the broken correlation across the regions will result in a significantly lower signal, compared to both the data and the predictions.
Instead, analyses should be carried out within the individual mock regions.
\textcolor{black}{ The peak statistics analysis described in \citet{2017arXiv170907678M} is protected against this, 
since the shear peaks are found from an aperture mass algorithm that works on individual camera pointings that each cover about 1 deg$^2$.
}


\begin{table}
   \centering
       \caption{Additional content of the mock galaxy catalogue at the KiDS-450 galaxy positions -- all columns from the KiDS-450 source catalogues described in Table \ref{table:cat_columns} are included as well.
       The XY are in the coordinate frame of the mask, and related to the RA-Dec with the {\sc WCSTools} sky2xy or xy2sky. 
       }

   \begin{tabular}{ccc} 
  \hline
 content & units & description \\
\hline
    X &  &  \multirow{2}{*}{$\Bigg\}$ sky coordinates}\\
    Y &    & \\
    w & & {\it lens}fit weight from the data \\
   \tt FieldPos & &  telescope pointing\\
  \hline
    \end{tabular}
   \label{table:cat_V1}
\end{table}

\section{More details on the GAMA HOD}
\label{sec:GAMA_HOD}

\begin{figure}
\begin{center}
\includegraphics[width=3.2in]{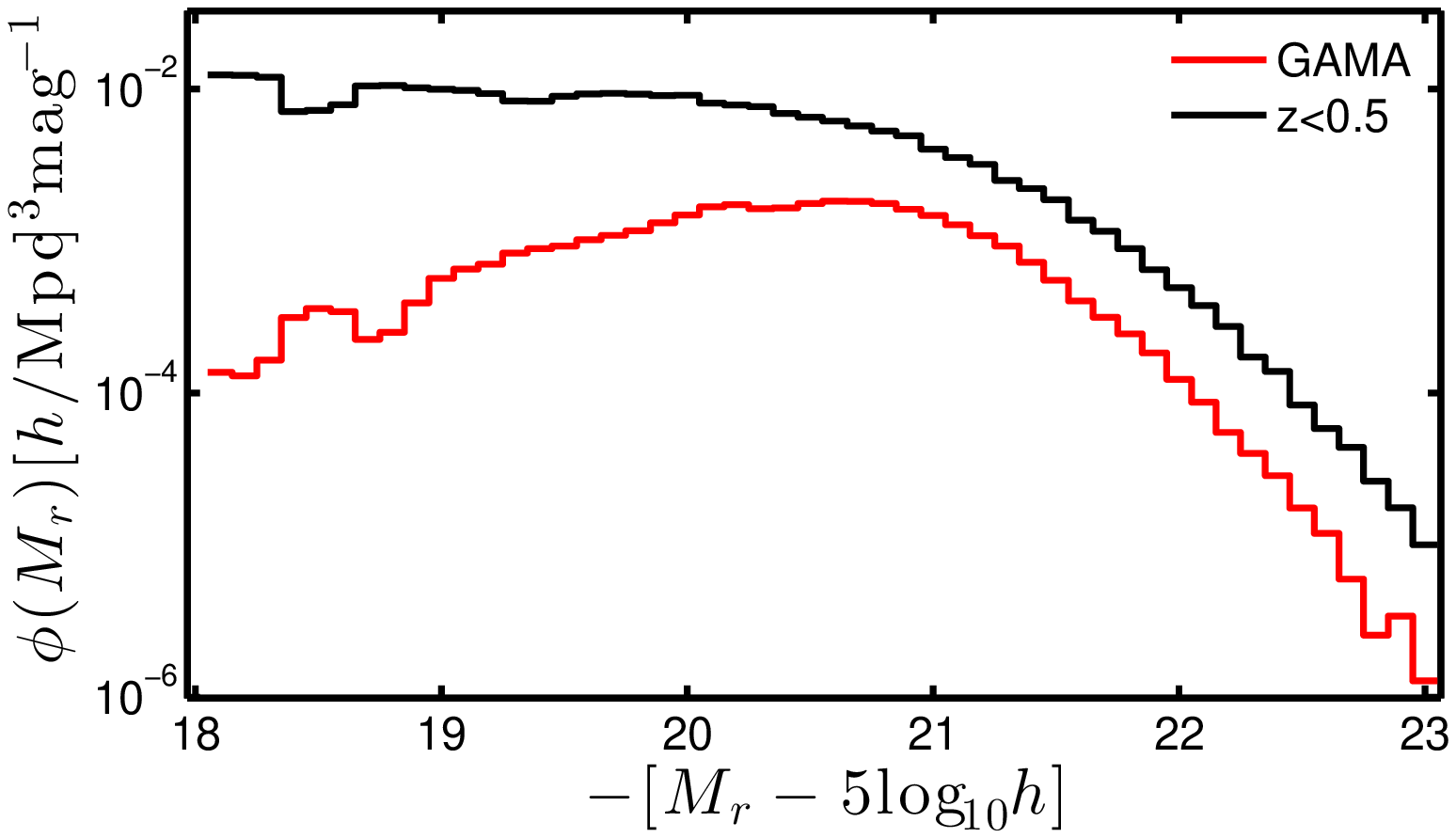}
\caption{Luminosity function of the GAMA mocks that includes redshift evolution of the HOD and $K$-correction. 
The black line represents the effect of removing the $m_r<19.8$ requirement, keeping otherwise all galaxies up to $z=0.5$.}
\label{fig:GAMA_CLF}
\end{center}
\end{figure}

We describe in this Appendix the ingredients that allow us to model the GAMA mock survey including the redshift- and luminosity-dependence of the HOD parameters. We closely follow the modelling of \citet{2017arXiv170106581S}, 
but include some details relevant to this mock production. This HOD is also used in the production of the KiDS-HOD and LSST-like HOD mocks, 
described in Section \ref{subsubsec:KiDS} and \ref{subsubsec:LSST-HOD}, respectively.
First, as noted explicitly in Eq. \ref{eq:GAMA_L_M}, the relation between luminosity and halo mass
changes with redshift, and its evolution is characterised by the parameter $Q$. Second, the dependence on luminosity 
requires the construction of relations between $L$,  $M_{\rm min}$ and $M_{1}'$, which are given by the same 
functional form as Eq. \ref{eq:GAMA_L_M}, but replacing some terms.
To establish the $M_{1}'(L)$ relation, we replace ($M_{\rm h}, L_\star A_{\rm t}, M_{\rm t}, \alpha_{\rm M} $) by ($M_{1}'(L), 3.70\times 10^9 ~h^{-2}L\odot, 4.78\times10^{12}~h^{-1}M_{\odot}, 0.306$), while for the $M_{\rm min}(L)$ we replace them by ($M_{\rm min}(L), 3.92\times 10^9 ~h^{-2}L_\odot, 3.07\times10^{11}~h^{-1}M_{\odot}, 0.258$).

Following the scaling relations from \citet{2017arXiv170106581S}, we next include a luminosity dependence of  $M_{0}(L)$, $\alpha(L)$ and $\sigma_{{\rm log}_{10}M}$ as:
\begin{eqnarray}
M_0(L) = 10^{1.78L - 5.98}
\end{eqnarray}
\begin{eqnarray}
\alpha(L) = {\rm log}_{10}\bigg[(0.0983L)^{80.3} + 10.0\bigg]
\end{eqnarray}
\begin{eqnarray}
      \sigma_{{\rm log}_{10}M} = 0.0258 + \frac{0.655}{1.0 + 2.5{\rm exp}\big[M_r + 21.05\big]}
\end{eqnarray}
After inspection, it is hard to reconcile this prescription with the satellite number in the data,
and we notice how the fit for $M_0$ in figure 4 of \citet{2017arXiv170106581S} is inaccurate at the faint end,
which otherwise best matches our observations. We improve the match by dividing the resulting $M_0$ by 100.0. 
Similarly, we also divide $M_{\rm min}$ by 50.0 to bring 
more galaxies into our selected sample and improve the clustering agreement.
The redshift evolution of the HOD is finally obtained by multiplying the three mass parameters $M_{0}(L)$, $M_{1}'(L)$ and $M_{\rm min}(L)$
by the function $f_z(M_r)$, which we extract from figure 6 of \citet{2017arXiv170106581S}. We interpolate the value of this function at the redshift of the host halo
when assigning galaxies to it.

The SLICS GAMA mocks do not include the hybrid SDSS/GAMA luminosity function described in  \citet{2017arXiv170106581S},
and our $K$-correction differs from their Table 1 as well. Our approach is instead to combine the uncertainty on the redshift evolution 
into an empirical $K$-correction that we apply to the mocks and fit to the  $K$-corrected data. Modelling the correction term $k(z)$ as:
\begin{eqnarray}
      k(z) = a_0 z^4 + a_1 z^3 + a_2 z^2 + a_3 z + a_4 \mbox{\hspace{2mm} and }  \hspace{2mm} m_r(z) = m_r + k(z),
\end{eqnarray}
we find $(a_0, a_1, a_2, a_3, a_4) = (-9.0, 8.4, 0.8, -1.5, 0.15)$.     
 This $K$-correction is applied to the apparent magnitude of every galaxy as a function of its spectroscopic redshift, 
 which shifts higher redshift galaxies to brighter apparent magnitude.
 This provides a better fit to the data when a  magnitude cut enters in the selection function. 
 The underlying ($K$-corrected) luminosity function is presented in Fig. \ref{fig:GAMA_CLF}, which matches reasonably well with the results from \citet[][their figure 9]{2017arXiv170106581S}.

\section{Ray-tracing vs. clustering coordinates}
\label{subsec:coordinates}

\begin{figure*}
\begin{center}
\includegraphics[width=3.5in]{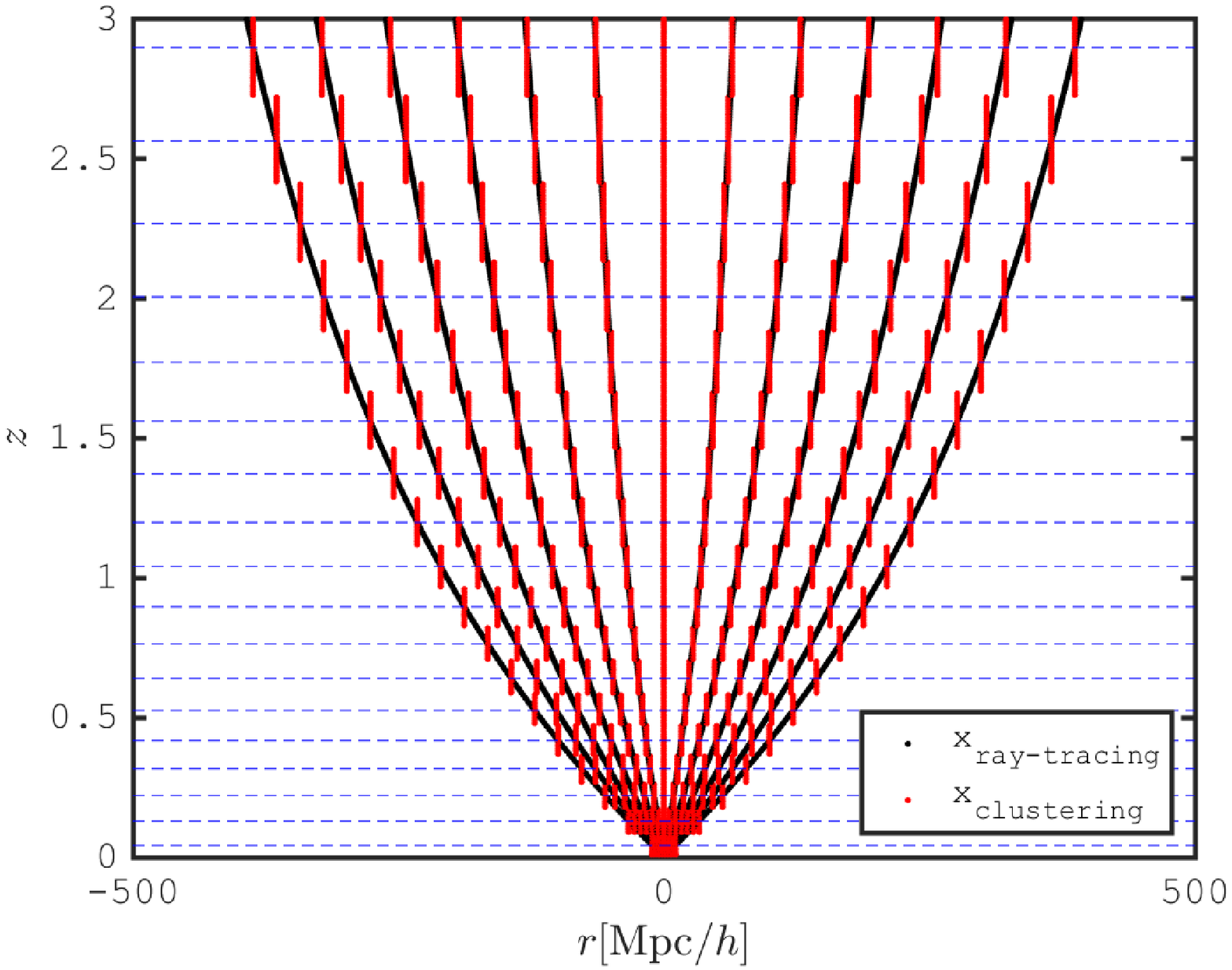}
\includegraphics[width=3.5in]{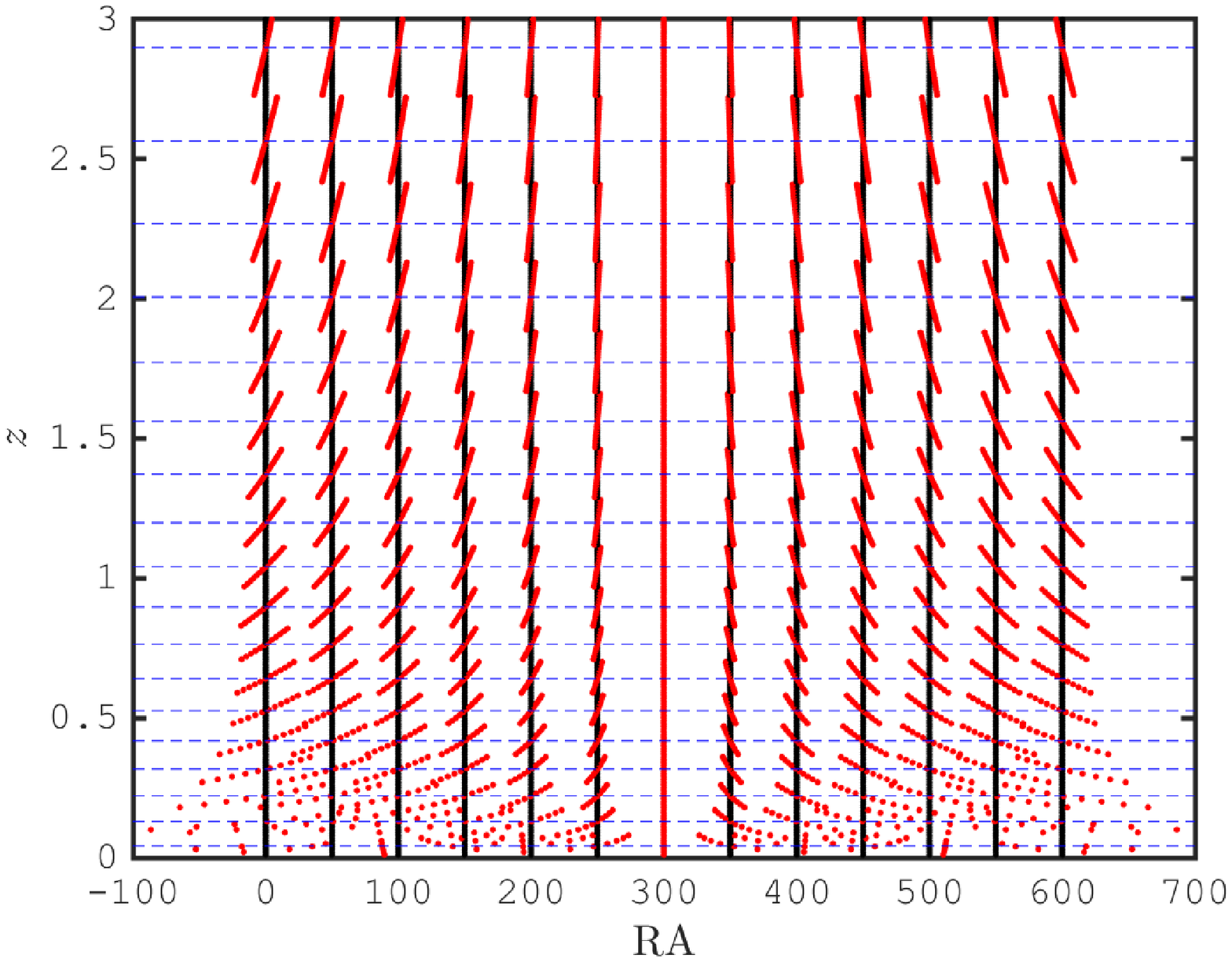}
\caption{Illustration of the two coordinate systems needed by the mock data. 
({\it left:}) The data points in each red `stick' represent example of objects that share the same ${\boldsymbol \theta}_{\rm ray-tracing}$ coordinates,
even though their ${\boldsymbol \theta}_{\rm clustering}$ coordinates differ. 
In the multiple lens technique, the photon trajectories descend along red sticks that are connected by a common black line.  
They are then assigned to pixels with coordinate ${\boldsymbol \theta}_{\rm ray-tracing}$, traced by the black lines.
In the far field limit, the red and black align.
 ({\it right:})  The ${\boldsymbol \theta}_{\rm clustering}$ coordinate of the same red points, as seen from lines of constant 
${\boldsymbol \theta}_{\rm ray-tracing}$. In this frame, the ${\boldsymbol \theta}_{\rm clustering}$ coordinates extend outside the $10\times10$ deg$^2$ patch. }
\label{fig:RA_correction}
\end{center}
\end{figure*}

\begin{figure}
\begin{center}
\includegraphics[width=3.2in]{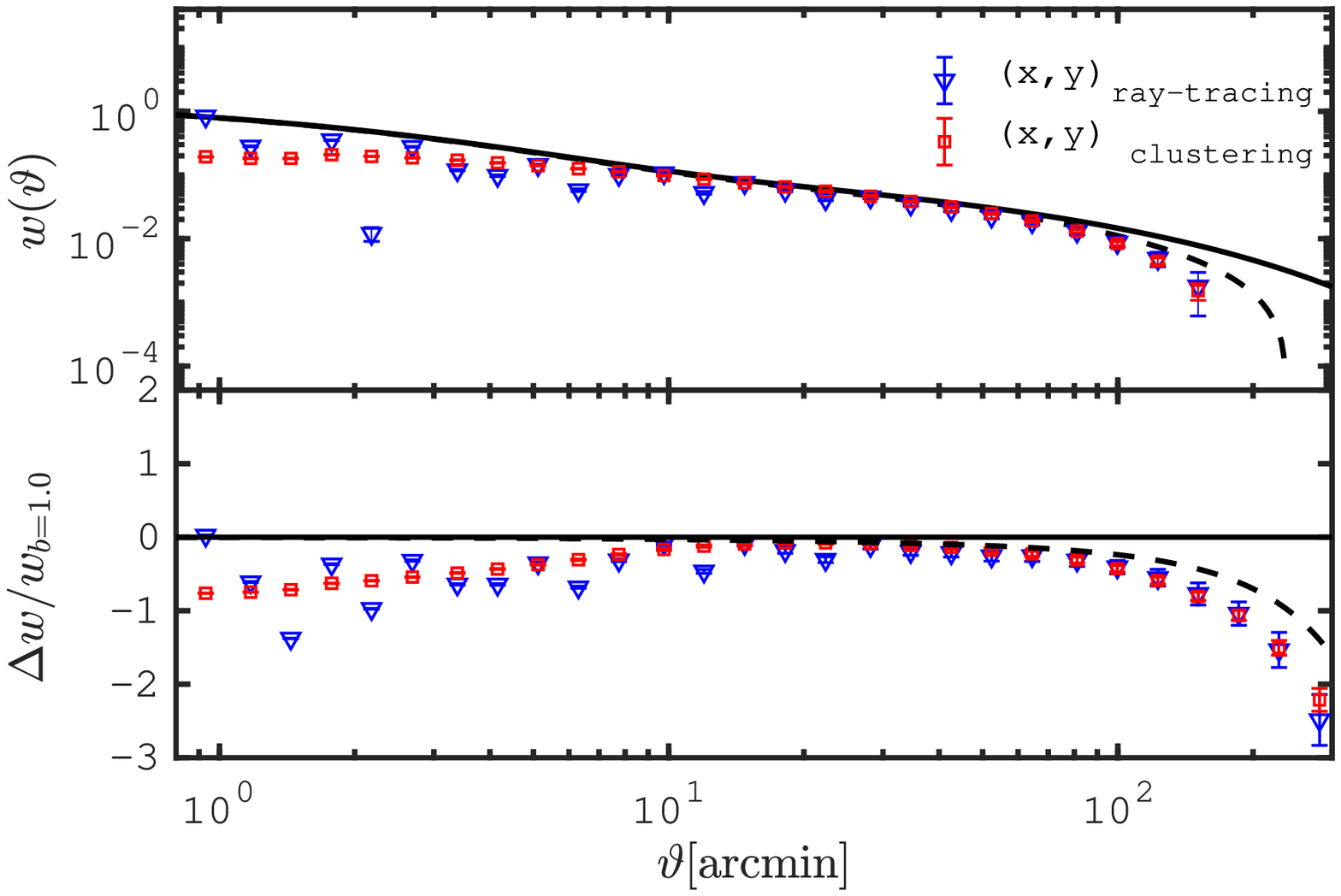}
\caption{Same as Fig.  \ref{fig:w_theta_halo}, but here including the measurements from the two coordinates described in the main text:
${\boldsymbol \theta}_{\rm clustering}$  (red) and ${\boldsymbol \theta}_{\rm ray-tracing}$
(blue).   Clustering measurements must use the former, lensing measurements the latter.
}
\label{fig:w_theta_halo_ray}
\end{center}
\end{figure}

As mentioned earlier, the SLICS mocks are based on the flat sky multiple plane geometry (described in Section \ref{subsec:planes}),
which is an excellent approximation for current cosmic shear analyses that probe the lensing signal out to angular scales as large as 10 degrees.
By construction, the cosmological volume that contributes to a pixel ${\boldsymbol \theta}$ in the $i^{\rm th}$ mass plane $\delta_{\rm 2D}(z^i_{\rm l},{\boldsymbol \theta})$ comes from the projection of half the simulation box (with thickness $L_{\rm box}/2 = 256.5 ~h^{-1}{\rm Mpc}$) along one of the Cartesian axis.
Assuming the flat sky and far-field limits, this axis is therefore identified as the radial direction and used thereafter in the assignment of both redshifts and comoving distances
for haloes and galaxies living in the light cones. This is no longer accurate for near field objects, or for projected quantities involving less than five parallel planes,
especially when looking at clustering of these low redshift lenses, and requires a correction that we describe here.
 Since we know the exact three-dimensional position of each halo and galaxy from the simulation, we can compute the correct angular coordinates of the objects
 (i.e. projecting radially, not along a Cartesian axis), and store these quantities as well.

For the sake of precision, there is thus a need for two coordinate systems to describe the lenses in our simulations. 
We define the `ray-tracing' coordinate, or ${\boldsymbol \theta}_{\rm ray-tracing}$, as the
mass projection coordinate. That is, all objects that contribute to the same pixel in the mass map (or shear map) share the same 
${\boldsymbol \theta}_{\rm ray-tracing}$ coordinate. Their true coordinate, which we refer to as the  `clustering coordinate',
or ${\boldsymbol \theta}_{\rm clustering}$, can be significantly different on account of the differences in the projection, especially for lower redshift objects. 
This is illustrated by the left panel of Fig. \ref{fig:RA_correction}. 
The thin horizontal lines represent the 18 lens planes listed in Table \ref{table:redshifts},
each subtending 10 degrees and 7745 pixels in both direction.
The vertical red `sticks' show how volume elements are projected at their centres, as part of the mass plane construction.
These red sticks represent the clustering coordinates, sampled at 13 angles, and clearly show the  discontinuities\footnote{Because of these discontinuities in redshift, the full three-dimensional correlation is broken across these boundaries, which will affect three dimensional clustering measurement such as $\xi(r)$ or $w(r_{\rm p})$.
This does not prevent the application of the SLICS to such data analyses, but might shape the data vector such as to impose similar selection cuts 
in the data.} that occurs between the mass planes. 

The black lines in the left panel of Fig. \ref{fig:RA_correction} show  the ${\boldsymbol \theta}_{\rm ray-tracing}$ coordinates of the same objects, 
which are continuous at all redshifts. 
These coordinates are not physical, and rather serve as a label that connects haloes with mass sheets.
Note that both coordinate systems coincide on the lens planes and at the very centre of the light cone. 
Their difference increases for objects that approach the edges of the light cone,  the junction redshifts,
and at lower redshift in general. 
We show in the right panel of Fig. \ref{fig:RA_correction} the ${\boldsymbol \theta}_{\rm clustering}$ and ${\boldsymbol \theta}_{\rm ray-tracing}$ coordinates of the same red sticks, but as seen in the ${\boldsymbol \theta}_{\rm ray-tracing}$ frame.
The black curved lines from the left panel become straight lines of constant RA, while the large differences between the two 
coordinate systems become even more apparent.



 We emphasise again that 
 ${\boldsymbol \theta}_{\rm clustering}$ corresponds to the actual position of the object in the simulation, and hence should be used for clustering measurements such as $w(\vartheta)$, $w(r_{\rm p})$, void-finding, etc. In contrast, ${\boldsymbol \theta}_{\rm ray-tracing}$ traces
 the projection used in the making of the mass sheets and should be used for lensing measurements ($\gamma_{\rm t}$, $\xi_{\pm}$, etc.).
As an example, we show in Fig. \ref{fig:w_theta_halo_ray}  the angular correlation function   $w(\vartheta)$ of all redshift $z \equiv 0.22$ haloes, previously presented in Fig. \ref{fig:w_theta_halo}. 
For this measurement to be accurate, it is critical to have random catalogues that properly capture the properties 
of the survey in absence of clustering. We discuss this further in the context of our light cone geometry in Section \ref{subsubsec:randoms}.
Shown in red is the clustering measurement from ${\boldsymbol \theta}_{\rm clustering}$, i.e. at their correct positions, 
compared with theoretical predictions that assume  a bias of 1.0.
In black is the same measurement carried out with ${\boldsymbol \theta}_{\rm ray-tracing}$ instead, which shows
clear unphysical features. 
This illustrates the importance of using the correct column in the mocks.

For example, in a mock joint-probe analysis involving cosmic shear from the KiDS-450, galaxy-galaxy lensing from KiDS-450 combined with CMASS, 
and clustering of CMASS, the measurement would involve: 
\begin{itemize}
\item ${\boldsymbol \theta}_{\rm ray-tracing}$ in the KiDS-450 mocks for the cosmic shear measurement,
\item ${\boldsymbol \theta}_{\rm ray-tracing}$  in the KiDS-450 mocks and ${\boldsymbol \theta}_{\rm ray-tracing}$  in the CMASS mocks for the tangential shear measurement, and 
\item ${\boldsymbol \theta}_{\rm clustering}$  in the CMASS mocks for the $w(\vartheta)$ measurement.
\end{itemize}

To make this easy for the user, we provide both coordinates in our halo and galaxy catalogues.
We also include simple codes to switch between these two coordinate systems, made available with the simulation products.

\section{Flat Sky Approximation}

In this Appendix, we verify the validity of the flat sky assumption in the SLICS simulations. The three-dimensional coordinates of the galaxies/haloes 
in the simulation box are first given in Cartesian coordinates, then transformed into angles and redshifts.
In this process, the third Cartesian axis is assumed to be equivalent to the radial direction, which is only valid in the far field limit.
The two angles are not affected by this approximation, but the redshift is. For example, a galaxy located  at a large angle (for example at  $X = Y = 5$ deg) 
{\it and} very close to the front of the simulation box (for example  at 15 $h^{-1}$Mpc) appears at redshift $z(\chi=15 h^{-1}{\rm Mpc}) = 0.005$.
However, its true distance to the observer is $\chi$ =  15.11 $h^{-1}$Mpc, which is a sub-percent effect.
Moreover, only  a minor fraction of objects at very low redshifts will suffer from error larger than 1\% coming from the the flat sky approximation.

\begin{figure}
\begin{center}
\includegraphics[width=3.2in]{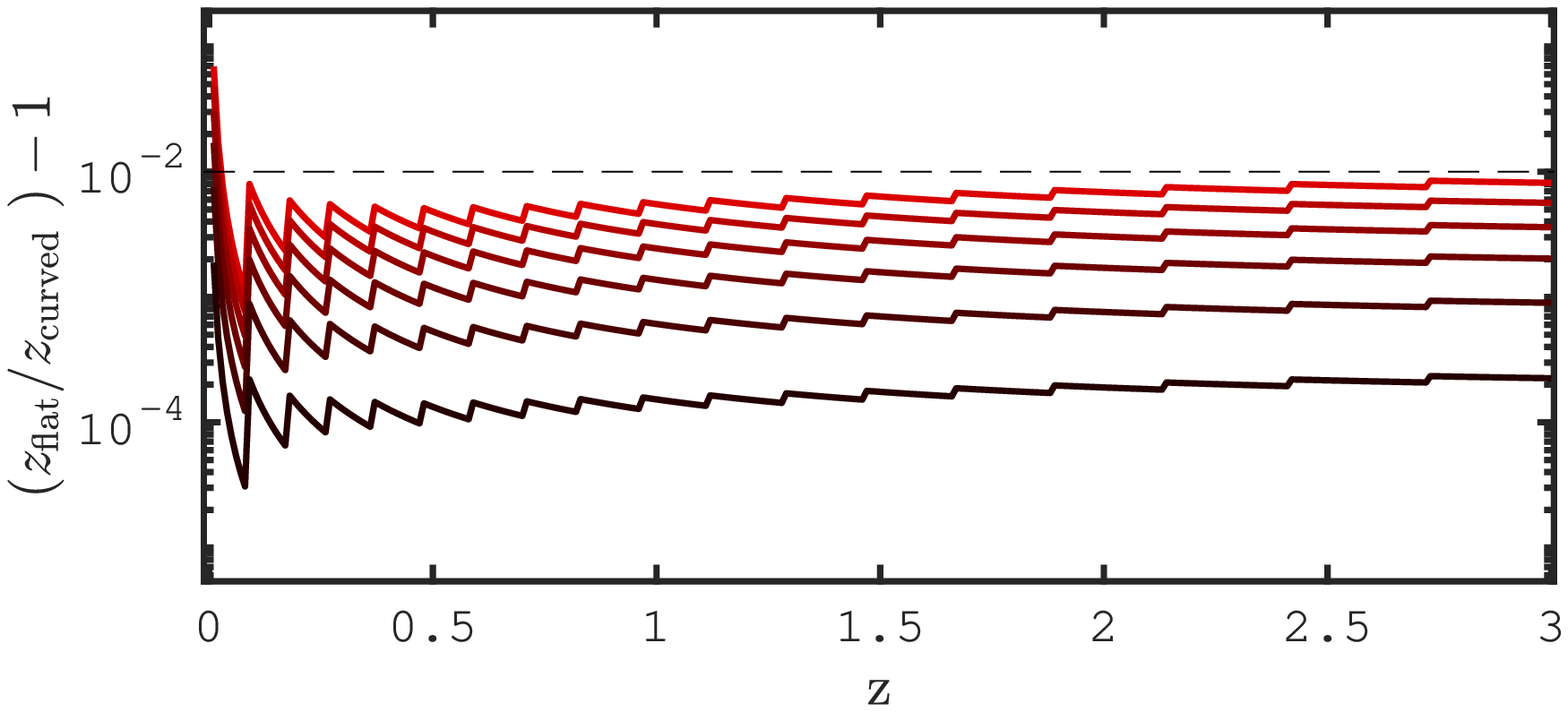}
\caption{Fractional error between flat sky and curved sky redshifts, for objects at different positions on the light cone. 
Objects shown with redder lines are closer to the edges of the simulation box, where the correction is more important.
The dashed line marks the 1\% error.}
\label{fig:FlatCurve}
\end{center}
\end{figure}

To show this, we  populate a light cone with a number of objects covering all angles and redshifts present in the mocks.
We then calculate the fractional effect of the approximation on the computed redshift at all these coordinates and show the results 
in Fig. \ref{fig:FlatCurve}. We recover that only the lowest redshifts are affected by this, which are heavily down-weighted in any lensing analysis,
hence conclude that this is not an issue for the science cases targeted by the SLICS simulations.

\label{lastpage}

\end{document}